%% file: main.tex
\documentclass[12pt, letterpaper, twoside]{article}

\usepackage[math,light]{iwona}
\usepackage[utf8]{inputenc}
\usepackage[T1]{fontenc}

\usepackage{geometry}
\usepackage{graphicx}
\usepackage{amsmath}
\usepackage{amssymb}
\usepackage{amsfonts}
\usepackage{bbm}
\DeclareMathAlphabet\mathbfcal{OMS}{cmsy}{b}{n}

\usepackage{tabularx,booktabs}
\usepackage{multirow}
\usepackage{todonotes}
\usepackage{enumitem}

\usepackage{fancyhdr}
\usepackage{lettrine}
\usepackage[explicit]{titlesec}
\usepackage{watermark}
\usepackage{color}
\usepackage[final]{pdfpages}
\usepackage{setspace}

\usepackage[sort, numbers, compress]{natbib}
\usepackage[nottoc]{tocbibind}

\usepackage{doi}

\usepackage[demo,abs]{overpic}
\newcommand{\igwithlabel}[3]{\begin{overpic}
  [#2]{#3}
  \put(5,10){\small #1}
\end{overpic}}

\usepackage{caption}
\usepackage{enumitem}
\setlist{nosep, label={\roman*)}, leftmargin=36pt, labelwidth=15pt, align=left, labelsep=3pt}

\numberwithin{equation}{section}
\numberwithin{figure}{section}
\numberwithin{table}{section}

\usepackage{afterpage}
\usepackage{bookmark}
\usepackage{hyperref}

\newcommand{\hlabel}{\phantomsection\label}

\definecolor{grey}{rgb}{0.5,0.5,0.5}
\definecolor{l-grey}{rgb}{0.8,0.8,0.8}
\definecolor{withe}{rgb}{1,1,1}
\definecolor{black}{rgb}{0,0,0}

\input{snp/nonumberSection.tex}

\newcommand{\subsubsubsection}[1]{\begin{center}
  \textit{#1}
\end{center}}

\newcommand{\quotes}[2]{
\vspace{-60pt}
\begin{minipage}{0.96\textwidth}
  \setstretch{1.0}
  \begin{flushright}
    {\it "#2"}

    #1
  \end{flushright}
\end{minipage}

\vspace{29pt}
}

\DeclareMathOperator{\tr}{tr}

\newcommand{\ket}[1]{\ensuremath{\vert{#1}\rangle}}
\newcommand{\ketw}[1]{\ensuremath{\vert{\widetilde{#1}}\rangle}}
\newcommand{\bra}[1]{\ensuremath{\langle{#1}\vert}}
\newcommand{\braw}[1]{\ensuremath{\langle{\widetilde{#1}}\vert}}
\newcommand{\braket}[2]{\ensuremath{\langle{#1}\vert{#2}\rangle}}
\newcommand{\braopket}[3]{\ensuremath{\bra{#1}{#2}\ket{#3}}}

\newcommand{\brawopketw}[3]{\ensuremath{\braw{#1}{#2}\ketw{#3}}}
\newcommand{\ketbra}[2]{\ensuremath{\vert{#1}\rangle{\!}\langle{#2}\vert}}
\newcommand{\ketopbra}[3]{\ensuremath{\vert{#1}\rangle{#2}\langle{#3}\vert}}
\newcommand{\expect}[1]{\ensuremath{\langle{#1}\rangle}}
\newcommand{\varian}[1]{\ensuremath{(\Delta #1 )^2}}
\newcommand{\varinv}[1]{\ensuremath{(\Delta #1 )^{-2}}}
\newcommand{\ver}[2]{\ensuremath{\genfrac{}{}{0pt}{}{#1}{#2}}}

\newcommand{\bs}[1]{\ensuremath{\boldsymbol{#1}}}
\newcommand{\dicke}[1]{\ensuremath{\textnormal{D}_{#1}}}
\newcommand{\bound}[1]{\ensuremath{\mathcal{B}_{\, #1 }}}
\newcommand{\coss}[1]{\ensuremath{\text{c}_{ #1 }}}
\newcommand{\sins}[1]{\ensuremath{\text{s}_{ #1 }}}
\newcommand{\tans}[1]{\ensuremath{\text{t}_{ #1 }}}
\newcommand{\abbrln}[2]{#1 & - & #2 \\}
\newcommand{\qfif}[1]{\qfi[{\textstyle #1}]}

\newcommand{\mydate}{March 02, 2017}

\def\be{\begin{equation}}
\def\ee{\end{equation}}
\def\mtxid{\mathbbm{1}}
\def\lpar{\left(}
\def\rpar{\right)}

\def\lcor{\left\{}

\def\qfi{\mathcal{F}_{\textnormal{Q}}}
\def\ghz{\textnormal{GHZ}}
\def\prob{\text{Pr}}

\def\db{\textnormal{dB}}

\begin{document}

\renewcommand{\thefootnote}{\fnsymbol{footnote}}

\pagestyle{fancy}
\renewcommand{\headrulewidth}{0pt}
\fancyhead{}
\fancyfoot{}
\pagenumbering{alph}
\bookmark[level=1,dest=title]{Cover}


\includepdf[pages=-, offset=72 -36]{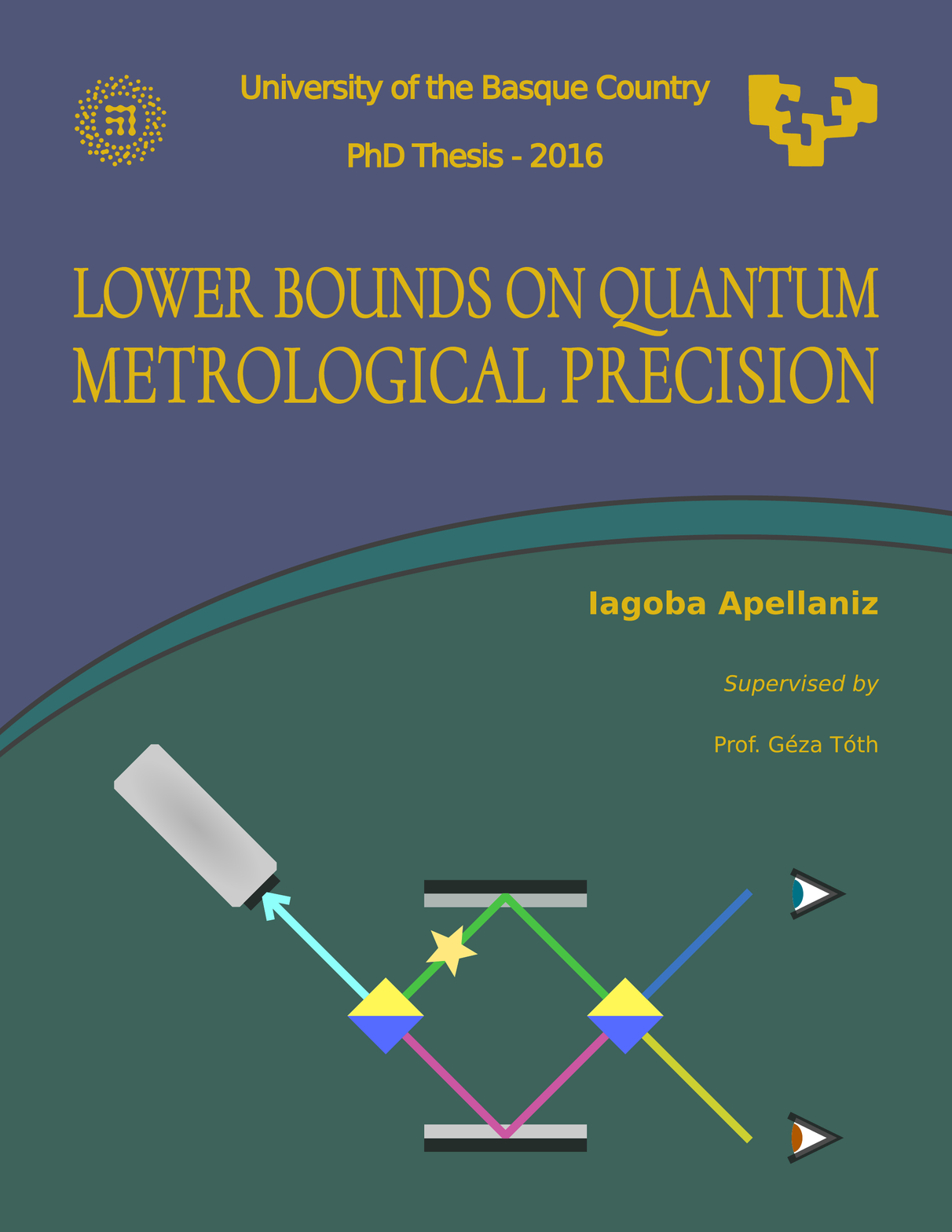}

\cleardoublepage

\vspace*{150pt}
\begin{center}
  \Huge
  Lower Bounds
  \vspace{20pt}

  on
  \vspace{20pt}

  Quantum Metrological Precision
\end{center}

\clearpage
\hypertarget{colophon}{}
\bookmark[level=1,dest=colophon]{Colophon}
\begin{center}
  Fisika Teorikoa eta Zientziaren Historia Saila\break
  Euskal Herriko Unibertsitatea (UPV/EHU)\break
  P. O. Box 644, 48080 Bilbao, Spain
  \vspace{10pt}

  Department of Theoretical Physics\break
  University of the Basque Country (UPV/EHU)\break
  P. O. Box 644, 48080 Bilbao, Spain
  \vspace{10pt}

  \textit{The \textnormal{\LaTeX}~source of the thesis with figures is at}\break
  \href{https://gitlab.com/i-apellaniz/PhD-Thesis}
  {https://gitlab.com/i-apellaniz/PhD-Thesis}
  \vspace{10pt}

  \textit{A PDF copy of this work can be found at}\break
  \href{https://gitlab.com/i-apellaniz/PhD-Thesis/tree/master/out}
  {https://gitlab.com/i-apellaniz/PhD-Thesis/tree/master/out}
\end{center}
\vfill

\begin{flushleft}
\noindent
This document was generated with the 2016 \LaTeX~distribution.\break
The plots and figures of this thesis were generated with Matplotlib,\break
and then edited in Inkscape and Blender.
\vspace{15pt}

\noindent
\includegraphics[height=20pt]{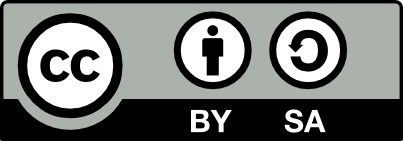}\break
2012-2017 Iagoba Apellaniz.
This work is licensed under the Creative Commons Attribution-ShareAlike 4.0 International License.
To view a copy of this license, visit\break
\href{http://creativecommons.org/licenses/by-sa/4.0/deed.en_US}
{http://creativecommons.org/licenses/by-sa/4.0/deed.en\_US}
\end{flushleft}
\clearpage

\input{00-0-prologe.tex}
\input{00-1-laburpena.tex}
\input{00-2-publications.tex}
\input{00-3-toc.tex}


\cleardoublepage

\pagenumbering{Roman}
\fancyfoot{}
\input{snp/title.tex}
\cleardoublepage

\pagenumbering{arabic}
\setcounter{page}{1}

\hypertarget{dedication}{}
\bookmark[level=1,dest=dedication]{Dedication}
\vspace*{100pt}
\begin{center}
\emph{To my parents, my family}

\emph{and to all the people}

\emph{I have had around these years.}
\end{center}

\cleardoublepage

\fancyfoot[LE,RO]{\thepage}
\phantomsection
\section*{Acknowledgments}
\addcontentsline{toc}{section}{Acknowledgments}

\lettrine[lines=2, findent=3pt,nindent=0pt]{I}{}would like to thank the people that has supported me in this endeavor.
Especially, I want to acknowledge my office and discussion mates, Giuseppe Vitagliano and Iñigo Urizar-Lanz as well as Phillip Hyllus, Matthias Kleinmann and Zoltán Zimborás.
A special thank to my director Géza Tóth for all the support, without whom my work at hand would not have been possible.
I would also like to thank more people from the department I belong to at the University of the Basque Country, the Department of Theoretical Physics and the History of Science.
I appreciate the effort done by people like Iñigo L. Egusquiza in promoting the Ph.D. program of the Department of Theoretical Physics, which has successfully formed plenty of researchers now pursuing cutting edge research projects world wide.
I would also like to acknowledge the Ph.D. students, named at the end, from different departments for the friendly atmosphere I could enjoy working here at \emph{our} university.

I have been visiting several research groups during these years as well as attending different conferences and workshops.
First, I would like to thank the TQO group at the University of Siegen as a whole, especially with whom I have discussed or collaborated, Sanah Altenburg, Sabine Wölk and Otfried Gühne.
Second, I would also like to thank the QSTAR group at the University of Florence, particularly Manuel Gessner, Luca Pezz\`e and Augusto Smerzi for inspiring discussions.
And finally, I would like to thank the people with whom I also discussed or collaborated a lot but have not had the chance to visit them, like Micha{\l} Oszmaniec, Jan Kolodynski, Morgan W. Mitchell and Antonio Acín from ICFO in Barcelona, and Bernd L\"ucke, Jan Peise, and Carsten Klempt from the University of Leibniz in Hannover.
\pagebreak

I take this opportunity to cheer all the rest I have in mind right now: Constantino, Roberta, Tobias, Frank, Roope, Nikolai, Mariami, Felix, Christina, Ali, Daniela, Marius, Jannik and Tristan which I met them in Siegen; Roberto, Fabian, Victor and Marco from Firenze; and last but not least Mikel, Pablo, Julen, Iraultza, Joanes, Jon, Laura, Peio, Borja, Sofia, Ariadna, Urtxi, Iñigo, Iker, Jo\~ao, Mattin, Olatz, Kepa, Carlos, Naiara, Diego, Unai, Paul, Maria, Irene, Raul, Lluc, Montse and Santi, from Bilbao.
I see us like the components of an $N$-dimensional vector, each one with its own direction, and at the same time pointing together as one!
Thank you guys for the nice and comfortable atmosphere!

\cleardoublepage


\renewcommand{\headrulewidth}{0.5pt}
\fancyfoot[LE,RO]{\thepage}
\fancyhead[LE]{\rightmark}
\fancyhead[RO]{\leftmark}

\input{snp/fancySection.tex}

\input{01-introduction.tex}
\input{02-background.tex}
\input{03-vicinity.tex}
\input{04-legendre.tex}
\input{05-gradient.tex}
\input{06-conclusions.tex}

\input{snp/nonumberSection.tex}
\fancyhead[RO]{APPENDICES}
\appendix
\input{appendix.tex}

\cleardoublepage
\fancyhead[RO]{\leftmark}
\phantomsection
\bibliography{biblio}

\null
\newpage
\renewcommand{\headrulewidth}{0pt}
\fancyhead{}
\fancyfoot{}
\null
\newpage
\pagenumbering{Alph}
\hypertarget{backcover}{}
\bookmark[level=0,dest=backcover]{Back-cover}
\includepdf[pages=-, offset=72 -36]{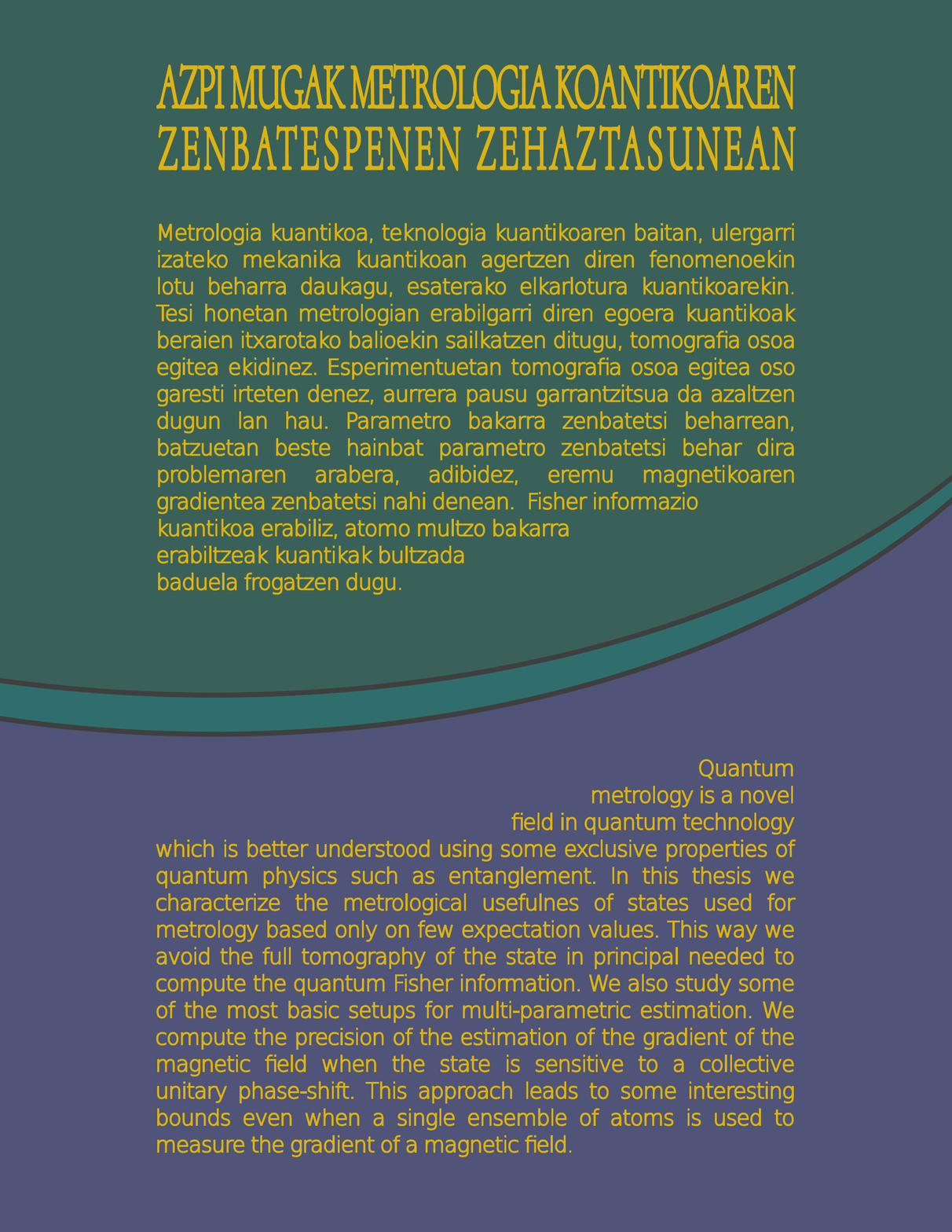}
\end{document}

%% file: snp/nonumberSection.tex
\titleformat{\section}[display]
{\vspace*{150pt}
\bf\Huge}
{{\textcolor{black}{\thesection}}. #1}
{0pt}
{#1}
[]
\titlespacing*{\section}{40pt}{10pt}{40pt}[40pt]

%% file: 00-0-prologe.tex
\hypertarget{prologue}{}
\section*{Prologue\hlabel{sec:pr}}
\bookmark[level=section,dest=prologue]{Prologue}
\setcounter{page}{1}
\pagenumbering{roman}
\fancyfoot[LE,RO]{\thepage}

\lettrine[lines=2, findent=3pt,nindent=0pt]{T}{his} work is based on the doctoral project on quantum information which I started in the summer of 2013.
It collects part of the research I have done during these years.
In page~\pageref{sec:la}, I include a brief summary of the thesis written in Basque.
It gave me the opportunity to explain what I did these past years in my tongue language.
I find it useful for the people I love most, my family and friends.

I will make an effort to be clear throughout the thesis in order to make it useful for an audience as wide as possible.
This way I hope it will be readable by any person with a bachelor in science, particularly in physics or at least with a partial knowledge in quantum mechanics.
With that in mind, the first and the second chapters will introduce the background to the reader as well as the basic notions of quantum metrology, the field our work belongs to.
Even though I write this thesis for a broad audience in mind, a basic notions in quantum physics and statistics is needed to follow it properly as I said before.
For instance, I will assume among other things that the reader knows what probability is and which its properties are, or what a quantum state is and what it represents.
I will give references where to find complementary material when necessary.

The thesis has been done within the research group in quantum information in which Prof. G\'eza T\'oth is the group leader and principal investigator.
I have to mention the rest of the members of the group Dr. Philipp Hyllus, Dr. Giuseppe Vitagliano, Dr. I\~nigo Urizar-Lanz, Dr. Zolt\'an Zinbor\'as and Dr. Matthias Kleinmann at the time I was working on the projects of this thesis.
Apart from the group of G\'eza T\'oth based in Bilbao, Spain, this thesis also collects some work done in collaboration with the Theoretical Quantum Optics group lead by Prof. Otfried G\"uhne at the University of Siegen, Germany, and the group of Prof. Carsten Klempt at the Leibniz University in Hannover, also in Germany.
The last one is an experimental group specialized on the creation of exotic quantum states with very many particles with a variety of applications in quantum technology.

In this thesis, first, we investigate the metrological usefulness of a family of states known as unpolarized Dicke states, which turn to be very sensitive to the magnetic field.
Quantum mechanics plays a central role in achieving such a high precision.
Second, we investigate possible lower bounds on the quantum Fisher information, a quantity that characterizes the usefulness of a state for quantum metrology, using the theory of Legendre transforms such that we obtain tight lower bounds based on few measurements of the initial quantum state that will be used for metrology.
And last but not least, we investigate gradient magnetometry, i.e., we develop a theory to study the sensitivity of some states of the change in space of the magnetic field.

\begin{flushright}
  Iagoba Apellaniz

  Bilbao, \mydate
\end{flushright}

%% file: 00-1-laburpena.tex
\afterpage{\hypertarget{laburpena}{}}
\section*{Laburpena\hlabel{sec:la}}
\bookmark[level=section,dest=laburpena]{Laburpena}

\lettrine[lines=2, findent=3pt,nindent=0pt]{L}{an} honek metrologia kuantikoaren baitan egindako hainbat ikerketa biltzen ditu tesi moduan.
Aurkezten dudan ikerlan hau, Zientzia eta Teknologia Kuantikoko masterra bukatu ondoren, azkeneko lau urte hauetan Prof. Géza Tóth irakaslearen lan-taldean burutua izan da.
Tesi honetan agertzen ez diren beste hainbat lan plazaratu ditugu nik eta elkarrekin lan egin dugun hainbat ikertzailek.
Publikatutako artikuluen lista tesi honen \pageref{sec:pu}.~orrialdean aurki daiteke.

Tesi honetan agertzen diren ikerketa lanak gauzatzeko ezinbestekoa izan da nazioarteko elkarlana.
Izandako elkarlanen artean Alemaniako Siegen hiriko unibertsitatean dagoen Otfried Gühnek zuzentzen duen TQO taldea dago.
Beste kolaborazio garrantzitsu bat Italiako Florentzian dagoen unibertsitateko Augusto Smerzik zuzentzen duen QSTAR taldea izan da.
Azkenik, Alemaniako Hannover hiriko unibertsitateko Carsten Klemptek zuzentzen duen ikerkuntza talde esperimental batekin izandako elkarlana azpimarratu nahi da.

\ref{sec:in}.~kapituluan teknologia kuantikoak eta metrologia kuantikoak duten garrantzia azpimarratzen da.
Teknologia kuantikoa prozesu kuantikoez baliatzen baita klasikoki lortu ezin diren hainbat helburu lortzeko.
Adibide gisa, ordenagailu kuantikoek hainbat posibilitate aldi berean aztertzeko izango luketen gaitasuna, edota simulazio kuantikoek modelo konplexu ezberdinak simulatzeko duten gaitasuna azpimarratzen dira.

Beste motatako teknologi kuantikoak alde batera utzita, metrologia kuantikoan oinarritzen da lan hau.
Metrologia kuantikoak zenbatetsi nahi diren parametroen errorea txikiagotzeko aukera ematen du metrologia klasikoarekin alderatuz gero.
Klasikoki diseinatutako aparailu batek $N$ proba egin ondoren errorea $\sqrt{N}$ aldiz txikitzea lortzen duen bitartean, metrologia kuantikoa erabiliz errorea $N$ aldiz txikitu daitekeela aski ezaguna da \cite{Giovannetti2004, Paris2009}.
Muga gaindiezin hauei "shot-noise scaling" deritze aparatu klasikoen kasurako eta "Heisenber scaling" egoera kuantiko orokorrek gainditzerik ez duenarentzako.

Ezaguna da baita ere elkarlotura kuantikoak, hau da, mekanika kuantikoaren propietatea eta klasikoki azalpenik ez duenak, zenbatespenean duen garrantzia.
Elkarlotura kuantikoa zenbatespena hobetzeko ezinbestekoa da, aldiz, elkarlotura kuantiko mota guztiak ez dute balio errorea txikitzeko.
Elkarlotura kuantikoak eta metrologia kuantikoak duten erlazioaren hainbat azterketa hurrengo erreferentzietan aurki daitezke \cite{Pezze2009, Louchet-Chauvet2010, Appel2009, Riedel2010, Gross2010, Luecke2011, Strobel2014, Hyllus2010}.

Honi guztiari azken urteotan metrologia kuantikoak piztu duen arreta gehitu behar zaio.
Kuantikak hobetutako metrologia erabiltzen da, adibidez, erloju atomikoetan \cite{Louchet-Chauvet2010, Borregaard2013, Kessler2014a}, zehaztasun handiko magnetometrian \cite{Wasilewski2010, Eckert2006, Wildermuth2006, Wolfgramm2010, Koschorreck2011, Vengalattore2007, Zhou2010}, edota uhin grabitazionalen detektagailuetan \cite{Schnabel2010, TheLIGOScientificCollaboration2011, Demkowicz-Dobrzanski2013}.

\ref{sec:bg}.~kapitulua metrologia kuantikoaren sarrera gisa uler daiteke.
Bertan estatistikan oinarritutako hainbat kontzeptu azaltzen dira.
Estatistika datuetatik ondorio ulerkorrak ateratzeko erabiltzen den zientzia matematikoa da.
Adibidez, datu lagin baten batezbestekoa kalkulatzeko erabiltzen den prozedura,
edota datu lagin baten bariantza kalkulatzeko erabilitako formulak azaltzen dira kapitulu honetan.
Datu lagin batek zenbatetsi nahi den parametroari buruzko informazioa izan lezake bere baitan.
Adibide bezala, pilota bat bosgarren pisutik jaurtitzerakoan lurra jotzeraino igarotako denbora neurtu da hainbat aldiz.
Denbora ezberdin guzti hauek erabiliz, bosgarren pisuraino dagoen altuera kalkula daiteke grabitateak pilotarengan duen eragina aldez aurretik ezaguna denean.

Estatistikaren baitan kokatzen da datu lagin batetik zenbatespena egiterakoan saihestezina den errorearen muga klasikoaren kalkulua.
Muga hau Fisher informazioan, hau da, laginaren probabilitate distribuzio funtzioa eta zenbatetsi nahi den parametroaren aldaketaren arteko korrelazioa neurtzen duen kantitatean, oinarrituta dago.

Kontzeptu hauek aztertu ondoren, mekanika kuantikoko tesi honetan erabilitako hainbat tresna aurkezten dira.
Tresna eta definizio hauek tesi hau hobeto ulertzeko azaltzen dira tesiaren hasierako kapitulu honetan.
Egoera kuantikoaren definizio eta propietateak azaltzen dira, baita operadore kuantikoenak ere.
Egoera kuantikoa matrize baten bitartez irudikatu daiteke gehienetan.
Matrize honen karratuarekin egoera kuantiko bera lortzen bada egoera purua dela esaten da.
Egoera kuantiko nahasiak aldiz, egoera puruen nahasketa baten bitartez adieraz daitezke.

Deskonposizio horien artean, egoera puruak beraien artean ortogonalak direnean deskonposizio propioa dela esaten da.
Egoera kuantiko bat beraz deskonposizio propio baten bitartez adieraz daiteke $\rho\equiv\sum_\lambda p_\lambda \ketbra{\lambda}{\lambda}$, non $p_\lambda$ probabilitate bat eta $\ket{\lambda}$ egoera puru bat adierazten duten.
Partikula multzo baten aurrean gaudenean, aurreko propietateez gain, beste propietate interesante batzuk agertzen dira.
Elkarlotura kuantikoa, adibidez, partikula multzoetan definitzen da.
Multzo osoaren egoera kuantikoa banakorra ez denean elkarlotuta daudela esaten da.

Tesi honetako operadore erabilienak momentu angeluarraren osagaiak dira, bai partikula bakarraren momentu angeluarrarenak, baita partikula guztien momentu angeluar kolektiboenak ere.
Momentu angeluarraren operadore hauek garrantzia handia daukate magnetometrian.
Partikula bakarraren spin operadoreak momentu angeluar operadoreak dira, eta spin operadoreen bitartez deskribatzen da partikulek eremu magnetikoekin daukaten interakzioa.
Oinarrizkotzat hartu daitekeen partikula bakarraren spin zenbakia gehienetan $\frac{1}{2}$ da.
Honi \emph{qubit} deitzen zaio kuantikoa.

Bestalde, momentu angeluarraren operadoreek deskonposizio propioan base berri bat sortzen dute.
Base honetan hainbat egoera kuantiko berezi topa daitezke, esaterako, Dicke egoera simetrikoak edota singletea.
Egoera hauek aztertzerakoan ikusten da magnetometrian edota beste teknologia kuantikoetan duten erabilgarritasuna.

Estatistikaren zenbatespen metodologia eta mekanika kuantikoa batzean metrologia kuantikoa sortzen da.
Metrologia kuantikoan zenbatespen prozesuaren errorearen mugak aztertzen dira askotan.
Tesi honetan aurkeztutako ikerkuntzekin muga hauen bilaketan aurrera pausu garrantzitsuak eman dira.
Fisher informazio kuantikoa da normalean erabiltzen den tresnarik esanguratsuena.
Zenbatespena egiteko erabiltzen den egoera kuantikoan eta interakzioak sortzen duen egoeraren eboluzioan oinarrituta dago Fisher informazio kuantikoa.
Beraz, ezinbestekoa da hasierako egoera kuantikoa ezagutzea Fisher informazio kuantikoa kalkulatzeko.
Hurrengo paragrafoetan, aldez aurretik egoera kuantikoa zein den jakin gabe, muga hauek bilatzeko garatu diren tresnak azaltzen dira.

\ref{sec:vd}.~kapituluan lehenbiziko ikerketa lana aurkezten da: Polarizatu gabeko Dicke egoeratik hurbil dauden egoera kuantiko nahasiek metrologian duten erabilgarritasuna.
Egoera kuantiko puruak gauzatzea oso zaila da praktikoki, eta egoera nahasiak lortzen dira gehienbat laborategietan.
Arrazoi honegatik, kapitulu honetan egoera nahasi hauek metrologian duten erabilgarritasunaren arabera sailkatzeko balio duen teknika aurkezten da.

Egoera ez polarizatuak egoera polarizatuak baino erabilgarriagoak izan daitezke magnetometrian.
Egoera polarizatuak erabiltzerakoan aldiz eremu magnetikoaren magnitudea zenbatestea nahiko zuzena da.
Egoerak denbora tarte batean eremu magnetikoaren pean polarizazioan jasandako errotazioa neurtzen da eta aldaketa honetatik eremu magnetikoaren zenbatespena egiten da.
Bestalde, egoera ez polarizatuak ezin dute teknika hau erabili, nahiz eta Fisher informazio kuantikoa kalkulatzerakoan magnetometriarako erabilgarriagoak direla argi dagoela ikusi.

Hau dela eta, Dicke egoera ez polarizatuek duten beste propietate bat erabiltzen da, polarizazioaren sakabanaketa.
Propietate hau polarizazioaren neurketetan lortzen den datuen sakabanaketa da.
Datuen sakabanaketa hau Heisenbergen ziurgabetasun printzipioarekin lotuta dago.
Dicke egoera ez polarizatuetan sakabanaketa hau txiki izatetik $N^2$-ko proportzioetara heltzen den magnitude bat da, beraz, $N^2$-ko proportzioetako aldaketa neurtuko da.
Polarizazioan oinarritutako zenbatespenak $N$-ko proportzioetara heltzen diren bitartean, sakabanaketan oinarritutakoak zenbatespenak "Heisenberg scaling" muga fisikotik hurbilago daude, ikusi \pageref{fig:in-magnetometry-totally-polarized}.~orrian dagoen \ref{fig:in-magnetometry-totally-polarized}.~irudia.
Polarizazioa gezi gorriak ematen duen bitartean, ziurgabetasuna zirkulu urdinak ematen du.

Guzti honetan oinarrituta, polarizazioaren sakabanaketaren aldaketa neurtzerakoan eremu magnetikoaren zenbatespena egin ahal da.
\ref{sec:vd}.~kapituluan erroreen hedapenaren formula aplikatuz, zenbatespenean gertatuko den errorea kalkulatzen dugu.
Errore hau hasierako egoeraren itxarotako balioen funtzio bezala idatzi ondoren, aski da lau behagarriren itxarotako balioak neurtzea.
Lau balore hauek neurtzearekin batera zenbatespenaren errorea lortuko dugu.
Errore hau Fisher informazio kuantikoaren gainetik egon arren, egoerak sailkatzen laguntzen du beraz.
Gainera, lau operadorereen neurketa partikula asko duten egoeren tomografia egitea baino askoz errazagoa da esperimentalki.

Kapitulua bukatutzat emateko, errorearen formulan oinarrituta, are gehiago sinplifikatzen dugu formula hau.
Oraingoan, operadore biren itxarotako balioan oinarritzen den beste ordezko ekuazio bat aurkezten dugu, honek dakarren abantaila esperimentala azpimarratuz.

\ref{sec:lt}.~kapituluan, Fisher informazio kuantikoaren mugak aztertzen ditugu egoera kuantiko baten operadore ezberdinek daukaten itxarotako balioen funtzio bezala.
Beraz, arazo berdinari egiten zaio aurre kapitulu honetan.
Praktikoki egoera kuantikoa zehatz-mehatz jakitea ezinezkoa denez, eta are gutxiago partikula asko duten sistemetan, kapitulu hau behagarriek egoera kuantikoan duten itxarotako balioetan oinarritzen da Fisher informazio kuantikoa mugatzerako orduan.

Oraingoan aldiz, problema honi beste ikuspuntu batekin aztertzeari ekiten zaio.
Legendreren transformazioan oinarritutako elkarloturaren neurketak egiteko metodo baten oinarrituta \cite{Guehne2007}, Fisher informazio kuantikoaren doitutako mugak topatzen dira.

Kapituluan zehar hainbat adibide garatzen dira.
Metodo honek edozein behagarri hartu eta beraren itxarotako balio bera duten egoera kuantiko guztien artean Fisher informazio kuantiko baxuenekoa aukeratzea ahalbidetzen du.
Metodo honek, bat bakarra beharrean, hainbat behagarri hartu ditzake.
Beraz, hainbat behagarri sorta hartu eta beraien itxarotako balioak aldez aurretik dakizkigula, metodoak emandako Fisher informazio kuantikoaren muga ezberdinak aztertzen ditugu.

Azkenik, adibide konkretu batzuei jarraituz, gure metodoa partikula askotako egoeretara nola luzatu daitekeen aztertzen dugu.
Datu esperimentalak erabiliz, aldez aurretik egindako esperimentuetarako zenbatespenaren mugak kalkulatzen ditugu.
Datu hauek \cite{Luecke2014} eta \cite{Gross2010}~erreferentziei jarraituz lortu ditugu.

Tesi honetan aurkezten dudan azkenengo ikerketa lana \ref{sec:gm}.~kapituluan topa daiteke: Eremu magnetikoaren gradientearen zenbatespenaren mugak atomo multzoak erabiltzerakoan, izenekoa.
Eremu magnetikoaren gradientea eremu magnetikoak espazioan daukan aldaketa adierazten du.
Aurreko kapituluetan ez bezala, honetan, Fisher informazio kuantikoa kalkulatzen da.

Eremu magnetikoa, beraz, parametro birekin zehaztuta dago, eremu magnetikoaren parte homogeneoa eta gradientea.
Ondorioz, parametro bat baino gehiago zenbatetsi behar ditugu, nahiz eta gradiente parametroan soilik interesatuta egon.
Parametro bat baino gehiagoko metrologia kuantikoaren oinarrizko problematzat hartu daiteke eremu magnetikoaren gradientearen zenbatespena.

Eremu magnetikoaren gradientea kalkulatzeko ezinbestekoa da egoera kuantikoak espazioan daukan izaera aztertzea, hau da, egoera kuantikoak espazioa nola betetzen duen jakitea.
Egoera kuantikoaren espazioaren partea, partikula puntualez osatutako egoera batera sinplifikatzen dugu, nahiz eta lortutako emaitzak bestelako kasuetara ere egokitzen diren.

Adibidez, lehenengo kasuan atomoak espazioko puntu ezberdinetan jartzen dira ilara zuzen bat sortuz.
Atomo ezberdinek eremu magnetikoaren intentsitate ezberdinak sumatuko dituzte.
Spin egoeraren arabera beraz, Fisher informazio ezberdinak kalkulatzen ditugu.
Bigarren kasuan atomo guztiak espazioko bi puntu ezberdinetan kokatuta daude, atomoen erdia puntu batean eta beste erdia bestean.
Kasu honetan topa daiteke eremu magnetikoaren zenbatespenerako spin egoerarik onena, Heisenbergen printzipioez mugatutako zenbatespena ematen duena.

Azkenengo kasuan, atomoak espazioan zehar era desordenatu baten sakabanatuta daude.
Esperimentu askotan topa daitekeen egoera da hau.
Adibidez, atomoak barrunbe batean daudenean.
Spin egoera ezberdinak aztertzen ditugu eta kasu bakoitzean beraien Fisher informazioa, zenbatespenean duten muga teorikoa, kalkulatzen dugu.

Ondorio gisa, lan honetan aurkeztutako azterketek zenbatespen kuantikoa dute jomuga.
Lehenengo ikerkuntza bietan, esperimentuen konplexutasuna sinplifikatzen da. Egoera kuantikoan oinarritu beharrean, behagarri batzuen itxarotako balioetan oinarritzen baita zenbatespenaren errorearen muga.
Metodo hauen inplementazio praktikoak aztertu ditugu aldez aurretik egindako esperimentuen datuak erabiliz.
Honek guztiak, etorkizunean egingo diren metrologia kuantikoko esperimentuetan, metodo hauek erabiltzea errazten du.
Bestalde, eremu magnetikoaren gradientearen azterketan topatu ditugun muga teorikoak Heisenbergen proportzionaltasuna ahalbidetzen dute.
Proportzionaltasun hau bi partikula multzo erabiltzen direnean eta baita multzo bakarra erabiltzen denean ere agertu daitekeela frogatu da.
Partikula multzo bakarra erabiltzerakoan beraz, partikula zenbakiarekin batera txikitzen da errorea, esperimentua eta ondoren etor daitekeen inplementazio praktikoa asko sinplifikatuz, eta Heisenbergen proportzionaltasuna oraindik ere mantenduz.

%% file: 00-2-publications.tex
\afterpage{\afterpage{\hypertarget{publications}{}}}
\section*{Publications\hlabel{sec:pu}}
\bookmark[level=section,dest=publications]{Publications}

\begin{minipage}{\linewidth}
  \setlength{\parindent}{0cm}
  \textbullet\hspace{5pt}G\'eza T\'oth and Iagoba Apellaniz.
  \textit{Quantum metrology from a quantum information science prespective.}
  Journal of Physics A: Mathematical and Theoretical, \textbf{47} 424006, (2014).
  \vspace{6pt}

  \textbullet\hspace{5pt}Iagoba Apellaniz, Bernd L\"ucke, Jan Peise, Carsten Klempt and G\'eza T\'oth.
  \textit{Detecting metrologically useful entanglement in the vicinity of Dicke states.}
  New Journal of Physics, \textbf{17} 083027, (2015).
  \vspace{6pt}

  \textbullet\hspace{5pt}Iagoba Apellaniz, Matthias Kleinmann, Otfried G\"uhne and G\'eza T\'oth.
  \textit{Optimal detection of metrologically useful entanglement.}
  Physical Review A, \emph{in press}, (2017).
  \vspace{16pt}

  {\large\bf In preparation}
  \vspace{6pt}

  \textbullet\hspace{5pt}Iagoba Apellaniz, Iñigo Urizar-Lanz, Zoltán Zimborás, Philipp Hyllus and G\'eza T\'oth.
  \textit{Precision bounds for gradient magnetometry with atomic ensembles.}
  \vspace{16pt}

  {\large\bf Out of the scope of this thesis}
  \vspace{6pt}

  \textbullet\hspace{5pt}Giuseppe Vitagliano, Iagoba Apellaniz, I\~nigo Luis Egusquiza and G\'eza T\'oth.
  \textit{Spin squeezing and entanglement for arbitrary spin.}
  Physical Review A, \textbf{89} 032307, (2014).
  \vspace{6pt}

  \textbullet\hspace{5pt}Giuseppe Vitagliano, Iagoba Apellaniz, Matthias Kleinmann, Bernd L\"ucke, Carsten Klempt and G\'eza T\'oth.
  \textit{Entanglement and extreme spin squeezing of unpolarized states.}
  New Journal of Physics, \textbf{19} 013027, (2017).
\end{minipage}

%% file: 00-3-toc.tex

\vspace*{100pt}
\leavevmode
\afterpage{\hypertarget{toc}{}}
\tableofcontents
\bookmark[level=section, dest=toc]{Table of contents}

\afterpage{\afterpage{\hypertarget{abbreviations}{}}}
\section*{Abbreviations, figures and tables\hlabel{sec:ab}}
\bookmark[level=section,dest=abbreviations]{Abbreviations, figures and tables}
\fancyfoot[LE,RO]{\thepage}
\subsection*{Abbreviations}
\hspace{7pt}
\begin{tabular}{l c l}
  \abbrln{BEC}{Bose-Einstein condensate}
  \abbrln{GHZ}{Greenberger-Horne-Zeilinger}
  \abbrln{PDF}{Pobability distribution function}
  \abbrln{POVM}{Positive-operator valued measure}
  \abbrln{QFI}{Quantum Fisher information}
\end{tabular}

\makeatletter
\newcommand\listoffigurename{Figures}
\newcommand\listoftablename{Tables}
\renewcommand\listoffigures{
  \subsection*{\listoffigurename}
  \@starttoc{lof}
}
\renewcommand\listoftables{
  \subsection*{\listoftablename}
  \@starttoc{lot}
}
\makeatother
\listoffigures
\listoftables

%% file: snp/title.tex

\thiswatermark{\centering
\put(0,-110){\includegraphics[height=2.5cm]{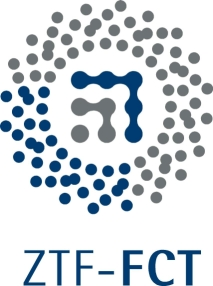}}
\put(430,-100){\includegraphics[height=2cm]{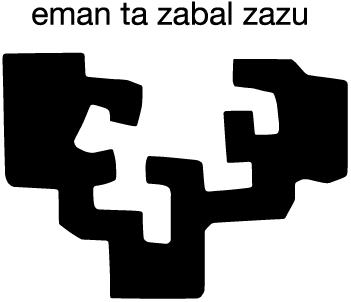}}
}
\hypertarget{thesis}{}
\bookmark[level=0,dest=thesis]{Thesis}

\begin{center}

\vspace*{20pt}
\textsc{\LARGE University of the Basque Country}

\vspace{20pt}
\textsc{\Large PhD Thesis}

\vspace{50pt}
\hrule

\vspace{16pt}
{\huge \bfseries Lower bounds on quantum metrological precision}
\vspace{16pt}

\hrule
\vspace{40pt}

\begin{minipage}{0.4\textwidth}
\begin{center}
  \Large
  \textbf{Iagoba \textsc{Apellaniz}}
  \vspace{20pt}

  \large
  \emph{Supervised by}
  \vspace{7pt}

  Prof. G\'eza \textsc{T\'oth}
\end{center}
\end{minipage}

\includegraphics[width=0.8\hsize]{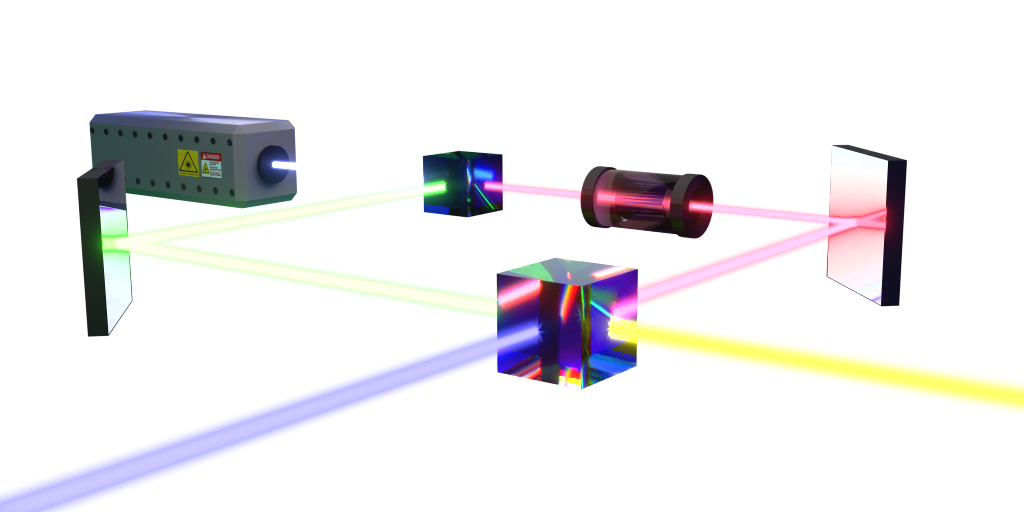}
\vfill

{\large \mydate}

\end{center}

\cleardoublepage

%% file: snp/fancySection.tex
\titleformat{\section}[display]
{\vspace*{150pt}
\bf\Huge}
{\hspace{-72pt}{\textcolor{grey}{\thesection}} \hspace{38pt} {\textcolor{white}{#1}}}
{0pt}
{}
[]
\titlespacing*{\section}{108pt}{10pt}{60pt}[20pt]

%% file: 01-introduction.tex
\section{Introduction}
\input{snp/singleLineWaterMark.tex}
\label{sec:in}

\quotes{Daniel M. Greenberger}{The story I am about to tell, while it is not directly about John Bell,\break could not have taken place without him.}

\lettrine[lines=2, findent=3pt,nindent=0pt]{M}{etrology} plays an important role in many areas of physics and engineering \cite{Glaser2010}.
With the development of experimental techniques, it is now possible to realize metrological tasks in physical systems that cannot be described well by classical physics, instead quantum mechanics must be used for their modeling.
Quantum metrology \cite{Giovannetti2004, Giovannetti2006, Paris2009, Gross2012} is the novel field, which is concerned with metrology using such quantum mechanical systems.

In quantum metrology, the quantumness of the system plays an essential role \cite{Demkowicz-Dobrzanski2015, Pezze2014}.
One can find bounds on the highest achievable precision of a metrological setup.
One of the usual methods to find such bounds is using the theory of the quantum Fisher information (QFI) \cite{Helstrom1969, Holevo1982, Braunstein1996, Petz2008}.
There have been efforts recently connecting quantum metrology to quantum information science, in particular, to the theory of entanglement \cite{Toth2010}.
Entanglement, a feature of quantum mechanics, lies at the heart of many problems has attracted an increasing attention in recent years.

There are now efficient methods to detect entanglement with a moderate experimental effort \cite{Horodecki2009, Guehne2009}.
However, in spite of intensive research, many of the intriguing properties of entanglement are not fully understood.
One of such puzzling facts is that, while entanglement is a sought after resource, not all entangled states are useful for some particular quantum information processing task.
For instance, it has been realized recently that entanglement is needed in very general metrological tasks to achieve high precision \cite{Pezze2009}.
Remarkably, this is true even in the case of millions of particles, which is especially important for characterizing the entanglement properties of cold atomic ensembles \cite{Louchet-Chauvet2010, Appel2009, Riedel2010, Gross2010, Luecke2011, Strobel2014}.
However, there are highly entangled pure states that are useless for metrology \cite{Hyllus2010}.

\begin{figure}[htp]
  \centering
  \igwithlabel{(a)}{scale=1.15}{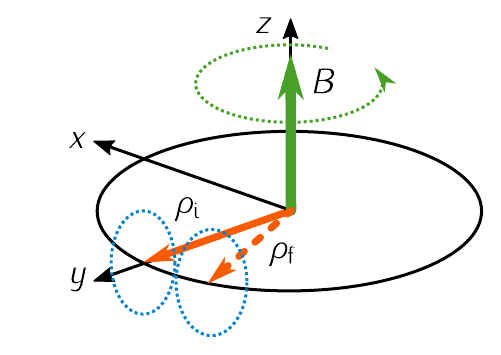}
  \igwithlabel{(b)}{scale=1.15}{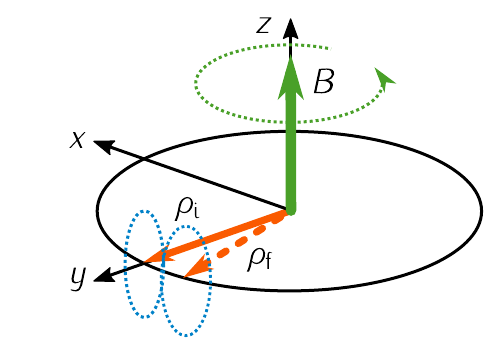}
  \caption[Magnetometry with polarized states]{
  (a) (red-arrow) Initial state $\rho_{\text{i}}$ pointing in the $y$-direction.
  (blue-dashed-circle) Uncertainty ellipse of the polarization perpendicular to the mean spin.
  The state is rotated with a speed proportional to the strength of the magnetic field $B$ (green-arrow).
  Hence, the magnetic field can be estimated from the final state $\rho_{\text{f}}$ (red-dashed-arrow).
  (b) When the uncertainty ellipse is reduced in one direction perpendicular to the polarization, the state is called a spin-squeezed state.
  If the direction in which the uncertainty is reduced coincides with the direction of the rotation, then it is easier to distinguish the final state from the initial state.
  Which turns into a better precision for the estimation of the magnetic field.}
  \label{fig:in-magnetometry-totally-polarized}
\end{figure}

In the light of these results, beside verifying that a quantum state is entangled, we should also show that it is useful for metrology.
One of the basic tasks of quantum metrology is magnetometry with an ensemble of spin-$j$ particles.
Magnetometry with a state completely polarized works as follows.
The total spin of the ensemble is rotated by a homogeneous magnetic field perpendicular to it.
We would like to estimate the rotation angle or phase $\theta$ based on some measurement.
Then, the phase can be used to determine the strength of the magnetic field,
see Figure~\ref{fig:in-magnetometry-totally-polarized}-(a).

In recent years, quantum metrology has been applied in many scenarios, from atomic clocks \cite{Louchet-Chauvet2010, Borregaard2013, Kessler2014a} and precision magnetometry \cite{Wasilewski2010, Eckert2006, Wildermuth2006, Wolfgramm2010, Koschorreck2011, Vengalattore2007, Zhou2010} to gravitational wave detectors \cite{Schnabel2010, TheLIGOScientificCollaboration2011, Demkowicz-Dobrzanski2013}.
There have been many experiments with fully polarized ensembles \cite{Gross2012, Wasilewski2010, Wildermuth2006, Vengalattore2007, Behbood2013, Koschorreck2011, Muessel2014}, in which the collective spin of the particles is rotated as in Figure~\ref{fig:in-magnetometry-totally-polarized}-(a) and the angle of rotation is estimated by collective measurements.
It has also been verified experimentally that spin squeezing can result in a better precision compared to fully polarized states \cite{Riedel2010, Gross2012, Wasilewski2010, Muessel2014, Fernholz2008, Hald1999, Julsgaard2001, Hammerer2010, Esteve2008} since spin-squeezed states are characterized by a reduced uncertainty in a direction orthogonal to the mean spin \cite{Kitagawa1993, Wineland1994, Sorensen2001, Ma2011}, see Figure~\ref{fig:in-magnetometry-totally-polarized}-(b).

Besides almost fully polarized states, there are also unpolarized states considered for quantum metrology.
Prime examples of such states are Greenberger-Horne-Zelinger (GHZ) states \cite{Greenberger1990}, which have already been realized experimentally many times \cite{Leibfried2004, Bouwmeester1999, Pan2000, Zhao2003, Lu2007, Gao2010, Sackett2000, Monz2011}.
Recently, new types of unpolarized states have been considered for metrology, such as the singlet state \cite{Urizar-Lanz2013, Behbood2014} and the symmetric unpolarized Dicke states  which have been realized in cold gases \cite{Luecke2011, Hamley2012, Krischek2011}.

In the present thesis we study how large precision can be achieved for realistic noisy systems \cite{Escher2011, Demkowicz-Dobrzanski2012}.
We also study how such states can be characterized with few measurements which reduce the experimental efforts considerably.
And finally, we discuss the multi-parameter estimation problem.
It turns out that some states not useful for homogeneous magnetometry may become useful when we use them for differential magnetometry, which is one of the most fundamental two-parameter estimation tasks.

%% file: snp/singleLineWaterMark.tex
\thiswatermark{\put(1,-241){\color{l-grey}\rule{84pt}{48pt}}
\put(84,-241){\color{grey}\rule{410pt}{48pt}}}

%% file: 02-background.tex
\section{Background in quantum metrology}
\input{snp/singleLineWaterMark.tex}
\label{sec:bg}

\quotes{Roger J. Barlow}{In the real world, doing real experiments, statistics began to matter.}

\vspace{0pt}
\lettrine[lines=2, findent=3pt,nindent=0pt]{I}{n} this chapter, we will study the basics of quantum metrology, metrology enhanced by quantum phenomena.
First of all, metrology, as the science of measuring, has played an essential role in the development of science and technology.
Metrology studies several aspects of the estimation process, for example, which strategy to follow in order to improve the precision of the estimation.
Metrology also covers all intermediate processes, from the design aspects of a precise measuring device, to the most basic mathematical concepts that arise from the formulation of the different problems.

Historically, with the discovery of quantum theory and the subsequent development of quantum mechanics, new opportunities emerged for advances in metrology in the early decades of the twentieth century.
Later with the arrival of concepts like \emph{qubit} and \emph{quantum cryptography}, quantum theory embraced the so-called quantum information theory, which merges the notions of theory of information and computer science with quantum mechanics.
Due to the rapid development of quantum technologies, quantum information attracted a lot of attention from the scientific community.
Moreover, those emerging fields rapidly became a very interesting interdisciplinary playground of science with many scientists as well as resources involved.

Next, we will focus on quantum metrology, for which the role of entanglement, an exclusive feature of quantum mechanics which cannot be described using classical probabilistic theories, is essential.
Entanglement connects quantum metrology with concepts like local realism or quantum teleportation.
Nevertheless, throughout the thesis we will focus mainly on the achievable precision for different systems and schemes.

On the other hand, we will also present some important tools and concepts of statistics, without which many descriptions of important physical findings would lack the rigorous interpretation needed.
It basically helps to analyze raw data to make it readable from the human perspective.

This chapter is organized as follows.
In Section~\ref{sec:bg-statistics-and-stimation}, we introduce the basic concepts of statistics and estimation theory, focusing on what is needed for the understanding of this thesis.
In Section~\ref{sec:bg-quantum-mechanics-for-metrology}, we present the necessary tools of quantum mechanics used throughout the text.
Finally in Section~\ref{sec:bg-quantum-metrology}, we arrive at the main set of tools used in the quantum metrology framework and by extension in the present work.

\subsection{Background in statistics and theory of estimation}
\label{sec:bg-statistics-and-stimation}

In this section, we will enumerate the basic concepts of statistics as well as the estimation process.
As we have said, the main mathematical tools used by metrology belong to statistics.
Moreover, we are especially interested in estimation theory, which shows how to estimate properly some quantity based on a data sample.
The data can be of any kind.
For instance, the data sample might be a set of the heights of members of a basketball team, or the outcomes of coin tosses, or even the wavelengths of photons coming out of some radioactive sample\footnote{
For a more detailed introduction to statistics, see Refs.~\cite{Riley2006, Barlow1989}}.

\subsubsection{Probability, data samples, average and variance}

The probability indicates the relative chance of an event to happen.
For instance, if there is a box with 10 red balls and 5 blue balls and assuming we extract on of them randomly, the probability of obtaining a blue ball is given by $\frac{n_{\text{blue}}}{n_{\text{total}}}=\frac{5}{10+5}=\frac{1}{3}$.
In the same way, the probability of obtaining a red ball is $\frac{n_{\text{red}}}{n_{\text{total}}}=\frac{2}{3}$.
Note that some properties of the probabilities arise from this simple example.
First, the probability of any event is always given by a number in between zero and one.
Second, the sum of the probabilities of all possible events, in this case either to extract a blue ball or a red ball, sum up to one.

Once we have a data sample, we normally face the task of analyzing the data to extract the relevant properties from it.
We assume that the data sample always comes from a population sample which represents all the available data before the measurements.
Hence, the data sample might not be complete in the most general case (one could lose some data in the measurement, or the data population might be so large that we can only obtain part of it).
The measurement itself might also induce some error to the data (measuring a continuous variable is always error prone, e.g., the height of people).
Hence, the data sample inherits a probability number for each of its data elements.
In the subsequent paragraphs, we will describe these relations between the data sample and the corresponding probabilities, and we will enumerate the most useful properties and formulas\footnote{The notation used in this section follows mainly the one used in Ref.~\cite{Riley2006}.}.

The magnitude we would like to measure from the data population is called a random variable.
When measuring some random variable $X$, a probability distribution function (PDF) gives the probability of $x$ to be the outcome of the measurement and it is denoted by $\prob(X{=}x)$.
Second, due to the random nature of the measurement, the $N$ elements of a data sample are considered outputs of $N$ different random variables with their corresponding values denoted by $\{X_i{=}x_i\}_{i=1}^N$, or $\bs{X}{=}\bs{x}$ for short.
The joint probability of those random variables is in general not separable.
This is due to the fact that the data sample elements could depend on the rest of outcomes or some other more complex relation that makes the most general case to be non-separable from the probabilistic point of view.
The PDF of the data sample is an $N$-variable function denoted by
\be
  \prob(\bs{X}{=}\bs{x}) \equiv \prob(X_1{=}x_1,X_2{=}x_2,\dots,X_N{=}x_N).
  \label{eq:bg-n-variable-prob}
\ee
In the case of separable probabilities, Eq.~\eqref{eq:bg-n-variable-prob} is written as
\be
  \prob(\bs{X}{=}\bs{x}) = \prod_{i=1}^N\prob(X_i{=}x_i),
  \label{eq:bg-separable-likelyhood}
\ee
which is the case in many relevant situations.

When some indirect property of the system is defined as a function of the measured random variables, the result is also a new random variable with another assigned PDF.
For example, we measure the position of a body at some moment.
If the system was at rest at the origin when $t=0$ and assuming that the acceleration is constant, then from the measured final position one could infer the value of the acceleration by using $A=2X/t^2$, where $X$ denotes the final position at instant $t$ and $A$ the acceleration.
If $X$ is a random variable, which is the general case when measuring the position of some physical system, then the probability assigned to the random variable $A$ is computed by
\be
  \prob(A{=}a) = \left.\frac{\text{d}X}{\text{d}A}\right|_{A=a}\prob(X{=}x)=\frac{2}{t^2}\prob(X{=}x).
\ee
In general, for multiple random variables, we must require the following identity
\be
  |\prob(\bs{X}{=}\bs{x})\,\text{d}x_1\text{d}x_2\dots\text{d}x_N|=|\prob(\bs{Y}{=}\bs{y})\,\text{d}y_1\text{d}y_2\dots\text{d}y_N|,
\ee
which leads to some interesting formulas we will discuss later.

We now stick to the simplest case in which the data consists of a collection of values describing the same physical one-dimensional data population.
We assume that all the outcomes are independent from each other.
Hence, the PDF is of the form of Eq.~\eqref{eq:bg-separable-likelyhood}.
Next, we need some definitions used in this thesis: namely the average, variance, and the statistical moments and central moments.
The arithmetic average  is computed as
\be
  \overline{x}=\frac{1}{N}\sum_{i=1}^N x_i.
\ee
Note that there are other types of averages one can find in the literature.
The variance, which is related to the spread of the data, is computed as
\be
  \sigma^2=\frac{1}{N}\sum_{i=1}^N (x_i-\overline{x})^2,
\ee
where $\sigma$ is the standard deviation.
Different statistical moments are computed by $\overline{x^r}=\frac{1}{N}\sum x_i^r$, while central moments, i.e., statistical moments that are invariant under a global translation of the parameter space, are of the form of $c_r=\frac{1}{N}\sum (x_i-\overline{x})^r$.
Note that the variance is equivalent to the second central moment.

For completeness, when each element of the data consists of more than a single magnitude the co-variance between two magnitudes is obtained as
\be
  V_{X,Y} = \frac{1}{N}\sum_{i=1}^N (x_i-\overline{x})(y_i-\overline{y}),
\ee
where it represents how both magnitudes influence each other.
As an example, we could measure two magnitudes simultaneously $(v_i,a_i)$, where $v_i$ is the velocity and $a_i$ the acceleration of a body, in a experiment to estimate the friction coefficient of the air.

The data population is represented in most cases by the probability distribution function.
While the mean values over the data sample of any function $g(x)$ are denoted with a bar, e.g., $\overline{x^r}$ for the $r$-th moment over the data sample or $\overline{g(x)}$ for the average of $g(x)$ itself, the mean values over the data population of any function $g(X)$ are denoted by $\text{E}[g(X)]$, e.g., the $r$-th moment over the data population is $\text{E}[X^r]$.
The data sample always consist of discrete data elements.
On the other hand, the data population can be represented by a PDF for discrete values or by a PDF for continuous values.
For completeness, here are expressed the two definitions for the mean value over the data population of a function $g(\bs{X})$ as
\be
  \label{eq:bg-expectation-value-of-any}
  \text{E}[g(\bs{X})] = \lcor
  \begin{aligned}
    &\int g(\bs{x}) \prob(\bs{X}{=}\bs{x})\,\text{d}^N \bs{x},\\
    &\sum_{i,j,\dots} g(\bs{x}) \prob(\bs{X}{=}\bs{x}).
  \end{aligned}
  \right.
\ee
I also write the variance of $g(X)$ over the data population as $\text{V}[g(X)] \equiv \text{E}[g(X)^2] - \text{E}[g(X)]^2$.

\subsubsection{Estimators and Fisher information}
\label{sec:bg-estimators}

Let us suppose that the data sample has encoded some parameters $\bs{a}\equiv(a_1,a_2,\dots)$ we are interested in.
Therefore, the underlying probability, in general also unknown, might be conditioned by the real values of the parameters $\bs{a}$.
The probability of the data sample is written as
\be
  \prob(\bs{X}{=}\bs{x}|\bs{a}),
\ee
where "$|\bs{a}$" indicates its dependency on these parameters.
Note that an estimate of one of the parameters must be based on the data sample elements.
A function of this type is called the \emph{estimator} and it is connected to the PDF of the data population.
Hence as mentioned before, the estimator of $a$, one of the unknown parameters, is denoted by $\hat{a}$ and the underlying PDF associated to it is computed from the PDF of the data population as
\be
  \prob(\hat{a}|\bs{a})\,\text{d}\hat{a} = \prob(\bs{x}|\bs{a})\,\text{d}x_1\text{d}x_2\dots\text{d}x_N.
\ee
For short, we have omitted writing "$\bs{X}{=}$".
The conditional joint probability of $N$ random variables $\bs{X}$ is written simply as $\prob(\bs{x}|\bs{a})$ then.

An estimator is a function of the data sample.
For example, one of such estimators can be the estimator of the mean value of the data population, which is in general unknown.
The mean value of the data population, which in general we do not have access to, is denoted usually by $\mu$.
Note that $\mu$ is in general different from the mean value of the data sample $\overline{x}$.
A valid estimator for the mean value $\mu$ would be the mean itself of the data sample, i.e., $\hat{\mu}=\overline{x}$.

An important notion of an estimator is its \emph{efficiency}.
The smaller the variance of the estimator the more efficient it is.
Remember that an estimator is considered a random variable so it must have a variance when our knowledge about the population is incomplete.

Focusing now on what is more interesting for this thesis, an estimator of any kind has a theoretical lower bound for its variance.
For the proof of the previous statement, which we will show only for continuous random variables without loss of generality, we start with the normalization formula for any given PDF
\be
  \int \prob(\bs{x}|\bs{a})\,\text{d}^N\bs{x} = 1.
\ee
Next, we compute the partial derivative over $a$ such that
\be
  \label{eq:bg-expectation-value-of-logarithm}
  \int {\partial_a}\prob(\bs{x}|\bs{a})\,\text{d}^N\bs{x} = \int \partial_a\big[\ln  \prob(\bs{x}|\bs{a})\big] \prob(\bs{x}|\bs{a}) \,\text{d}^N\bs{x} = 0,
\ee
where for the second equality we used the identity for logarithmic derivatives.
From Eq.~\eqref{eq:bg-expectation-value-of-any}, it turns out that Eq.~\eqref{eq:bg-expectation-value-of-logarithm} is the expectation value of $\partial_a(\ln\prob)$.
Finally, if we have an \emph{unbiased} estimator, i.e., an estimator for which the expectation value $\text{E}[\hat{a}]$ coincides with true value $a$, the partial differentiation of $\text{E}[\hat{a}]$ over $a$ must be equal to one.
Therefore, we apply similar identities as in Eq.~\eqref{eq:bg-expectation-value-of-logarithm} to arrive at
\be
  \label{eq:bg-expectation-of-estimator-derivated}
  \begin{split}
    \partial_a\text{E}[\hat{a}] &= \partial_a a
    = \partial_a \int \hat{a} \prob(\bs{x}|\bs{a})\,\text{d}^N\bs{x} \\
    &=  \int \hat{a} \, \partial_a\prob(\bs{x}|\bs{a})\,\text{d}^N\bs{x} =  \int \hat{a} \, \partial_a\big[\ln  \prob(\bs{x}|\bs{a})\big] \prob(\bs{x}|\bs{a}) \,\text{d}^N\bs{x} = 1,
  \end{split}
\ee
where we have used the definition of the expectation value for continuous variables Eq.~\eqref{eq:bg-expectation-value-of-any} and we use the fact that the estimator is not a function of the parameter $a$.

At this point, we invoke the Schwarz inequality for two real multidimensional functions $g(\bs{x})$ and $h(\bs{x})$ such that $(\int g h \,\text{d}^N\bs{x})^2\leqslant (\int g^2 \,\text{d}^N\bs{x}) (\int h^2 \,\text{d}^N\bs{x})$.
With this, we can obtain a lower bound for the variance of a general estimator.
First the definition of the variance over the data population of $\hat{a}$ looks like
\be
  V[\hat{a}] = E[(\hat{a}-a)^2] = \int (\hat{a} - a)^2\prob(\bs{x}|\bs{a})\,\text{d}^N\bs{x},
\ee
for unbiased estimators.
Second, since $a$ is not a function of $\bs{x}$, subtracting $a$ times Eq.~\eqref{eq:bg-expectation-value-of-logarithm} to Eq.~\eqref{eq:bg-expectation-of-estimator-derivated} we have that
\be
  \int (\hat{a}-a)\,\partial_a\big[\ln  \prob(\bs{x}|\bs{a})\big] \prob(\bs{x}|\bs{a})\,\text{d}^N\bs{x} = 1.
\ee
Hence, using the Schwarz inequality we can write
\be
  \label{eq:bg-classical-cr-bound-and-fi}
  \text{V}[\hat{a}] = \int (\hat{a} - a)^2\prob(\bs{x}|\bs{a})\,\text{d}^N\bs{x} \geqslant \frac{1}{\int \lpar\partial_a\big[\ln \prob (\bs{x}|\bs{a}) \big]\rpar^2 \prob (\bs{x}|\bs{a})\,\text{d}^N\bs{x}},
\ee
which is also known as the Cram\'er-Rao bound or the Fisher inequality.
The denominator in the right hand-side is called generally the Fisher information or simply information.

With this review of the most interesting properties of "classical" metrology, from the point of view of this thesis, we conclude this section.
In the next section, we will discuss some properties of quantum mechanics and then we will follow with another section which presents the basics of quantum metrology.

\subsection{Quantum mechanics from metrology perspective}
\label{sec:bg-quantum-mechanics-for-metrology}

The ubiquitous probabilistic nature of quantum mechanics makes us to work with probabilities on a regular basis.
Moreover, if one studies fields connected to experiments or some short of physical realizations, this probabilistic nature of quantum mechanics becomes even more visible.
On the other hand, exotic features such as entanglement arise from quantum mechanics, which are directly connected with the probabilistic properties of quantum system.
The present section is intended to describe quantum systems from the point of view of metrology.

\subsubsection{The quantum state, multiparticle state and entanglement}
\label{sec:bg-the-quantu-state}

A formal mathematical description of the quantum state is given next.
This also allows us to introduce some notation used throughout the thesis.
A \emph{state} in quantum mechanics lives in a Hilbert space, $\mathcal{H}$.
The state $\rho$ has the following properties:
\begin{enumerate}
  \item
  It is Hermitian, so it is invariant under the complex transposition, $\rho=\rho^\dagger$ and all its eigenvalues are real.
  \item Its trace is equal to one, $\tr(\rho)=1$.
  \item It is positive semi-definite, i.e, all its eigenvalues are larger or equal to zero, $\rho=\sum_{\lambda}p_\lambda \Pi_\lambda$ where $p_\lambda\geqslant 0$ and $\Pi_\lambda\equiv\ketbra{\lambda}{\lambda}$ is the projector to the eigenstate $\ket{\lambda}$, a vector state satisfying the following eigenvalue equation $\rho\ket{\lambda}=p_{\lambda}\ket{\lambda}$.
  From (ii), it follows that $\sum_\lambda p_\lambda = 1$.
  \item If all $p_\lambda$ are zero except one, the state is called a pure state and is equivalent to the projector to the corresponding eigenstate $\rho=\Pi_\lambda=\ketbra{\lambda}{\lambda}$.
  \item Using the properties (iii) and (iv), it follows that the quantum states form a convex set, where the extremal states are pure states.
  \item An expectation value of an observable $A$ is computed as $\expect{A}=\tr(A\rho)$.
  To make a connection with the previous section, the state might represent the so-called data population. Hence, using the notation in Sec.~\ref{sec:bg-statistics-and-stimation}, the following equivalence holds, $\expect{\mathcal{O}}\equiv\text{E}[\mathcal{O}]$.
\end{enumerate}

The composite system of $N$ different parties live in the Hilbert space $\mathcal{H} = \mathcal{H}^{(1)}\otimes\mathcal{H}^{(2)}\otimes\cdots\otimes\mathcal{H}^{(N)}$ or $\mathcal{H} = \bigotimes_{i=1}^N\mathcal{H}^{(i)}$ for short, where $\otimes$ stands for tensor product.
For instance, this composite Hilbert space could be used to represent a many-particle system, in this case $N$ particles.
A \emph{separable} state in this Hilbert space can be written as \cite{Werner1989}
\be
  \rho_{\text{sep}} = \sum_{i}p_i\rho_i^{(1)}\otimes\rho_i^{(2)}\otimes\cdots\otimes\rho_i^{(N)},
  \label{eq:bg-separable-state-definition}
\ee
where $p_i$ are convex weights that add up to one and are equal to or larger than zero.
If a state cannot be written like Eq.~\eqref{eq:bg-separable-state-definition}, the state is \emph{entangled}.
Note that Eq.~\eqref{eq:bg-separable-state-definition} is a formal description of the separable states, for more details see Refs.~\cite{Guehne2009, Luis2004}.
One may note at this moment, that relaxing the requirements of Eq.~\eqref{eq:bg-separable-state-definition}, one can lead to different classifications of the states.
Concepts like genuine multipartite entanglement, $k$-producible states, or entanglement depth, among others, arose from weaker constraints than Eq.~\eqref{eq:bg-separable-state-definition} \cite{Guehne2009, Luis2004}.

\begin{figure}[htp]
  \centering
  \includegraphics[scale=1.15]{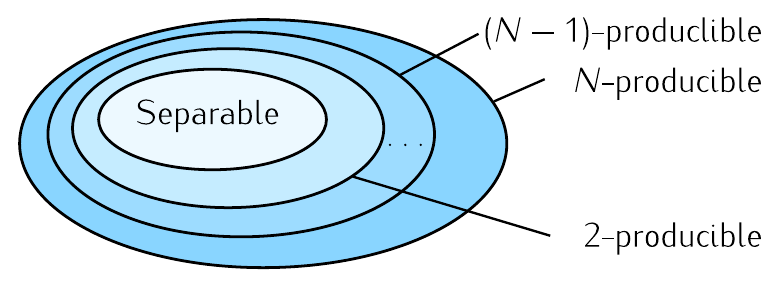}
  \caption[Diagram for $k$-producible sets]{
  Hierarchic sets for $k$-producible states, were $k$-producible set contains $(k{-}1)$-producible states.
  Based on the Eq.~\eqref{eq:bg-k-producible-estate}, one can see that a state that cannot be written as $k$-producible must be $(k{+}1)$-entangled or equivalently it must have $k{+}1$ entanglement depth.
  A separable state can always be written as 1-producible state by definition.}
  \label{fig:bg-separability-k-producibility-circle}
\end{figure}

It is important to describe one of such classifications in order to characterize the different levels or multipartite entanglement followed in this work.
We call a state $k$-producible \cite{Guehne2005}, if it can be written as a mixture of the tensor product of different multipartite states with at most $k$ parties in each,
\be
  \label{eq:bg-k-producible-estate}
  \rho_{k\text{-pro}} = \sum_i p_i\rho^{(\alpha,\dots,\beta)}_i \otimes \rho^{(\gamma,\dots,\delta)}_i\otimes \dots,
\ee
where superscript indexes between parenthesis go from 1 to $N$ and denote to which parties belong to the state, and where each index appears once in each sum element.
For instance, a separable state like Eq.~\eqref{eq:bg-separable-state-definition} is 1-producible.
If a state cannot be written as $k$-producible, then it must be $(k{+}1)$-entangled.
This defines the entanglement depth, see Figure~\ref{fig:bg-separability-k-producibility-circle}.
Later on, these concepts of entanglement and entanglement depth will arise naturally on the metrological framework \cite{Toth2014}.

\subsubsection{Angular momentum operators on multiparticle systems}
\label{sec:bg-angular-momentum-operators}

We now present a set of operators that will appear many times in all chapters, namely the angular momentum operators.
Again these definitions allow us to introduce much of the notation used on this thesis.
For a single-particle with $d$ discrete levels, and therefore with spin $j=(d-1)/2$, the eigenvalue equation for the angular momentum projection operators are
\be
  j_l^{(n)}\ket{m}_l^{(n)} = m \ket{m}_l^{(n)}
  \label{eq:bg-single-particle-am-components}
\ee
for $m=-j,\dots,+j$ and where $l=x,y,z$.
It is usual to omit the subscript of $\ket{m}_l^{(n)}$ when $l=z$ because it is typically the preferred direction and the superscript ($n$) can also be omitted in some cases, e.g., $\ket{\psi} = \ket{m_1}\otimes\ket{m_2}\otimes\dots\otimes\ket{m_N}$.
In many cases, it is also usual to merge all single-particle states into a single ket such that $\ket{\psi}=\ket{m_1, m_2,\dots, m_N}$\footnote{
This notation is normally used in many-body quantum mechanics, see Refs.~\cite{Cohen-Tannoudji1977, Sakurai2010}.}.

The square of the total angular momentum, $\bs{j}^2=j_x^2+j_y^2+j_z^2$, for a single-party $(n)$ acts on any state $\rho$ simply as
\be
  (\bs{j}^2)^{(n)}\rho=j_n(j_n+1)\rho,
\ee
where the Hilbert space where $\rho$ is defined must contain $\mathcal{H}^{(n)}$ and where $j_n$ is the spin number of such a particle.
Note that in order to distinguish the spin number $j_n$ and the operators $j_l^{(n)}$ we may use the fact that the operators are attached to a Hilbert space with a superscript or even we can use a "hat" to denote which of them is an operator like $\hat{j_l}$ and which is not.

The collective angular momentum projection operators $J_l$ are defined as the sum of their respective single-party spin operators $j_l^{(n)}$ such that they are extended to the remaining of the Hilbert spaces by tensor products of the identity operators defined in the rest of subspaces,
\be
  \label{eq:bg-total-angular-momentum-progection-operators}
  J_l = \sum_{i=1}^N \mtxid^{(1,\dots, i-1)} \otimes j_l^{(i)} \otimes \mtxid^{(i+1,\dots,N)} \equiv \sum_{i=1}^N j_l^{(i)},
\ee
where $\mtxid$ stands for the identity operator and where in the last equality the identity operators are omitted for simplicity.
On the other hand, note that the squares of the different projections of total angular momentum are not equal to the sum of the square angular momentum projections of each of the parties.
The square angular momentum components are
\be
  J_l^2 = \sum_{i,j}^N j_l^{(i)} j_l^{(j)} = \sum_{i=1}^N (j_l^2)^{(i)}+\sum_{i\neq j}^N j_l^{(i)} j_l^{(j)},
\ee
for $l=x,y,z$.
Therefore, neither the square of the total angular momentum is the sum of the square of all single-party angular momenta but
\be
  \bs{J}^2 = \sum_{i}^N \bs{j}^{(i)} + \sum_{l=x,y,z}\sum_{i\neq j}^N j_l^{(i)} j_l^{(j)},
\ee
where we separated the sum into two parts. The first one corresponds to the sum of all single-party square angular momentum and the second corresponds to the product of angular momentum projection operators of two distinct parties summed for all $l=x,y,z$.
Many more combinations of these single-party operators may arise in different contexts.
In the Appendix~\ref{app:angular-subspaces}, we discuss in more detail the addition of the angular momenta and the block structure that arises in this new basis, e.g., the symmetric subspace or the singlet subspace.

\subsubsection{Dynamics of quantum systems}

The most basic evolution of the state is represented by unitary evolution operators denoted by $U$ and those are the only ones appearing throughout the thesis.
On the other hand, there are other types of dynamics involving particle losses, entropy changes in the system and open quantum systems in general.
These transformations are governed by master-equations such as the Lindblad equation \cite{Lindblad1976, Nielsen2000, Breuer2002}.

For the understanding of this thesis it is enough to present the unitary evolution operators.
We also restrict ourselves to the case in which the generator $G$ is constant, so are in general the Hamiltonians of the metrological setups.
The unitary evolution operator is defined as
\be
  U = \exp(-i \alpha G)=\sum_{n=0}^{\infty} \frac{(-i \alpha G)^n}{n!},
\ee
for the amount of change $\alpha$ and where we use the matrix exponentiation in the last equality.
When a constant Hamiltonian $H$ acts on an initial state $\rho$, the state evolves in time as
\be
  \rho(t) = U\rho U^{\dagger} = e^{-itH/\hbar} \rho e^{+itH/\hbar}.
\ee
Note that all information we can extract from the system comes in the form of expectation values of different operators at different times, $\expect{A}(t) = \tr(A\rho(t))$.
When the state evolves in time but the operators are constant the picture of the system is called the Schr\"odinger-picture.
Using the cyclic property of the trace, $\tr(ABC) = \tr(CBA)$, the Heisenberg-picture, a dual interpretation of the same physical system emerges, in which the state remains the same while the operators evolve in time.
It is well known that the operators in this picture evolve as $A(t) = U^{\dagger} A U$, where $A$ is the initial operator, see Refs.~\cite{Cohen-Tannoudji1977, Sakurai2010}.

\subsection{Quantum metrology}
\label{sec:bg-quantum-metrology}

In this section, we summarize important recent advances in quantum metrology.
Quantum metrological setups encode some unknown parameter into the quantum state.
Then, from the readout of the final state one could in principal estimate this parameter.
The basic ideas of quantum metrology emerge when one applies the notions of estimation theory to the intrinsic probabilistic nature of quantum mechanics.
Merging the probabilistic features of quantum mechanics and the estimation theory is not trivial.
Nevertheless, with the initial pioneering works of C. W. Helstrom, W. K. Wootters, and S. L. Braunstein and C. M. Caves, in 1969, 1981 and 1992-1994 respectively \cite{Helstrom1969, Wootters1981, Braunstein1992, Braunstein1994}, until the works of V. Giovannetti {\it et al}, and M. G. Paris, roughly two decades later \cite{Giovannetti2004, Paris2009}, quantum metrology got firm foundations.
Later, advanced works in quantum metrology appeared \cite{Hyllus2010, Hyllus2012, Hyllus2010a, Kolodynski2010, Kolodynski2013}
together with experimental realizations \cite{Behbood2013, Koschorreck2011, Luecke2011}, which raised the interest in this topic.
In this section, we will highlight the most important aspects of this field and with this we will conclude this introductory chapter.

\begin{figure}[htp]
  \centering
  \includegraphics[scale=1.15]{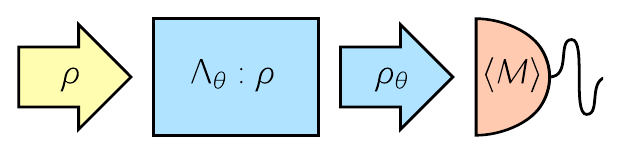}
  \caption[The estimation process in quantum metrology]{
  Sequence of the different steps of the estimation process in quantum metrology.
  First, an input state $\rho$ enters the region in which the unknown parameter $\theta$ is imprinted on it, which is represented with $\Lambda_{\theta}$ for the most general case.
  The encoded parameter $\theta$ is inferred from the measured quantity $\expect{M}$ over the final state $\rho_{\theta}$.}
  \label{fig:bg-preparation-encoding-estimation}
\end{figure}

The most basic scheme for a metrological setup in the present context is the following.
First, a state $\rho$ is prepared followed by a general evolution represented by a mapping $\Lambda_{\theta}$ in which the unknown parameter $\theta$ is imprinted on the state.
Finally, the outgoing state is characterized by some measured quantity $\expect{M}$, which allows us to infer the value of the parameter $\theta$.
Figure~\ref{fig:bg-preparation-encoding-estimation} illustrates the main steps of quantum metrology.

In the many-particle case, most of the metrology experiments have been done for systems with simple Hamiltonians that do not contain interaction terms.
Such Hamiltonians cannot create entanglement between the particles.
A typical situation is that we rotate our many particle state by some angle and we want to estimate the rotation angle $\theta$.
It has been shown that particles exhibiting quantum correlations, or more precisely, quantum entanglement \cite{Guehne2009, Luis2004}, provide a higher precision than an ensemble with non-entangled particles.
The most important question is how the achievable precision of the angle estimation $\varian{\theta}$ scales with the number of particles $N$.
Very general derivations lead to, at best,
\be
  \label{eq:bg-shot-noise-scaling}
  \varian{\theta}\sim \frac{1}{N}
\ee
for non-entangled particles.
The equation above is called \emph{shot-noise} or standard scaling, the term originating from the shot-noise in electronic circuits, which is due to the discrete nature of the electric charge.
On the other hand, quantum entanglement makes it possible to reach
\be
  \label{eq:bg-heisenberg-scaling}
  \varian{\theta}\sim \frac{1}{N^2}
\ee
which is called the \emph{Heisenberg} scaling.
Note that if the Hamiltonian of the dynamics has interaction terms, then these bounds can be surpassed \cite{Luis2004, Napolitano2011, Boixo2007, Braun2011, Roy2008, Choi2008, Rey2007}.

It is time to mention that the calculations above have been carried out for an ideal situation.
When an uncorrelated noise is present in the system, it turns out that for large enough particle number the scaling becomes shot-noise scaling \cite{Demkowicz-Dobrzanski2012}.
The possible survival of a better scaling under correlated noise, under particular circumstances, or depending on some interpretation of the metrological task, is at the center of attention currently.
All these are strongly connected to the question of whether strong multipartite entanglement can survive in a noisy environment.

Finally, note that often, instead of $\varian{\theta}$ one calculates its inverse, which is large for high precision.
It scales as $\varinv{\theta}\sim N$ for shot-noise scaling and as $\varinv{\theta}\sim N^2$ for the Heisenberg scaling.

\subsubsection{Quantum magnetometry}
\label{sec:bg-quantum-magnetometry}

Without loss of generality, we present in this section the characterization of the precision of one of the simplest metrological tasks, namely the estimation of a homogeneous magnetic field based on the interaction between the system and the field.
In this section, we will mostly study the interaction of a system with a homogeneous magnetic field in the $z$-direction.
In the Chapter~\ref{sec:gm}, we will show a different situation in which the magnetic field changes linearly with the position of the system.
Coupling the magnetic moment of the state and the magnetic field, we imprint the magnetic field strength on the evolved state.
Finally, measuring how the state has changed one could in principle infer on the strength of the magnetic field.

In general, we will say that the magnetic moment of the state comes exclusively from the spin angular momentum, neglecting any possible contribution from the orbital angular momentum.
This way the physics is simpler.
This is justified in the sense that most of the recent experiments in this context have been carried out with ion-traps, Bose-Einstein condensates (BEC) or at most cold atomic ensembles, which have indeed a negligible orbital angular momentum.

Beside this considerations, the interaction Hamiltonian can be written as
\be
  H = - \bs{\mu} \cdot \bs{B},
\ee
Now in the simplest case we will choose the magnetic field to be pointing in the $z$-direction as ${B}=B\bs{k}$, where $\bs{k}$ is the unitary vector pointing to the $z$-direction.
This way one does not need to determine the direction of the magnetic field.

The magnetic moment of the system is proportional to the total angular momentum, $\bs{\mu}=-\mu_\text{B} g_{\text{s}}\hbar \bs{J}$, where $\mu_{\text{B}}$ and $g_{\text{s}}$ are the Bohr magneton and the anomalous gyromagnetic factor respectively, and where we multiply it by the Plank reduced constant $\hbar$.
Note that throughout all the thesis the angular momentum component were defined with $\hbar=1$ for simplicity, e.g., see Eq.~\eqref{eq:bg-single-particle-am-components}.
Hence, one can rewrite the interaction Hamiltonian as
\be
  \label{eq:bg-hamiltonian-homogeneous-field}
  H = \gamma B J_z.
\ee
where $\gamma = \mu_\text{B} g_{\text{s}}\hbar$ and we have used that $\bs{J}\cdot\bs{k}=J_z$.
Finally, the unitary operator leading the evolution of the system can be written as
\be
  \label{eq:bg-unitary-homogeneous-field}
  U = \exp(-i\theta J_z),
\ee
where the magnetic field strength is encoded into the phase-shift $\theta=-\mu_\text{B} g_\text{s} t B$ and $t$ is the evolution time.

We have to mention that for large particle ensembles, typically only collective quantities can be measured in order to characterize the state in different phases of the metrological sequence, see Figure~\ref{fig:bg-preparation-encoding-estimation}.
Such collective quantities are in this case the total angular momentum components defined in Eq.~\eqref{eq:bg-total-angular-momentum-progection-operators} and their linear combinations.
More concretely, we can measure the expectation values of any direction
\be
  \label{eq:bg-total-angular-momentum-projector-arbitrary-direction}
  J_{\bs{n}}:=\sum_{l=x,y,z} n_l J_l,
\ee
where $\bs{n}=(n_x,n_y,n_z)$ is a unit vector describing the direction of the component.

\subsubsection[Metrology with almost polarized states]{Metrology with almost polarized states, including spin-squeezed states}
\label{sec:bg-metrology-with-almost-polarized}

\begin{figure}[htp]
  \centering
  \includegraphics[scale=1.15]{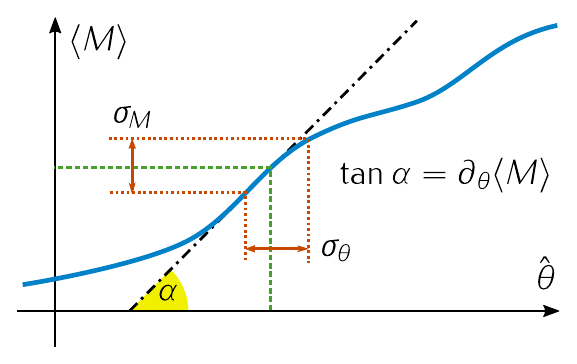}
  \caption[Error-propagation formula]{
  (Blue solid) Functional relation between the expectation value $\expect{M}$ and the estimator of the wanted parameter $\hat\theta$.
  (Green dashed) One to one correspondence when the estimator $\hat{\theta}$ is based on $\expect{M}$. (Red dotted) Obtaining the error of the estimate is based on Eq.~\eqref{eq:bg-error-propagation-formula}.
  The slope of the curve at that point, denoted with $\tan\alpha$, directly relates the uncertainty $\sigma_M$ on the measured quantity $\expect{M}$ and the error on the estimation $\sigma_\theta$.}
  \label{fig:bg-expect-m-evo}
\end{figure}

Let us present one of the basic approaches to calculate the metrological precision of a quantum setup.
In order to estimate the phase-shift $\theta$ we measure the expectation value of a Hermitian operator, which we will denote by $M$ in the following.
If the evolution time is a constant, then estimating $\theta$ is equivalent to estimating the magnetic field strength $B$ appearing in Eq.~\eqref{eq:bg-hamiltonian-homogeneous-field}.
The precision of the estimation can be characterized with the error propagation formula as
\be
  \label{eq:bg-error-propagation-formula}
  \varian{\theta} = \frac{\varian{M}}{|\partial_\theta\expect{M}|^2},
\ee
where $\expect{M}$ is the expectation value of the operator $M$.
We used that $\expect{M}$ is a random variable with probability $\prob(\expect{M}=m|\theta)$ for the outcomes of the measurements, see the Section~\ref{sec:bg-estimators}.
For a recent review discussing this approach in detail, see Ref.~\cite{Kolodynski2010}.
Based on the formula Eq.~\eqref{eq:bg-error-propagation-formula}, one can see that the larger the slope $|\partial_\theta \expect{M}|$, the higher the precision.
On the other hand, the larger the variance $\varian{M}$, the lower the precision.
Figure~\ref{fig:bg-expect-m-evo} helps to interpret the quantities appearing in Eq.~\eqref{eq:bg-error-propagation-formula}.

We focus on multi-partite systems of spin-$\frac{1}{2}$ particles, widely known as qubits.
Let us start with an almost polarized state.
The state totally polarized along the $l$-direction is written as
\be
  \ket{+j}_l^{\otimes N}\equiv \ket{+j}_l^{(1)}\ket{+j}_l^{(2)}\cdots\ket{+j}_l^{(N)},
  \label{eq:bg-totally-polarized}
\ee
For $l=y$, the state Eq.~\eqref{eq:bg-totally-polarized} is almost all the times one of the states that saturates the shot-noise limit, which will be defined later in this section.
In the present case, the spin-vector of the ensemble, originally pointing into the $y$-direction, will be rotated.
The rotation after the evolution is used to estimate the strength of the magnetic field.
A way to measure the rotation of the system is to measure the expectation value $\expect{J_x}$, which is zero at the beginning.

For small angles of $\theta$ and using the Eq.~\eqref{eq:bg-error-propagation-formula} after substituting $M$ by $J_x$, we have that
\be
  \label{eq:bg-error-propagation-measuring-jx}
  \varinv{\theta} = \frac{|\partial_\theta \expect{J_x}|^2}{\varian{J_x}},
\ee
for a state almost completely polarized along the $y$-axis.
Next, we have to compute these values for small angles of $\theta$.

To compute them we use the Heisenberg picture of the operator $J_x$, which after applying the unitary operator Eq.~\eqref{eq:bg-unitary-homogeneous-field}, is written as
\be
  J_x(\theta) = U^{i\theta J_z} J_x U^{-i\theta J_z} = \coss{\theta}J_x -  \sins{\theta} J_y,
\ee
where we introduce a notation for trigonometric functions such that $\coss{x}=\cos(x)$, $ \sins{x}=\sin(x)$ and $\tans{x}=\tan(x)$.
The square of $J_x$ is simply
\be
  J_x^2(\theta) = \coss{\theta}^2J_x^2 -2 \coss{\theta}\sins{\theta}\{J_x,J_y\}+  \sins{\theta}^2 J_y^2,
\ee
where $\{A,B\}=AB+BA$ is the anti-commutator.
We also compute the derivative with respect to $\theta$ of $\expect{J_x}$ as
\be
  \partial_{\theta}\expect{J_x}(\theta)=\partial_{\theta}\coss{\theta}\expect{J_x} - \partial_{\theta}\sins{\theta}\expect{J_y} = -\sins{\theta}\expect{J_x}-\coss{\theta}\expect{J_y}.
\ee

Hence, assuming that we are in the small angle limit we can neglect the terms proportional to $\sin(\theta)$ when computing the Eq.~\eqref{eq:bg-error-propagation-measuring-jx}.
Finally, the precision of the estimation is given by
\be
  \label{eq:bg-error-propagation-measuring-jx-computed}
  \varinv{\theta} = \frac{|\coss{\theta}\expect{J_y}|^2}{\coss{\theta}^2\expect{J_x^2} - (\coss{\theta}\expect{J_x})^2} = \frac{\expect{J_y}^2}{\varian{J_x}},
\ee
when we measure $\expect{J_x}$ to estimate the angle $\theta$.

For the state totally polarized along the $y$-axis, the initial expectation values $\expect{J_x}$, $\expect{J_x^2}$ and $\expect{J_y}$ needed to compute the precision are
\begin{subequations}
  \begin{align}
    \expect{J_x}_{\text{tp}}  & = 0,\\
    \expect{J_x^2}_{\text{tp}}  & = \tfrac{N}{4},\\
    \expect{J_y}_{\text{tp}}  & = \tfrac{N}{2}.
  \end{align}
\end{subequations}
Thus, we obtain a precision that scales linearly with $N$,
\be
  \varinv{\theta}_{\text{tp}} = \frac{N^2/4}{N/4} = N.
\ee
Note that the totally polarized state Eq.~\eqref{eq:bg-totally-polarized} is a separable pure state with all particles pointing into the $y$-direction.
Hence, we obtained the shot-noise scaling, even with very simple qualitative arguments.

A way of improving the precision is considering that the variances of the angular momentum components are bounded by the Heisenberg uncertainty relation as \cite{Kitagawa1993}
\be
  \varian{J_x}\varian{J_z} \geqslant \frac{1}{4}|\expect{J_y}|^2.
\ee
When decreasing the variance $\varian{J_x}$ to obtain a better precision, our state fulfills
\be
  \varian{J_x}<\frac{1}{2}|\expect{J_y}|,
\ee
where the main spin points along the $y$-axis and where we would obtain $\varian{J_x}=\frac{1}{2}|\expect{J_y}|$ for totally polarized states.
Such states are called \emph{spin-squeezed} states \cite{Kitagawa1993, Wineland1994, Sorensen2001, Ma2011} and they can in principal overcome the shot-noise scaling.

Next, we can ask, what the best possible phase estimation precision is for the metrological task considered in this section.
For that, we have to use the following inequality based on general principles of angular momentum theory
\be
  \label{eq:bg-total-angular-momentum-saturability}
  \expect{J_x^2+J_y^2+J_z^2} \leqslant \frac{N(N+2)}{4}.
\ee
Note that Eq.~\eqref{eq:bg-total-angular-momentum-saturability} is saturated only by symmetric multiparticle states, see Appendix~\ref{app:angular-subspaces}.
Together with the identity connecting the second moments, variances and expectation values
\be
  \varian{J_l} + \expect{J_l}^2 = \expect{J_l^2},
\ee
Eq.~\eqref{eq:bg-total-angular-momentum-saturability} leads to a bound on the uncertainty for the angular momentum component in the anti-squeezed direction, in this case the $z$-direction, as
\be
  \varian{J_z} \leqslant \frac{N(N+2)}{4} - \expect{J_y}^2 = \frac{N}{2} + \frac{N^2}{4}\lpar 1 - \frac{\expect{J_y}^2}{J_{\max}^2}\rpar,
\ee
where $J_{\max}=N/2$ is the maximum value an angular momentum component can take.
This leads to a bound in the precision when measuring $\expect{J_x}$ as
\be
  \label{eq:bg-unpolarize-states-are-better}
  \varinv{\theta} = \frac{\expect{J_y}^2}{\varian{J_x}}\leqslant 4\varian{J_z}\leqslant 2N + N^2\lpar 1 - \frac{\expect{J_y}^2}{j_{\max}}\rpar,
\ee
which indicates that the precision is limited for almost completely polarized states to the shot-noise scaling and that less polarized states can overcome the shot-noise limit.
The bound above is not optimal, as for the fully polarized states we would expect $N$ and we obtain $2N$.

\subsubsection{The quantum Fisher information}
\label{sec:bg-qfi}

In this section, we review the theoretical background of the Fisher information, the Cram\'er-Rao bound and we introduce the quantum Fisher information.

First of all, we have already introduced the Fisher information and the Cram\'er-Rao bound in Section~\ref{sec:bg-estimators}.
The formula Eq.~\eqref{eq:bg-classical-cr-bound-and-fi} tells us that if we measure an operator, say $M$, and we know its probability distribution function $\prob(\expect{M}=m|\theta)$, we can bound the achievable precision.
In quantum mechanics the PDF of a state $\rho_\theta$ when measuring the operator $M$ is given by
\be
  \prob(m|\theta) = \tr({\Pi_{m}\rho_\theta}),
\ee
where $\Pi_{m}$ are the projector operators or the eigenstates on which $M$ is expanded, $M = \sum m \Pi_m$.
Following arguments found on Refs.~\cite{Giovannetti2004, Paris2009}, one can arrive at a universal bound valid for any measurement.
This bound is called the quantum Cram\'er-Rao bound given as
\be
  \label{eq:bg-quantum-cr-bound}
  \varinv{\theta} \leqslant \qfif{\rho,J_z},
\ee
where $\qfif{\rho, J_Z}$ denotes the quantum Fisher information for the initial state $\rho$ and the unitary evolution generated by $J_z$.
In principle, it might be difficult to find the operator that leads to the best estimation precision just by trying several operators.
Fortunately, the theory tells us that the bound Eq.~\eqref{eq:bg-quantum-cr-bound} can be saturated \cite{Helstrom1976, Holevo1982}.

As a direct consequence, based on Eq.~\eqref{eq:bg-error-propagation-formula} for any given $M$, we have that
\be
  \qfif{\rho,J_z} \geqslant \frac{|\partial_{\theta}\expect{M}|^2}{\varian{M}}
\ee
is an upper bound for the precision obtained by measuring $\expect{M}$.
For example, when the rotation of the system is estimated by measuring $\expect{J_x}$, based on the Eq.~\eqref{eq:bg-error-propagation-measuring-jx-computed} the quantum Fisher information is bounded from below by \cite{Pezze2009}
\be
  \label{eq:bg-pezze-bound}
  \qfif{\rho,J_z} \geqslant \frac{\expect{J_y}^2}{\varian{J_x}}.
\ee

The quantum Fisher information can be computed by closed formulas such as
\begin{align}
  \label{eq:bg-qfi-definition-eigen-decomposition}
  \qfif{\rho,J_z} &= 2\sum_{\lambda,\nu} \frac{(p_\lambda-p_\nu)^2}{p_\lambda+p_\nu} |\braopket{\lambda}{J_z}{\nu}|^2, \\
  \label{eq:bg-qfi-definition-convex-roof}
  \qfif{\rho,J_z} &= \inf_{p_k,\ket{\psi_{k}}} 4 \sum_{k}p_k \varian{J_z}_{\ket{\psi_k}},
\end{align}
for the case of unitary evolution Eq.~\eqref{eq:bg-unitary-homogeneous-field}.
In the Eq.~\eqref{eq:bg-qfi-definition-eigen-decomposition}, the state is decomposed in its eigenbasis as $\rho=\sum_{\lambda}p_\lambda\ketbra{\lambda}{\lambda}$.
On the other hand, in the Eq.~\eqref{eq:bg-qfi-definition-convex-roof}, the state is arbitrarily decomposed as $\rho=\sum_k p_k \ketbra{\psi_k}{\psi_k}$.
Hence, the first equation is based on the eigen-decomposition of the state whereas the second is the convex-roof of $4\varian{J_z}$ \cite{Paris2009, Toth2013, Yu2013}.
Both formulas yield the same expression for pure states
\be
  \label{eq:bg-qfi-for-pure-states}
  \qfif{\ket{\psi}, J_z} = 4\varian{J_z}.
\ee
Using the identity $(a-b)^2 = (a+b)^2 - 4 ab$ and that $p_\lambda = \delta_{\lambda,1}$ (we choose the pure state to be $\ket{1}$ without loss of generality) in Eq.~\eqref{eq:bg-qfi-definition-eigen-decomposition}, we arrive at the result
\be
  \begin{split}
    \qfif{\ket{\psi},J_z} &= 2\sum_{\lambda, \nu} \lpar p_\lambda + p_\nu - \frac{4p_\lambda p_\nu}{p_\lambda+p_\nu} \rpar|\braopket{\lambda}{J_z}{\nu}|^2\\
    &=4\tr(J_z\rho) - 8 \sum_{\lambda,\nu}\frac{p_\lambda p_\nu}{p_\lambda+p_\nu} |\braopket{\lambda}{J_z}{\nu}|^2\\
    &= 4\varian{J_z}.
  \end{split}
  \label{eq:bg-rewrite-qfi}
\ee
From the Eq.~\eqref{eq:bg-qfi-definition-convex-roof}, we also obtain the same result since the convex-roof over a pure state must be computed in the state itself.

Some interesting properties of the quantum Fisher information emerge from these expressions and we summarize them in the following list:
\begin{enumerate}
  \item The QFI is convex in states, as it is directly shown in Eq.~\eqref{eq:bg-qfi-definition-convex-roof}. It can also be proven if one starts from Eq.~\eqref{eq:bg-qfi-definition-eigen-decomposition},
  \be
    \qfif{p\rho_1+(1-p)\rho_2, J_z}\leqslant p\qfif{\rho_1, J_z}+(1-p)\qfif{\rho_2, J_z}.
  \ee
  \item Based on Eq.~\eqref{eq:bg-qfi-definition-convex-roof}, it has recently been shown, that the quantum Fisher information is the largest convex function that fulfills Eq.~\eqref{eq:bg-qfi-for-pure-states} \cite{Toth2013, Yu2013}.
  \item For pure states $\qfif{\ket{\psi}, J_z} = 4\varian{J_z}$, as it has been already shown.
  \item For all states, the QFI is smaller or equal to four times the variance \be
    \qfif{\rho,J_z} \leqslant 4\varian{J_z}_\rho.
  \ee
\end{enumerate}
It can be proven that the variance is the concave-roof of the variance itself \cite{Toth2013}.
Hence, the main relation between the quantum Fisher information and the variance can be summarized as follows.
For any decomposition $\{p_k, \ket{\psi_k}\}$ of a state $\rho$ we have
\be
  \frac{1}{4}\qfif{\rho,J_z}\leqslant \sum_{k}p_k \varian{J_z}_{\ket{\psi_k}}\leqslant \varian{J_z}_{\rho},
\ee
where the upper bound and the lower bound are both tight in the sense that there are decompositions that saturate the first inequality, and there are others that saturate the second one.

After the discussion relating the quantum Fisher information to the variance, and examining its convexity properties, we list some further useful relations for the QFI.
In this case, we substitute the generator of the phase shift $J_z$ by some more general Hermitian operators.
From Eq.~\eqref{eq:bg-qfi-definition-eigen-decomposition}, we can obtain directly the following identities:
\begin{enumerate}
  \item The formula Eq.~\eqref{eq:bg-qfi-definition-eigen-decomposition} does not depend on the diagonal elements of the generator written in the eigenbasis of the state.
  Hence,
  \be
    \qfif{\rho, A} = \qfif{\rho, A+D},
  \ee
  where $D$ is an arbitrary matrix that is diagonal in the eigenbasis of the state, i.e., $[\rho,D]=0$.
  \item The following identity holds for all unitary dynamics $U$ as it could be expected from the Schrodinger vs Heisenberg pictures,
  \be
    \qfif{U\rho U^\dagger, A} = \qfif{\rho, U^\dagger A U}.
  \ee
  In particular the QFI does not change for unitary dynamics of the type $U=e^{-iB}$ when $[A,B]=0$.
  \item The quantum Fisher information is additive under tensor product as
  \be
    \qfif{\rho^{(1)}\otimes \rho^{(2)}, A^{(1)}\otimes \mtxid^{(2)}+\mtxid^{(1)}\otimes A^{(2)}} = \qfif{\rho^{(1)},A^{(1)}}+ \qfif{\rho^{(2)},A^{(2)}}.
    \label{eq:bg-qfi-additive-for-tensor-product}
  \ee
  For $N$-fold tensor product of the system, we obtain an $N$-fold increase in the quantum Fisher information as
  \be
    \qfif{\rho^{\otimes N},\textstyle\sum_{n=1}^{N}A^{(n)}}=N\qfif{\rho, A},
  \ee
  where $A^{(n)}\equiv A$ for all $n$.
  \item The quantum Fisher information is additive under a direct sum \cite{Liu2014}
  \be
    \qfif{\bigoplus_{k}p_k \rho_k,\bigoplus_{k} A_k} = \sum_{k}p_k\qfif{\rho_k,A_k},
  \ee
  where $\sum_k p_k = 1$.
  The above equation is relevant, for instance, for experiments where the particle number variance is not zero, and the $\rho_k$ corresponds to density matrices with a fixed particle number \cite{Hyllus2010, Hyllus2012}.
\end{enumerate}

\subsubsubsection{Entanglement and the quantum Fisher information}

The quantum Fisher information is strongly related to entanglement.
In this section, we discuss now this relation and we review some important facts concerning it.
We will show that entanglement is needed to overcome the shot-noise sensitivity in very general metrological tasks.
Moreover, not only entanglement but multipartite entanglement is necessary for a maximal sensitivity.
All these statements will be derived in a very general framework, based on the quantum Fisher information.
We will also briefly discuss the question whether inter-particle entanglement is an appropriate notion for our systems.

Let us first examine the upper bounds on the QFI for general quantum states and for separable states.
Due to the Cram\'er-Rao bound \eqref{eq:bg-quantum-cr-bound}, they are also bounds for the sensitivity of the phase estimation.
Entanglement has been recognized as an advantage for several metrological tasks (see, e.g., Refs. \cite{Sorensen2001, Boixo2008}).
For a general relationship for linear interferometers, we can take advantage of the properties of the quantum Fisher information discussed before.
Since for pure states the quantum Fisher information equals four times the variance, for pure product states we can have at most
\be
  \qfif{\rho,J_z} = 4\varian{J_z} = 4 \sum_{n=1}^N \varian{j_z^{(n)}} \leqslant 4Nj^2,
\ee
where $j$ is the spin of the particles.
Note that for spin-$\frac{1}{2}$ we recover the usual threshold one can find for instance in Ref.~\cite{Giovannetti2006}.
For the second equality, we used the fact that for a product state the variance of a collective observable is the sum of the single-particle variances.
Due to the convexity of the QFI, this upper bound is still valid for all separable states of the form Eq.~\eqref{eq:bg-separable-state-definition} and we obtain \cite{Pezze2009}
\be
  \label{eq:bg-shot-noise-limit}
  \qfif{\rho, J_z} \leqslant 4Nj^2,
\ee
a bound for not entangled states.
All states violating Eq.~\eqref{eq:bg-shot-noise-limit} are entangled.
The entangled states make it possible to surpass this bound, the shot-noise limit, and some might be more useful than separable states for the metrological tasks at hand.

The maximum achievable precision for general states, called the Heisenberg limit, can be obtained evaluating the Eqs.~\eqref{eq:bg-qfi-definition-eigen-decomposition} or \eqref{eq:bg-qfi-definition-convex-roof} for pure states only.
Therefore, similarly we have
\be
  \label{eq:bg-heisenberg-limit}
  \qfif{\rho, J_z} = 4 \varian{J_z}\leqslant 4N^2j^2,
\ee
which is a valid bound still for mixed states due to the convexity of the QFI.
Note that such a bound has the Heisenberg scaling Eq.~\eqref{eq:bg-heisenberg-scaling}.
Note also that our derivation is very simple, and does not require any information about what operator we measure to estimate $\theta$.
Equation~\eqref{eq:bg-shot-noise-limit} has been used already to detect entanglement based on the metrological performance of the quantum states in Refs.~\cite{Krischek2011, Luecke2011}.

At this point one might ask whether all entangled states can provide a sensitivity larger than the shot-noise limit.
This would show that entanglement is equivalent to metrological usefulness.
Concerning linear interferometers, it has been proven that not all entangled states violate the shot-noise limit Eq.~\eqref{eq:bg-shot-noise-limit}, even allowing local unitary transformations.
Thus not all entangled states are useful for phase estimation \cite{Hyllus2010}.
It has been shown that there are even highly entangled pure states that are not useful.
Hence, the presence of entanglement is a necessary but rather than a sufficient condition.

The quantum Fisher information can be used to define an entanglement parameter that characterizes the metrological usefulness as
\be
  \chi = \frac{\qfif{\rho, J_z}}{N},
  \label{eq:bg-entanglement-criterion-qfi}
\ee
which is not larger than one for separable states and it can take at most the value of $N$ as is deduced from Eq.~\eqref{eq:bg-heisenberg-limit}.

Based on methods similar to the ones used to find the bound for separable states in Eqs.~\eqref{eq:bg-shot-noise-limit} and \eqref{eq:bg-heisenberg-limit}, for $k$-producible states the QFI is bounded from above as \cite{Hyllus2012, Toth2012}
\be
  \begin{split}
    \chi &\stackrel{\phantom{k\ll N}}{\leqslant} \frac{sk^2 + (N-sk)^2}{N}\stackrel{k\ll N}{\approx} k,
  \end{split}
  \label{eq:bg-entanglement-depth-for-qfi}
\ee
where $s$ is the integer part of $\frac{N}{k},$ and we write out explicitly the bound for
 $k\ll N$.
It is instructive to write the equation above for the case in which $N$ is exactly divisible by $k$ as
\be
  \chi\leqslant{k}.
\ee
Thus, the bounds reachable by $k$-producible states are distributed linearly in $k$.

It is also instructive to define a QFI averaged over all possible directions.
Simple calculations show that
\be
  \label{eq:bg-average-qfi}
  \underset{\bs{n}}{\text{avg}}\,\qfif{\rho, J_{\bs{n}}} \equiv  \int_{\bs{n}=(\coss{\varphi}\sins{\vartheta},\sins{\varphi}\sins{\vartheta},\coss{\vartheta})} \qfif{\rho,J_{\bs{n}}}\sin(\vartheta)\,\text{d}\varphi\text{d}{}\vartheta = \frac{1}{3}\sum_{l=x,y,z} \qfif{\rho,J_l},
\ee
where we used the spherical coordinates such that $\varphi$ and $\vartheta$ are the azimuthal and polar angles respectively.
Therefore, bounds similar to Eqs.~\eqref{eq:bg-shot-noise-limit} and \eqref{eq:bg-heisenberg-limit} for separable and general states can be obtained for the average quantum Fisher information as
\begin{subequations}
\begin{align}
  \underset{\bs{n}}{\text{avg}}\,\qfif{\rho, J_{\bs{n}}} &\leqslant \frac{2}{3}N, \\
  \underset{\bs{n}}{\text{avg}}\,\qfif{\rho, J_{\bs{n}}} &\leqslant \frac{1}{3}N(N+2).
\end{align}
\end{subequations}
Similarly to Eq.~\eqref{eq:bg-entanglement-criterion-qfi}, an entanglement criterion can be constructed for the averaged QFI.
Similar to Eq.~\eqref{eq:bg-entanglement-criterion-qfi} an entanglement criterion can be constructed for the averaged QFI.

Finally, we also mention that bound entangled states can also be detected with the entanglement criteria based on the quantum Fisher information.
Bound entanglement is a weak type of entanglement, which is not distillable with local operations and classical communication \cite{Horodecki2009, Guehne2009}.
Ref.~\cite{Hyllus2012} presented states that were detected as bound entangled based on the criterion for the average QFI Eq.~\eqref{eq:bg-average-qfi}.
Ref.~\cite{Czekaj2015} presented states that violate the criterion based on a bound for the quantum Fisher information similar to Eq.~\eqref{eq:bg-entanglement-criterion-qfi}.

%% file: 03-vicinity.tex
\section{Metrology in the vicinity of Dicke states}
\input{snp/doubleLineWaterMark.tex}
\label{sec:vd}

\quotes{Robert H. Dicke}{An experimentalist should not be unduely inhibited by theoretical untidyness.}

\lettrine[lines=2, findent=3pt,nindent=0pt]{I}{n} this chapter we will present recent results regarding the metrological usefulness of a family of unpolarized states.
Such states can be used as trial states to estimate the homogeneous magnetic field strength, see Section~\ref{sec:bg-quantum-magnetometry} for references about magnetometry.
It turns out that unpolarized states are the most adequate states to reach the Heisenberg limit, as it was shown in the Section~\ref{sec:bg-quantum-metrology}.
Hence, these states have attracted considerable interest.

One of the figures of merit of these states is the so-called unpolarized Dicke state \cite{Dicke1954} given in the $l$-basis, which consists of an equal number of qubits pointing in the $l$-direction and pointing in the opposite direction while the whole state is symmetrized, and where $l=x,y,z$.
It can be written as
\be
   \ket{\dicke{N}}_l\equiv \ket{\dicke{N,N/2}}_l:= \binom{N}{N/2}^{-\frac{1}{2}}
  \sum_{k\in \sigma_\text{s}}
  \mathcal{P}_{k} ( \ket{0}_l^{\otimes N/2} \ket{1}_l^{\otimes N/2} ),
  \label{eq:vd-unpolarized-dicke}
\ee
where $\mathcal{P}_k$ denote all the distinct permutations in $\sigma_\text{s}$.
$\ket{0}_l$ and $\ket{1}_l$ are single-particle quantum states of a spin-$\frac{1}{2}$ system in the $l$-basis, see Appendix~\ref{app:angular-subspaces} for more details.
Note that in Eq.~\eqref{eq:vd-unpolarized-dicke} we omit the subscript giving the number of $\ket{1}$'s which is the notation we will follow in this chapter.
Such a state is known to be highly entangled \cite{Toth2007, Toth2009a} and can reach Heisenberg scaling when used for magnetometry \cite{Holland1993}.

One of the characteristics of state Eq.~\eqref{eq:vd-unpolarized-dicke} is that it is an eigenstate of the collective operator $J_l$ with corresponding eigenvalue equal to zero.
At the same time, it lives in the subspace where the collective total spin is maximum, i.e, $\expect{\bs{J}^2}=N(N+2)/4$.
Based on these and the fact that the state is unpolarized, we can see that has a very large uncertainty for the collective spin operators perpendicular to $J_l$.

For metrology, we chose the magnetic field to be pointing in the $z$-axis.
Hence, the Dicke state we choose must be an eigenstate of a perpendicular component of the angular momentum operator such as $J_x$.
Hence, we will consider a scheme in which the state is rotated around the $z$-direction and the rotation angle must be estimated based on collective measurements.
A criterion to detect the metrological usefulness of states of this type has been derived in Ref.~\cite{Zhang2014}.

In this chapter, we present a condition for metrological usefulness for the case when the second moment of a total angular momentum component is measured to obtain an estimate for the rotation angle.
Our method is expected to simplify the experimental determination of metrological sensitivity since it is much easier to measure the collective operators of the state than carrying out the metrological procedure and measure directly the sensitivity.
We also test our approach using the experimental results of Refs.~\cite{Luecke2011, Krischek2011}, which realize parameter estimation with a Dicke state.
Thus, our work is expected to be useful for similar experiments in the future.

The chapter is organized as follows.
In Section~\ref{sec:vd-unpolarized-states-magnetometry}, we review some important concepts behind the theory of metrology with unpolarized states.
In Section~\ref{sec:vd-evolution-of-the-expectation-values}, we present our criterion.
In Section~\ref{sec:vd-comparison-with-qfi}, we compare our findings to sensitivity bounds obtained from the quantum Fisher information.
In Section~\ref{sec:vd-testing-with-experimental-data}, we show how to apply our criterion to experimental results.

\subsection{Unpolarized Dicke states for magnetometry}
\label{sec:vd-unpolarized-states-magnetometry}

In Eq.~\eqref{eq:bg-unpolarize-states-are-better} we have shown that unpolarized states may overcome the shot-noise limit, i.e., the precision achieved by completely polarized states.
While the quantum Fisher information would give us directly the performance of the state, we typically cannot compute it because a complete knowledge of the state would be necessary, see Eq.~\eqref{eq:bg-qfi-definition-eigen-decomposition}.
On the other hand, we can use the error propagation formula Eq.~\eqref{eq:bg-error-propagation-formula} to obtain a bound on the achievable precision which at the same time bounds the QFI.

\begin{figure}[htp]
  \centering
  \includegraphics[scale=.65]{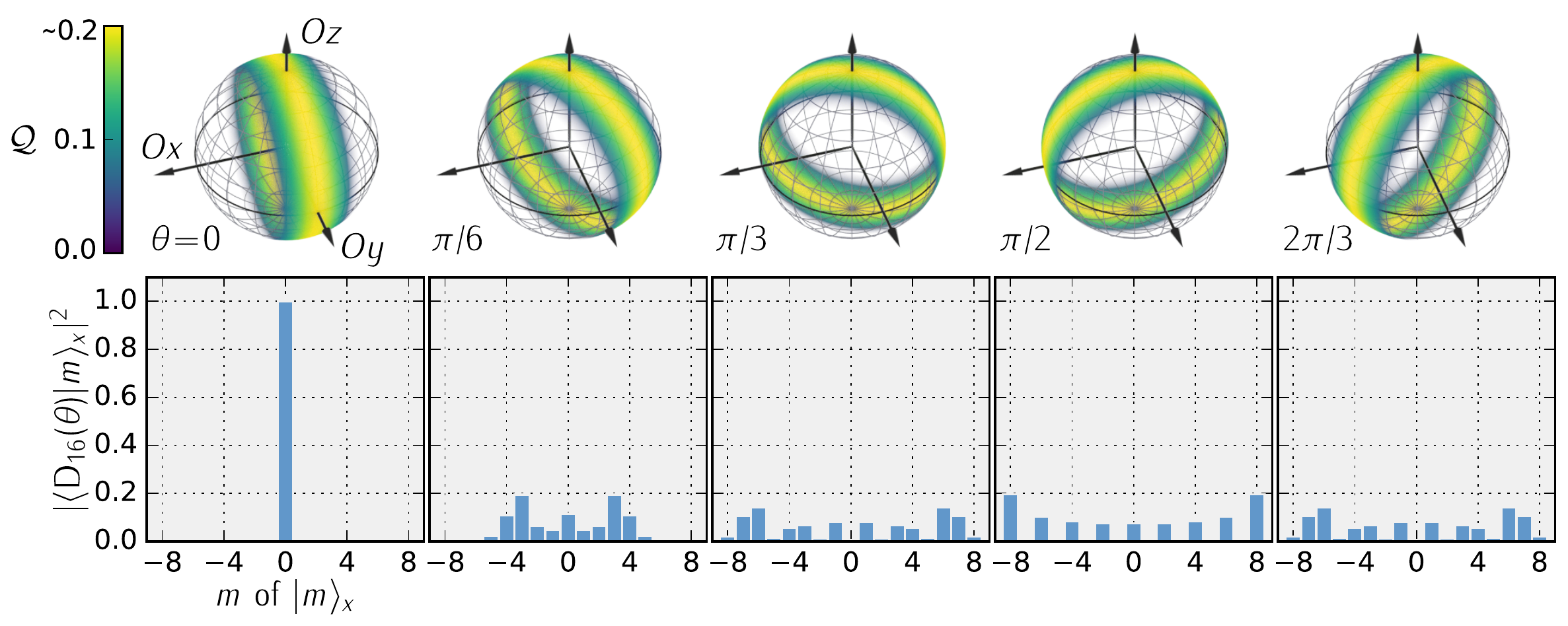}
  \caption[Sequence of Dicke state evolution]{
  Sequence of the evolution of an unpolarized Dicke state of 16 qubits for $\theta=\{i\pi/6\}_{i=0}^4$.
  (sequence-above) Bloch spheres representing the Husimi quasi-probabilistic distribution $\mathcal{Q}$ of the state, see Appendix~\ref{app:husimi-representation} for details.
  (sequence-below) PDF of the $J_x$ positive-operator valued measure (POVM) for each step of the sequence.}
  \label{fig:vd-secuence-evo}
\end{figure}

As one can see in Figure~\ref{fig:vd-secuence-evo}, a pure unpolarized Dicke state $\ket{\dicke{N}}_x$ of 16 qubits, an eigenstate of the $J_x$ operator, is rotated around the $z$-axis.
The state is unpolarized so the expectation value of any component of the total angular momentum remains zero.
It turns out that measuring the evolution of the second moment of $J_x$ allows the estimation of rotation angle $\theta$, and therefore, the magnetic field.
The expectation value $\expect{J_x^2}$ is initially zero for a pure unpolarized Dicke state, and it increases rapidly as it can be seen in the Figure~\ref{fig:vd-secuence-evo}.
Another observation is that for $\theta=\pi/2$ the value of $\expect{J_x^2}$ will be at its maximum proportional to $\expect{\bs{J}^2}$ or equivalently to $\mathcal{J}_{N/2}$, see Eq.~\eqref{eq:app-maximum-total-angular-momentum}.
Hence, the change in the second moment over the phase shift must be in this case proportional to $N^2$.
We are lead to the conclusion that one only needs to measure the second moment of the collective spin $J_x$ to achieve Heisenberg scaling for the estimation.

In certain situations, it is better to use Eq.~\eqref{eq:bg-error-propagation-formula} rather than Eq.~\eqref{eq:bg-quantum-cr-bound} for calculating the achievable precision, since it gives the precision for a particular operator to be measured in an experimental setup.
This is reasonable, since in a typical experiment, only a restricted set of operators can be measured.
In this work, we will consider many-particle systems in which the particles cannot be accessed individually, and only collective quantities can be measured.
As we said, measuring the second moment of $J_x$ is a valid choice to estimate the rotation angle.
In the following equation, we show the error propagation formula when measuring the second moment of the $J_x$ total angular momentum component,
\be
  \varinv{\theta} = \frac{|\partial_{\theta} \expect{J_x}|^2}{\varian{J_x^2}}.
  \label{eq:vd-error-propagation}
\ee

Since Eq.~\eqref{eq:vd-error-propagation} is always smaller than $\qfi$, and as a consequence of Eq.~\eqref{eq:bg-entanglement-criterion-qfi}, if
\be
  \frac{|\partial_{\theta} \expect{J_x}|^2}{\varian{J_x^2}}\geqslant N
\ee
holds, then the system is entangled.
Hence again, entanglement is required for a large metrological precision.
Based on Eq.~\eqref{eq:bg-entanglement-depth-for-qfi}, we can bound the entanglement depth from below of the systems as follows.
Similarly to the previous paragraph, if for a quantum state
\be
  \frac{|\partial_{\theta} \expect{J_x}|^2}{\varian{J_x^2}}\geqslant kN
\ee
holds, then it is at least $(k+1)$-entangled.

\subsection{Precision based on the error propagation formula when $M=J_x^2$}
\label{sec:vd-evolution-of-the-expectation-values}
With the aim of obtaining the precision, Eq.~\eqref{eq:vd-error-propagation}, we will compute the dependence on $\theta$ of the expectation value of the operator $J_x$ and higher order moments.
We will use the Heisenberg picture, where the operators evolve in time while the state remains the same.
The operator $J_x$ can be written as a function of $\theta$ in the following way
\be
  J_x(\theta) = e^{i \theta J_z} J_x(0) e^{-i \theta J_z} = J_x(0) \text{c}_\theta - J_y(0) \text{s}_{\theta},
\ee
where $J_l(0)$ for $l=x,y,z$ are the collective angular momentum operators at time equal zero.
We will denote them by $J_l$ from now on.
The notation for the trigonometric functions $\text{c}_\theta$ and $\text{s}_\theta$ was introduced in Section~\ref{sec:bg-metrology-with-almost-polarized}.

We need to compute the second and the fourth moments of $J_x$ as it is required by the Eq.~\eqref{eq:vd-error-propagation}.
But before any calculation we will make a simplifying assumption which turns out to be true in the most common situations.
The assumption is that both expectation values are even functions of $\theta$, so
\be
  \begin{split}
    \expect{J_x^2(\theta)} &=\expect{J_x^2(-\theta)}, \\
    \expect{J_x^4(\theta)} &=\expect{J_x^4(-\theta)}
  \end{split}
  \label{eq:vd-even-f-constraint}
\ee
holds.
This way we can omit the terms that are odd in $\theta$.
In Section~\ref{sec:vd-testing-with-experimental-data}, we will see that unitary dynamics of some experimentally prepared states have this property.
The assumption Eq.~\eqref{eq:vd-even-f-constraint} is needed to obtain a closed formula for the precision of the phase estimation.

The square of $J_x$ in the Heisenberg picture is written as
\be
  J_x^2(\theta)= J_x^2 \text{c}_\theta^2 + J_y^2 \text{s}_\theta^2
  - (J_xJ_y + J_yJ_x) \text{c}_\theta\text{s}_\theta.
  \label{eq:vd-evolution-of-jx2}
\ee
Hence, due to the first constraint of Eq.~\eqref{eq:vd-even-f-constraint} and Eq.~\eqref{eq:vd-evolution-of-jx2}, we require that
\be
  \expect{\{J_x,J_y\}} = 0.
  \label{eq:vd-init-2nd-constraint}
\ee
Eq.~\eqref{eq:vd-init-2nd-constraint} is based on expectation values of the initial state state.
The condition Eq.~\eqref{eq:vd-init-2nd-constraint} is fulfilled in a typical experiment.

As we have done with the expectation value of the square of $J_x$, now we  do the same for $J_x^4$.
This way one will be able to distinguish which other expectation value of combination of operators must vanish in order to have Eq.~\eqref{eq:vd-even-f-constraint} guarantied.
The fourth power of $J_x$ can be written in the Heisenberg picture as
\begin{multline}
  J_x^4(\theta)= J_x^4 \text{c}_\theta^4 + J_y^4 \text{s}_\theta^4
  + (J_x^2J_y^2 + J_xJ_yJ_xJ_y + J_xJ_y^2J_x + J_yJ_xJ_yJ_x + J_yJ_x^2J_y + J_y^2J_x^2) \text{c}_\theta^2\text{s}_\theta^2 \\
  -(J_x^3J_y+J_x^2J_yJ_x+J_xJ_yJ_x^2+J_yJ_x^3)\text{c}_\theta^3\text{s}_\theta
  -(J_xJ_y^3+J_yJ_xJ_y^2+J_y^2J_xJ_y+J_y^3J_x)\text{c}_\theta\text{s}_\theta^3.
\end{multline}
And again assuming that the expectation value of $J_x^4(\theta)$ must be an even function of $\theta$, we see that the terms multiplied by $\coss{\theta}^3\sins{\theta}$ and $\coss{\theta}\sins{\theta}^3$, respectively, must be zero.
So, the expectation value of $(J_x^3J_y+J_x^2J_yJ_x+J_xJ_yJ_x^2+J_yJ_x^3)$ and $(J_xJ_y^3+J_yJ_xJ_y^2+J_y^2J_xJ_y+J_y^3J_x)$ must vanish.
Hence, the second constraint of the Eq.~\eqref{eq:vd-even-f-constraint} can be fulfilled if
\be
  \begin{split}
    \expect{\big\{J_x^2 , \{ J_x,J_y\}\big\}}=0,\\
    \expect{\big\{J_y^2 , \{ J_x,J_y\}\big\}}=0.
  \end{split}
\ee

Finally, we can write the evolution of second and fourth moments of the $J_x$ operator as
\begin{subequations}
\begin{align}
  \expect{J_x^2(\theta)}=\; &\expect{J_x^2} \text{c}_\theta^2 + \expect{J_y^2} \text{s}_\theta^2
  \label{eq:vd-evo-2nd-moment}\\
  \begin{split}
    \expect{J_x^4(\theta)}=\; &
    \expect{J_x^4}\text{c}_\theta^4 + \expect{J_y^4} \text{s}_\theta^4 \\
    & + \expect{\{J_x^2,J_y^2\}+\{J_x,J_y\}^2} \text{c}_\theta^2\text{s}_\theta^2.
  \end{split}
\end{align}
\end{subequations}
From here, we are able to write the evolution of the variance of the second moment when Eq.~\eqref{eq:vd-even-f-constraint} is fulfilled.
We obtain
\be
  \begin{split}
    \varian{J_x^2(\theta)} &= \expect{J_x^4(\theta)} -\expect{J_x^2(\theta)}^2 \\
    &= \expect{J_x^4}\text{c}_\theta^4 + \expect{J_y^4} \text{s}_\theta^4
    + \expect{\{J_x^2,J_y^2\}+\{J_x,J_y\}^2} \text{c}_\theta^2\text{s}_\theta^2
    - \big(\expect{J_x^2} \text{c}_\theta^2 + \expect{J_y^2} \text{s}_\theta^2\big)^2\\
    &= \big(\expect{J_x^4}-\expect{J_x^2}^2\big)\text{c}_\theta^4
    + \big(\expect{J_y^4}-  \expect{J_y^2}^2\big)\text{s}_\theta^4
    + \big(\expect{\{J_x^2,J_y^2\}+\{J_x,J_y\}^2} - 2 \expect{J_x^2}\expect{J_y^2}\big)
    \text{c}_\theta^2\text{s}_\theta^2\\
    &=\varian{J_x^2}\text{c}_\theta^4 + \varian{J_y^2} \text{s}_\theta^4+ \big(\expect{\{J_x^2,J_y^2\}+\{J_x,J_y\}^2} - 2 \expect{J_x^2}\expect{J_y^2}\big)\text{c}_\theta^2\text{s}_\theta^2.
  \end{split}
\ee

In order to compute the Eq.~\eqref{eq:vd-error-propagation}, we also need the modulus square of the derivative of the second moment of the $J_x$ operator.
Using Eq.~\eqref{eq:vd-evo-2nd-moment} for the expression of the evolution of the second moment, the numerator of Eq.~\eqref{eq:vd-error-propagation} follows
\be
  \begin{split}
    |\partial_\theta \expect{J_x^2(\theta)}|^2 & = |-2\expect{J_x^2}\text{c}_\theta\text{s}_\theta+2\expect{J_y^2}\text{c}_\theta\text{s}_\theta|^2\\
    & = 4\expect{J_y^2-J_x^2}^2\text{c}_\theta^2\text{s}_\theta^2.
  \end{split}
\ee

From the equations above directly follows expression for the precision of $\theta$,
\be
\begin{split}
  \varian{\theta} & = \frac{\varian{J_x^2}\text{c}_\theta^4 + \varian{J_y^2} \text{s}_\theta^4+ \big(\expect{\{J_x^2,J_y^2\}+\{J_x,J_y\}^2} - 2 \expect{J_x^2}\expect{J_y^2}\big)\text{c}_\theta^2\text{s}_\theta^2}
  {4\expect{J_y^2-J_x^2}^2\text{c}_\theta^2\text{s}_\theta^2}\\
  & = \frac{\varian{J_x^2}\text{t}_\theta^{-2} + \varian{J_y^2} \text{t}_\theta^2+ \expect{\{J_x^2,J_y^2\}+\{J_x,J_y\}^2} - 2 \expect{J_x^2}\expect{J_y^2}}
  {4\expect{J_y^2-J_x^2}^2}.
\end{split}
\label{eq:vd-result-before-simp}
\ee
To this calculations further computations follow mainly regarding to the following expectation value $\expect{\{J_x^2,J_y^2\}+\{J_x,J_y\}^2}$.
This calculus is left for the Appendix~\ref{app:simplification-of-4th-moments}.
Finally, the expression Eq.~\eqref{eq:vd-result-before-simp} can be written as
\be
  \varian{\theta} = \frac{\varian{J_x^2}\text{t}_\theta^{-2} + \varian{J_y^2} \text{t}_\theta^2 + 4\expect{J_y^2} - 3 \expect{J_z^2} - 2\expect{J_x^2}(1+\expect{J_y^2}) + 6\expect{J_xJ_y^2J_x}}
  {4\expect{J_y^2-J_x^2}^2}.
  \label{eq:vd-precision-as-theta}
\ee

\begin{figure}[htp]
  \centering
  \igwithlabel{(a)}{scale=.65}{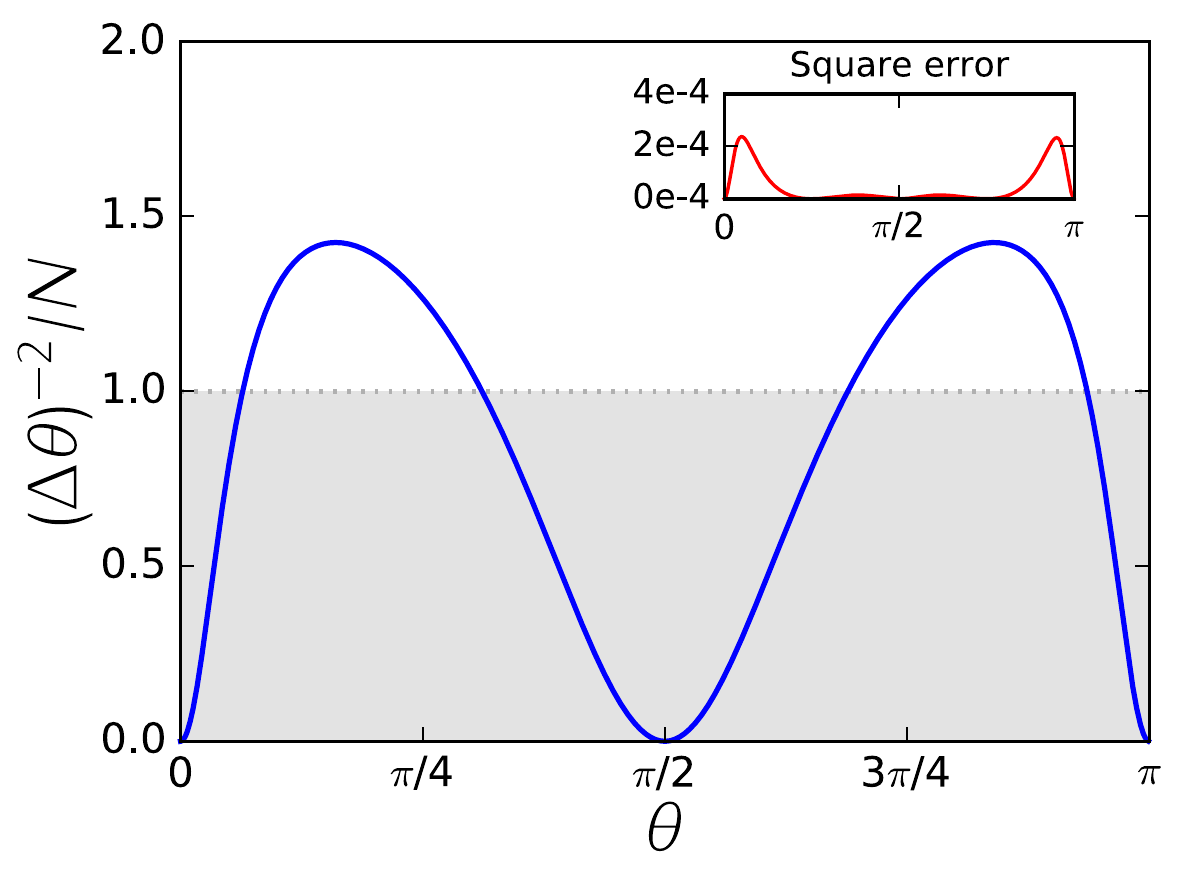}
  \igwithlabel{(b)}{scale=.65}{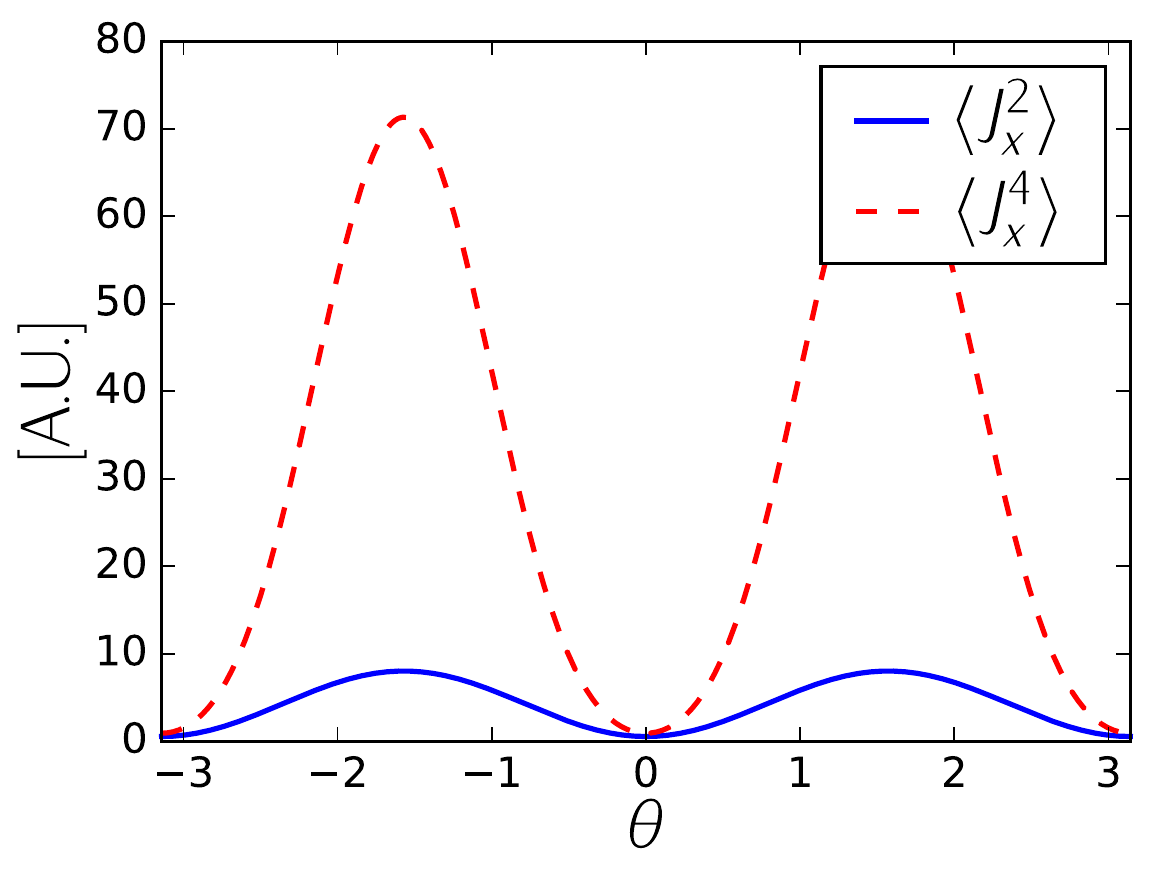}
  \caption[(a) Evolution of $\varinv{\theta}/N$. (b) Evolution of the expectation values]{
  (a) Evolution of the precision $\varinv{\theta}/N$ for 6 qubits based on the simulation of the system $\ket{\text{GS}}$ and its expectation values.
  The agreement with the Eq.~\eqref{eq:vd-precision-as-theta} is shown in the inset plot where the square of the difference between two approaches are plotted, the analytically obtained result and the simulation.
  The difference is more or less two orders of magnitude below the actual value for the relevant points, which is mainly because of computing the derivative near the points at which the expectation value $\expect{J_x^2}$ and $\varian{J_x^2}$ are both close to zero.
  (b) With the system at hand, we verified the parity with respect to $\theta$ of the expectation values of the second and the fourth moment, so to fulfill the constraint Eq.~\eqref{eq:vd-even-f-constraint}.}
  \label{fig:vd-evolution-of-precision}
\end{figure}

We have verified the correctness of our analytic formula Eq.~\eqref{eq:vd-precision-as-theta} comparing it with a numerical simulation of the Eq.~\eqref{eq:vd-error-propagation} for the ground-state of $H=J_x^2+J_y$ for 6 qubits, $|\text{GS}\rangle$.
We computed the evolution of the expectation values of the second and the fourth moments of the operator $J_x$ for $\theta \in [0,\pi]$, for thousand of equidistant points, from which we obtained the bound, see Figure~\ref{fig:vd-evolution-of-precision}-(a).
Finally, we have also checked that the constraints assumed at the beginning of this section are fulfilled.
For that, we considered the range $\theta \in [-\pi,\pi]$ and we have computed the expectation values, see Figure~\ref{fig:vd-evolution-of-precision}-(b).
We can conclude saying that our formula Eq.~\eqref{eq:vd-precision-as-theta} reproduces exactly the evolution of the error propagation formula, Eq.~\eqref{eq:vd-error-propagation}.

\subsubsection{The optimal precision}
First of all, note that all the dependence on the phase shift $\theta$ is in the first two terms of the numerator of Eq.~\eqref{eq:vd-precision-as-theta}.
Hence, one can minimize the sum on the first two terms in order to find where the precision is best.
So it follows that for the optimal angle
\be
  \label{eq:vd-optimal-phase}
  \tan^2(\theta_{\text{opt}}) = \sqrt{\frac{\varian{J_x^2}}{\varian{J_y^2}}}
\ee
holds.
By substituting Eq.~\eqref{eq:vd-optimal-phase} into Eq.~\eqref{eq:vd-precision-as-theta}, we obtain the optimal bound as
\be
  \varian{\theta}_{\text{opt}} = \frac{\sqrt{\varian{J_x^2} \varian{J_y^2} } + 4 \expect{J_y^2} - 3 \expect{J_z^2} - 2\expect{J_x^2}(1+\expect{J_y^2}) + 6\expect{J_xJ_y^2J_x}}
  {4\expect{J_y^2-J_x^2}^2}.
  \label{eq:vd-precision}
\ee.

We conclude this section checking our bound for pure unpolarized Dicke state aligned with the $x$-axis, $\ket{\dicke{N}}_x$, whose precision bound is well known using the QFI for pure states Eq.~\eqref{eq:bg-qfi-for-pure-states},
\be
  \qfif{\ket{\dicke{N}}_x, J_z} = 4\varian{J_z}_{\ket{\dicke{N}}_x}=\frac{N(N+2)}{2}.
  \label{eq:vd-qfi-for-pure-dicke}
\ee
With this aim we compute all the expectation values needed for the Eq.~\eqref{eq:vd-precision} which almost all of them are trivial, $\expect{J_xJ_y^2J_x}=\expect{J_x^4}=\expect{J_x^2}=0$ since the state is an eigenstate of $J_x$ with an eigenvalue zero.
The last expectation value is obtained as
\be
  \expect{J_y^2} = \expect{J_z^2} = \frac{N (N+2)}{8}.
  \label{eq:vd-2moment-pure-dicke}
\ee
Note that for the Eq.~\eqref{eq:vd-2moment-pure-dicke} we use that the state is invariant under rotations over the $x$-axis, the sum of all the second moments must give $\expect{\bs{J}^2} = \frac{N (N+2)}{4}$, and $\expect{J_x^2}=0$.
Hence, the Eq.~\eqref{eq:vd-2moment-pure-dicke} holds.

From the Eq.~\eqref{eq:vd-2moment-pure-dicke} using the expression for the optimal precision Eq.~\eqref{eq:vd-precision} and substituting all the terms that are zero, one arrives at the following formula for the precision of the phase shift for a pure unpolarized Dicke state,
\be
  \varian{\theta}_{\text{opt}} = \frac{2}{N(N+2)},
\ee
which coincides exactly with the inverse of the quantum Fisher information for such state Eq.~\eqref{eq:vd-qfi-for-pure-dicke} \cite{Luecke2011}.
Hence for the ideal Dicke state, the Cramér-Rao bound Eq.~\eqref{eq:bg-quantum-cr-bound} is saturated, which means that estimating the phase shift $\theta$ using the measurement of $\expect{J_x^2}$ is optimal.
Based on Eq.~\eqref{eq:vd-optimal-phase}, we add that the optimal angle for the ideal Dicke state is $\theta_{\text{opt}}=0$.

\subsection{Testing the formula against some known states}
\label{sec:vd-comparison-with-qfi}

In this section, we will compare our criteria based on few expectation values against the corresponding quantum Fisher information obtained for some known states.
We find that our formula gives a good lower bound on the quantum Fisher information, which is the best achievable precision when any measurement is allowed.
However note that the Cramér-Rao bound might be impractical.

\begin{figure}[htp]
  \centering
  \igwithlabel{(a)}{scale=.65}{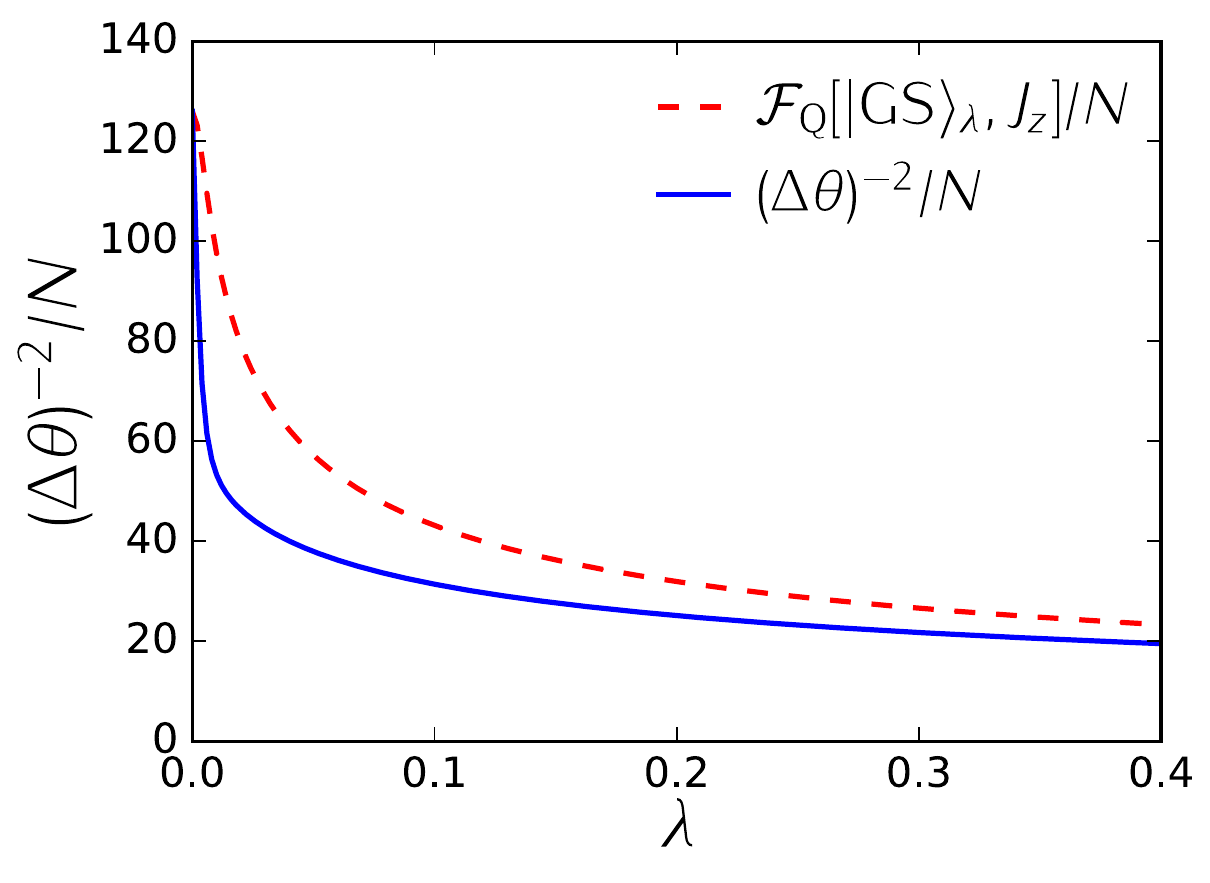}
  \igwithlabel{(b)}{scale=.65}{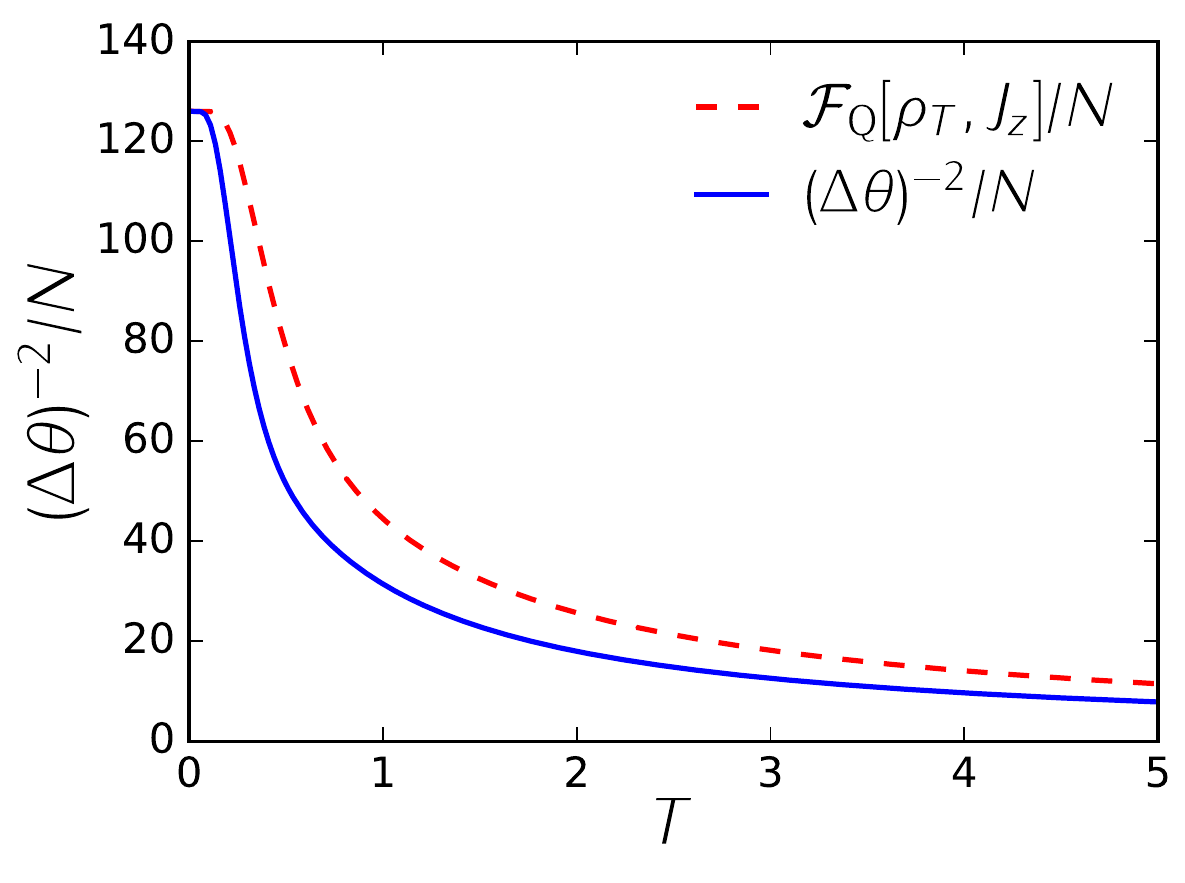}
  \caption[Comparing our bound and known QFIs.]{
  Comparison between our formula for the precision and the QFI for different states. (a) Comparison with ground states of $H_\lambda$. (b) Comparison with thermal mixture of Dicke states.}
  \label{fig:vd-comparing-the-bounds}
\end{figure}

Let us consider first the spin-squeezed states.
Those states will be defined as the ground states $\ket{\text{GS}}_\lambda$ of the spin-squeezing Hamiltonian given as
\be
  H_\lambda = J_x^2 - \lambda J_y,
  \label{eq:vd-ss-hamiltonian}
\ee
see Appendix~\ref{app:spin-squeezing-hamiltonian}.
For $\lambda>0$, the ground state is unique, and it is in the symmetric subspace.
Hence, we can restrict our attention to this subspace for our computations, and hence we can model larger systems.
For $\lambda\rightarrow\infty$, the ground state is the state totally polarized in the $y$-direction Eq.~\eqref{eq:bg-totally-polarized}.
And for $\lambda\rightarrow 0^{+}$, it is the Dicke state Eq.~\eqref{eq:vd-unpolarized-dicke}.
Note that for $\lambda=0$ the eigenvalue is degenerate, so there are more than one ground states.
On the other hand, we still can use a limit in which $\lambda$ tends to zero from the positive axis to arrive at the Dicke state.
Figure~\ref{fig:vd-comparing-the-bounds}-(a) shows the sensitivity we obtained together with the QFI for the same states.
Our bound is close to the QFI when the state is well polarized.
It also coincides with the bound in the $\lambda\rightarrow 0^{+}$ limit, when the ground state is close to the unpolarized Dicke state.

The second family of states we use to test our formula are the Gaussian mixture of Dicke states around the unpolarized Dicke state,
which have the following form as function of $T$ as
\be
  \rho_{T} \propto \sum_{n=0}^{N} e^{- \frac{(n+N/2)^2}{T}} \ket{\dicke{N,n}}_x\bra{\dicke{N,n}}_x
\ee
for even $N$, where $\ket{\dicke{N,n}}$ is defined in Eq.~\eqref{eq:app-definition-of-dicke-states}.
It can be used to model a noisy or thermal unpolarized Dicke state.
For $T=0$, we obtain the pure unpolarized Dicke state.
For $T>0$, other symmetric Dicke states in the vicinity of the unpolarized one are also populated.
The result can be seen in Figure~\ref{fig:vd-comparing-the-bounds}-(b).
Again, our bound seems to be quite close to the corresponding QFI.

Note also that in Figure~\ref{fig:vd-comparing-the-bounds}, based on Eq.~\eqref{eq:bg-entanglement-depth-for-qfi}, if the bound turns to be greater than $k$ integer, then a metrologically useful $(k+1)$-particle entanglement is detected in the system.
Note that this is true whenever $k$ is a divisor of $N$, or $k\ll N$.

Although showing how the optimal precision formula behaves compared with the quantum Fisher information for those two families of states, we will now prove that they indeed fulfill the constraints appearing in Eq.~\eqref{eq:vd-even-f-constraint}.
Hence, we compute the Eq.~\eqref{eq:vd-even-f-constraint} for the spin-squeezed states $\ket{\text{GS}}_{\lambda}$.
For that it is enough to know that since those states are non-degenerate ground states, they must have the same symmetries as the Hamiltonian.
In this case, the Hamiltonian Eq.~\eqref{eq:vd-ss-hamiltonian} is invariant under $x\leftrightarrow -x$ and $z\leftrightarrow -z$, thus it must be invariant under $r\pi$ angle rotations under the $y$-axis for $r$ integer, for more references about symmetries in quantum mechanics see Refs. \cite{Sakurai2010, Cohen-Tannoudji1977}.
Hence, we can write for the evolution of the expectation value of any power $m$ of $J_x$ such that
\be
  \tr(e^{+i\theta J_z}J_x^m e^{-i\theta J_z} \rho_{\lambda}) = \tr(e^{+i\theta J_z}J_x^m e^{-i\theta J_z} e^{-i r\pi J_y}\rho_{\lambda}e^{+i r\pi J_y}),
\ee
where $\rho_{\lambda}=\ketbra{\text{GS}}{\text{GS}}_{\lambda}$.
Then for the case $r=1$, we can use the cyclic property of the trace to arrive at
\be
\begin{split}
  \tr(e^{+i\pi J_y}e^{+i\theta J_z}J_x^m e^{-i\theta J_z} e^{-i\pi J_y}\rho_{\lambda}) & =
  \tr(e^{+i \theta (-1)J_z}(-1)^m J_x^m e^{-i\theta (-1) J_z}\rho_{\lambda})\\
  & = \tr(e^{-i \theta J_z}(-1)^m J_x^m e^{+i\theta J_z}\rho_{\lambda}),
\end{split}
\ee
or equivalently
\be
  \expect{J_x^m(\theta)}_{\rho_\lambda}=\expect{(-1)^m J_x^m(-\theta)}_{\rho_\lambda},
\ee
which implies that for even $m$, and specially for $m=2,4$, the expectation values are an even function of $\theta$, and that for odd $m$ the expectation values are odd functions of $\theta$, which proves the Eq.~\eqref{eq:vd-even-f-constraint} for the present case.

On the other hand, the eigenstates of the thermal state $\rho_T$ are also eigenstates of $J_x$, and hence the state invariant under rotations around the $x$-axis.
This still holds for the entire state, since it is a statistical mixture of states invariant under rotations around the $x$-axis.
Moreover, it is also invariant if the state is rotated around the $x$-axis by an angle $\pi.$
Hence, we have for the evolution of the expectation values of $J_x^m$ that
\be
  \tr(e^{+i\theta J_z}J_x^m e^{-i\theta J_z}\rho_T) = \tr(e^{+i\theta J_z}J_x^m e^{-i\theta J_z} e^{-i \pi J_x} \rho_T e^{+i \pi J_x})
\ee
holds for any $m$.
Finally, using again the cyclic properties of the trace, we flip the signs of the of angular momentum components orthogonal to $J_x$, so in this case $J_y \rightarrow - J_x$ and $J_z \rightarrow -J_z$, and we arrive at
\be
  \tr(e^{+i \pi J_x} e^{+i\theta J_z}J_x^m e^{-i\theta J_z} e^{-i \pi J_x} \rho_T) =  \tr(e^{-i\theta J_z}J_x^m e^{+i\theta J_z} \rho_T).
\ee
We conclude that for the thermal state this case all the moments of the $J_x$ operator are even functions of $\theta$ for the thermal state, i.e., $\expect{J_x^m(\theta)}_{\rho_{T}}=\expect{J_x^m(-\theta)}_{\rho_{T}}$, which proves that the Eq.~\eqref{eq:vd-even-f-constraint} holds for this case too.

\subsection{Using our method with real experimental data}
\label{sec:vd-testing-with-experimental-data}

In reference \cite{Luecke2014}, a state is produced in the laboratory with the proper characteristics of an unpolarized Dicke state, small variance in one direction, say $x$, and a very large variance in the directions perpendicular to the $x$-axis.
In the cited experiment with $N$ qubits, it is possible to determine the operator $J_x$ as the population imbalance of the two levels as
\be
  J_x = \frac{1}{2}(N_{1,x}-N_{0,x}),
\ee
where $N_{m,x}$ is the number of particles in the state $\ket{m}_x$.
Hence, measuring the population imbalance and collecting the statistics of the measurements, the expectation values of all moments of $J_x$ can be obtained.
In practice, it is possible to measure the lower order moments like $\expect{J_x^2}$ and $\expect{J_x^4}$, while higher order moments need too many repetitions of the experiment to collect enough statistics.

The other two global operators $J_y$ and $J_z$ are obtained by rotating the system using a $\frac{\pi}{2}$ microwave coupling pulse before the measurement of the population imbalance.
Whether $J_y$ or $J_z$ is obtained depends on the relation between the microwave phase and the phase of the initial BEC.
The condensate phase represents the only possible phase reference in analogy to the local oscillator in optics.
Intrinsically, it has no relation to the microwave phase, such that it homogeneously average over all possible phase relations.
From another point of view, one can also say that the fluctuation of the magnetic field results in a random rotation of the spin around the $z$-axis.
Hence, what is obtained in this case is
\be
  J_\alpha = \sin(\alpha) J_y + \cos(\alpha) J_z,
\ee
where $\alpha$ is an angle, and we need to consider the average over all possible angles.
Effectively, the state has the following form
\be
  \rho = \frac{1}{2\pi}\int e^{-i\alpha J_x} \rho_0 e^{i\alpha J_x}\, \text{d}\alpha,
  \label{eq:vd-rotational-invariant-state}
\ee
where $\rho_0$ is what we would obtain if we would have access to the phase reference.
Note that the integration over the rotation angle $\alpha$ does not create entanglement.
If the state $\rho$ is entangled then $\rho_0$ has to be also entangled.

Let us see the consequences of our state having the form  Eq.~\eqref{eq:vd-rotational-invariant-state}.
For the density matrix $\rho$, since it is invariant under rotations around the $x$-axis, we have
\be
  \expect{J_\alpha^m}=\expect{J_y^m}=\expect{J_z^m}
\ee
for all $m$.
Hence the expectation values of $\expect{J_y^m}$ and $\expect{J_z^m}$ can be obtained from the statistics of measuring $J_\alpha$.
Moreover, for states of the form Eq.~\eqref{eq:vd-rotational-invariant-state} the unitary dynamics will fulfill the condition Eq.~\eqref{eq:vd-even-f-constraint}.

There is a single remaining term in the expression for the achievable precision Eq.~\eqref{eq:vd-precision}, the expectation value for $\expect{J_xJ_y^2J_x}$, which can be bounded as
\be
\begin{split}
  \expect{J_xJ_y^2J_x} &= \frac{\expect{J_x (J_y^2 + J_z^2)J_x}}{2}
  =\frac{\expect{J_x (J_x^2 + J_y^2 + J_z^2 )J_x} - \expect{J_x^4}}{2} \\
  & \leqslant \frac{N(N+2)}{8} \expect{J_x^2} - \frac{\expect{J_x^4}}{2},
\end{split}
\label{eq:vd-simplification-of-last-4th-moment}
\ee
where the last inequality is due to that for all states $\expect{\bs{J}^2}\leqslant \mathcal{J}_{N/2}$, see Eq.~\eqref{eq:app-maximum-total-angular-momentum}, while symmetric states saturate the inequality in Eq.~\eqref{eq:vd-simplification-of-last-4th-moment}.
Note that obtaining $\expect{J_xJ_z^2J_x}$ can be hard experimentally.
In any case, this simplification can only make our estimation of the precision worse while for symmetric states the equality holds.
Hence, the lower bound for the achievable precision can be written as
\be
  \varian{\theta}_{\text{opt}} \leqslant \frac{\sqrt{\varian{J_x^2} \varian{J_y^2} } + \expect{J_y^2} + \frac{3N(N+2)-8}{4} \expect{J_x^2} - 2\expect{J_x^2}\expect{J_y^2} - 3\expect{J_x^4}}
  {4\expect{J_y^2-J_x^2}^2},
\ee
where some terms were reordered and further simplified.

It is worth to study the case appearing in Ref.~\cite{Krischek2011} and apply our methods such that we obtain conclusions about the metrological usefulness of the state.
The system under consideration has around $N=7900$.
Note that using the expectation value of the particle number, in our case $\expect{N}=7900$, cannot overestimate any lower bound on the precision.
For a discussion about entanglement criteria in systems with particle number fluctuations see Ref.~\cite{Hyllus2012a}.
The measured data for the system yields
\be
\begin{aligned}
  \expect{J_x^2} & = 112 \pm 31, \\
  \expect{J_x^4} & = 40 \times 10^3 \pm 22 \times 10^3,
\end{aligned}
\quad
\begin{aligned}
  \expect{J_y^2} & = 6 \times 10^6 \pm 0.6 \times 10^6, \\
  \expect{J_y^4} & = 6.2 \times 10^{13} \pm 0.8 \times 10^{13}.
\end{aligned}
\label{eq:vd-experimental-values}
\ee
Hence, we obtain the maximum precision as
\be
  \frac{(\Delta \theta)^{-2}_{\text{opt}}}{N} \geqslant 3.7 \pm 1.5.
  \label{eq:vd-precision-for-experiment}
\ee
The statistical uncertainties of Eqs.~\eqref{eq:vd-experimental-values} and \eqref{eq:vd-precision-for-experiment} have been obtained through bootstraping, while the direct substitutions of expectation values would yield to 3.3 of gain over the shot-noise limit $\varinv{\theta}=N$.
Based on Eq.~\eqref{eq:bg-entanglement-criterion-qfi}, this proves the presence of metrologically useful entanglement \cite{Pezze2009}.
Based on Eq.~\eqref{eq:bg-entanglement-depth-for-qfi}, it even demonstrated that the quantum state had metrologically useful 4-particle entanglement.
Assuming an error of a standard deviation, Eq.~\eqref{eq:vd-precision-for-experiment} still proves 3-particle entanglement.

\begin{figure}[htp]
  \centering
  \includegraphics[scale=.65]{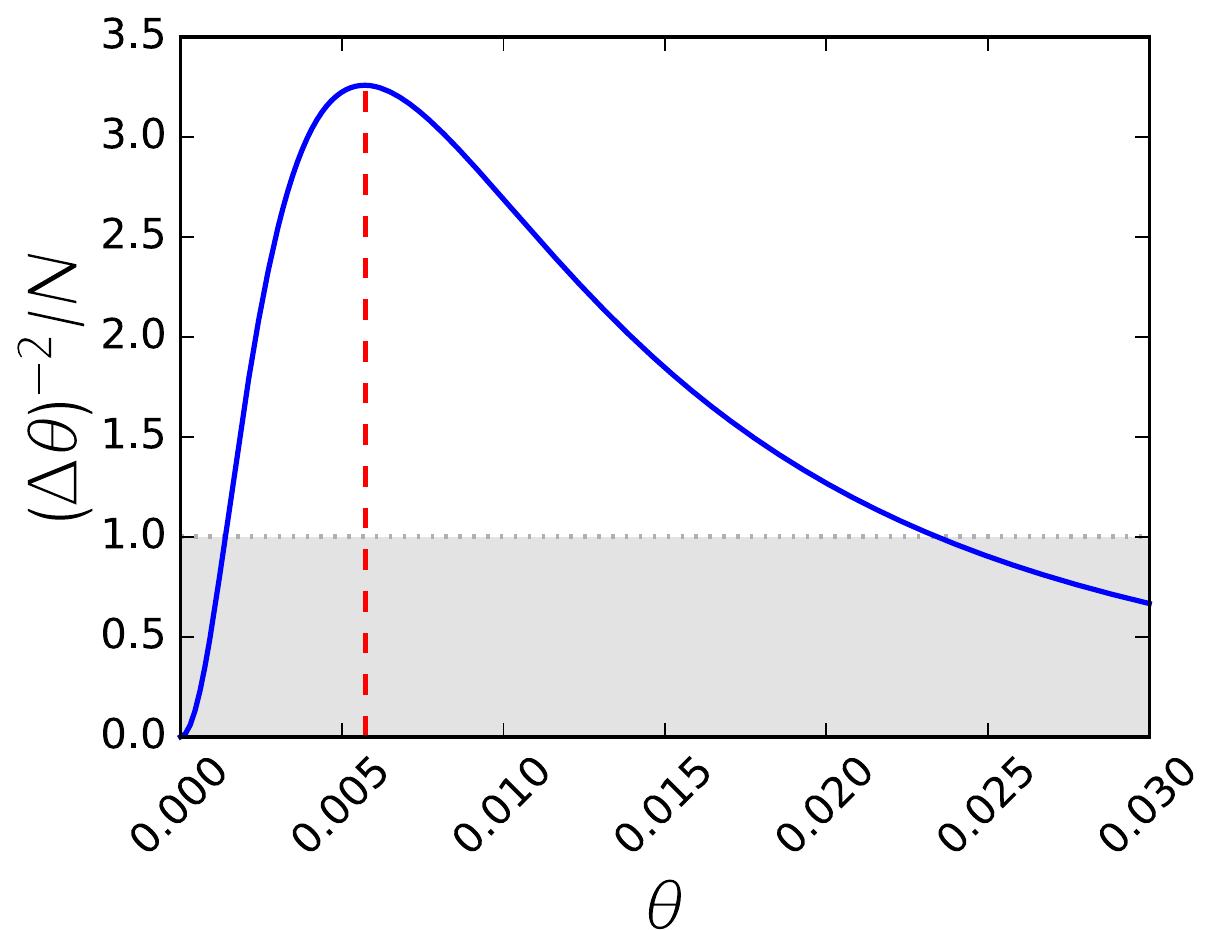}
  \caption[Evolution in $\theta$ of the precision.]{
  (solid) The precision as a function of the parameter $\theta$ given by Eq.~\eqref{eq:vd-precision-as-theta} varies through the evolution.
  Note that for the initial moment the precision is zero.
  (dashed) We highlight where the precision reaches its maximum at $\theta \approx 0.0057$.
  (gray-area) It represent the region where the precision is below the shot-noise limit.}
  \label{fig:vd-precision-theta-experiment}
\end{figure}

Next we plot the value for the precision substituting directly the experimental data into Eq.~\eqref{eq:vd-precision-as-theta}, see Figure~\ref{fig:vd-precision-theta-experiment}.
Since we cannot obtain the expectation value $\expect{J_xJ_y^2J_x}$ we approximate it with the right-hand side of Eq.~\eqref{eq:vd-simplification-of-last-4th-moment}.
With that we underestimate $\varinv{\theta}$.

Thus, we could detect metrological usefulness by measuring the second and fourth moments of the collective angular momentum components.
For future applications of our scheme, it is important to reduce further the number of quantities we need to obtain a lower bound for the precision.
In practice, one can easily avoid the need for determining $\expect{J_y^4}$.
Note that if we measure $J_y$ then the distribution of the values obtained is strongly non-Gaussian.
The values $\pm N/2$ appear most frequently, and the value 0 appears least frequently \cite{Luecke2011}.
See also the distribution of a pure Dicke state when $\theta=\pi/2$ in the Figure~\ref{fig:vd-secuence-evo}.
The state has more overlap with the eigenstates of the edges than in the middle.
Based on $\expect{AB}\leqslant\lambda_{\max}(A)\expect{B}$, where $\lambda_{\max}(A)$ is the largest eigenvalue of $A$, for two commuting positive-semidefinite observables,
\be
  \expect{J_y^4}\leqslant\frac{N^2}{4}\expect{J_x^2}.
\ee
Since even for a noisy Dicke state $\expect{J_y^2}$ is very large, the above equation is a very good upper bound.
Substituting it into the Eq.~\eqref{eq:vd-precision}, we will underestimate $\varinv{\theta}$.

It is also possible to approximate $\expect{J_x^4}$ with $\expect{J_x^2}$ in the sense that it is small and that mainly its value comes from technical noise,
\be
  \expect{J_x^4} \approx \beta \expect{J_x^2}^2.
\ee
This approximation, even if it is not a strict bound on the precision, can be very useful in order to characterize the metrological usefulness of the state based only on second statistical moments of only two angular momentum components, namely $\expect{J_y^2}$ and $\expect{J_x^2}$.
Those two expectation values are related with how  thin is the state in one direction and how wide in the perpendicular ones.
So in this case we use $\beta=3$ assuming that the distribution function has a Gaussian shape.

\begin{figure}[htp]
  \centering
  \igwithlabel{(a)}{scale=.65}{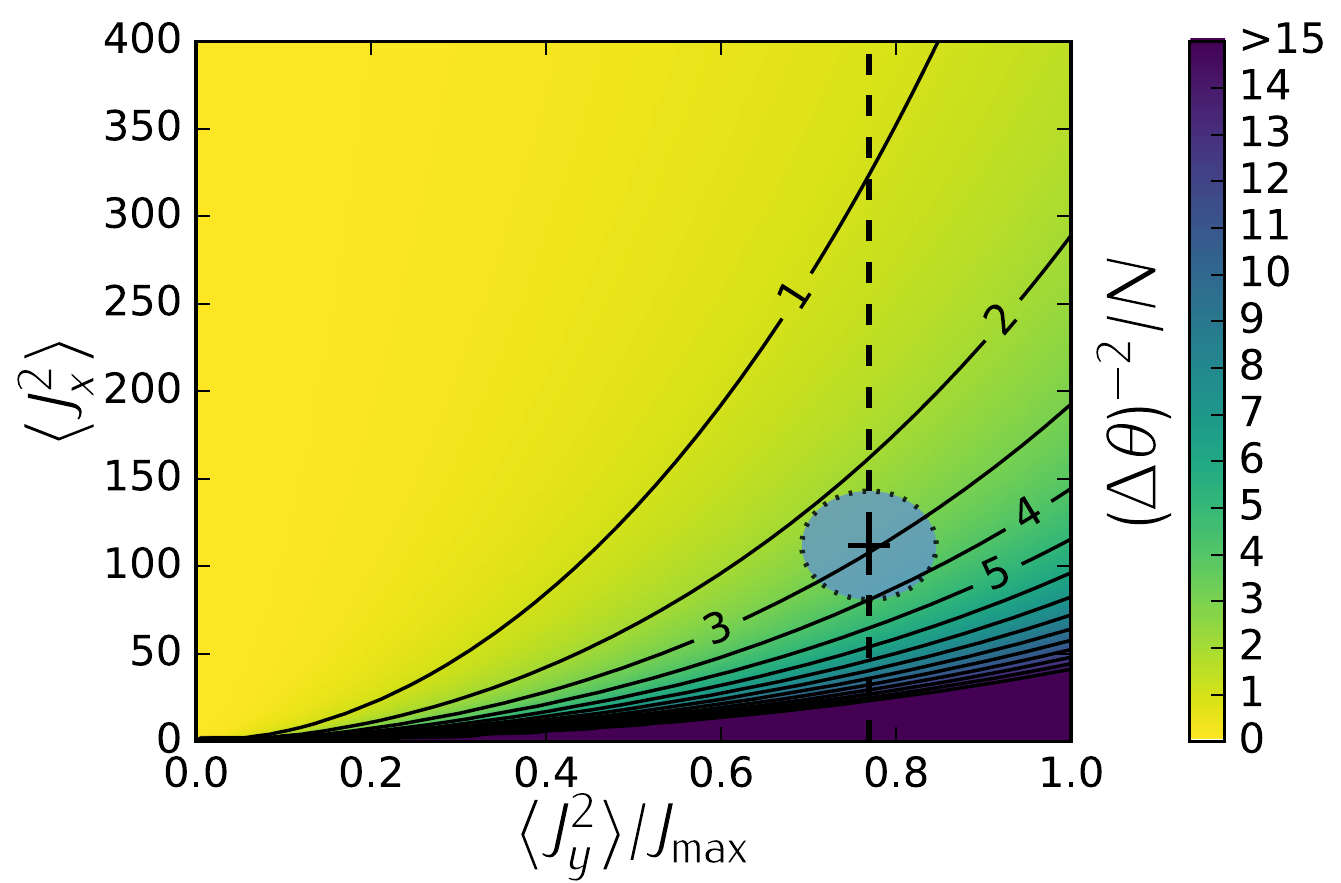}
  \igwithlabel{(b)}{scale=.65}{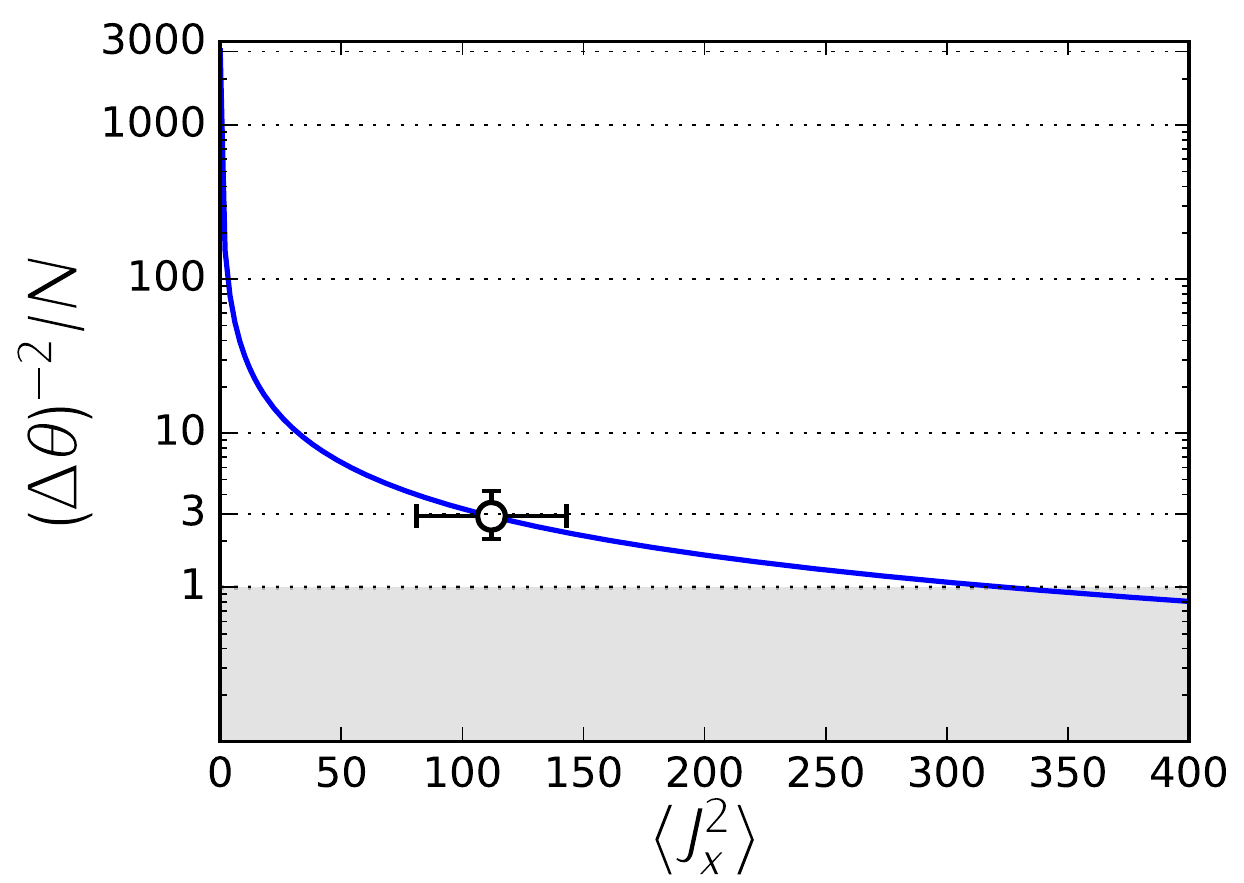}
  \caption[(a) Precision bound for $\expect{J_y^2}$ and $\expect{J_x^2}$. (b) Slice in constant $\expect{J_y^2}$.]{
  (a) Precision bound as a function of $\expect{J_y^2}$ and $\expect{J_x^2}$.
  The expectation value $\expect{J_y^2}$ is normalized with $\mathcal{J}_{N/2}$.
  (solid) Different boundaries for metrologically useful entanglement depths.
  (cross) Experimental data.
  (blue-ellipse) Region with one $\sigma$ confidence from the experimental data.
  (vertical-dashed) Constant $\expect{J_y^2}$ cross section plotted in (b).
  (b) Constant $\expect{J_y^2}$ cross section for the precision bound.
  (solid) Precision bound based on the Eq.~\eqref{eq:vd-precision-bound-for-second-moments}.
  One can see that decreasing further the uncertainty in $\expect{J_x^2}$ can improve the bound significantly.
  (white-point) Experimental data with the corresponding errors bars.
  (gray-area) Shot-noise limit.
  The point is slightly below the 4-particle entanglement level.}
  \label{fig:vd-experimental}
\end{figure}

From these considerations we are able to write a second bound with fewer expectation values for the optimal precision such that
\be
  \varian{\theta}_{\text{opt}} \leqslant \frac{\expect{J_y^2} + \frac{3N(N+2)-8}{4} \expect{J_x^2} + \Big(\sqrt{\frac{N^2}{2\expect{J_y^2}}-2} - 2 \Big) \expect{J_y^2}\expect{J_x^2} - 9\expect{J_x^2}^2}
  {4\expect{J_y^2-J_x^2}^2}.
  \label{eq:vd-precision-bound-for-second-moments}
\ee
We have used this formula to compute the bound for the optimal precision with the measured data shown on Eq.~\eqref{eq:vd-experimental-values}, $(\Delta \theta)^{-2}_{\text{opt}} \geqslant 2.9N$, see Figure~\ref{fig:vd-experimental}.
It turns out that even this way, 3-particle metrologically useful entanglement is detected.

In Figure~\ref{fig:vd-experimental}-(a), we show the two-dimensional plot that is obtained based on these considerations.
The regions with various levels of multipartite entanglement can clearly be identified.
The ideal Dicke state corresponds to the bottom-right corner, where $\expect{J_x^2}=0$ and $\expect{J_y^2}$ is largest.
In Figure~\ref{fig:vd-experimental}-(b), the cross section of the two-dimensional plot is shown.

Summarizing, we have shown that characterizing the metrological usefulness of noisy Dicke states can be done using few experimental data.
Furthermore, the precision bound is computed with the expectation values of the second and fourth moments of the angular momentum in the directions perpendicular to the magnetic field.
We have shown some simplifying techniques that reduce the experimental efforts needed to characterize the state.

%% file: snp/doubleLineWaterMark.tex
\thiswatermark{\put(1,-282){\color{l-grey}\rule{84pt}{88pt}}
\put(84,-282){\color{grey}\rule{410pt}{88pt}}}

%% file: 04-legendre.tex
\section[Witnessing metrologically useful entanglement]
{Witnessing metrologically useful {\color{grey} X} entanglement}
\input{snp/doubleLineWaterMark.tex}
\label{sec:lt}

\quotes{Adrien-Marie Legendre}{All the truths of mathematics are linked to each other,\break and all means of discovering them are equally admissible.}

\lettrine[lines=2, findent=3pt,nindent=0pt]{T}{ypically}, one has no access to the density matrix of the system which is used for metrology or for some other quantum information processing task.
Moreover, for systems in which the particle number is very large, which is the case when one would like to do metrology with quantum states, the details of the density matrix cannot be known due to practical reasons.
Since the quantum Fisher information is based on the complete knowledge of the density matrix, methods to avoid the complete tomography must be developed as we have shown a practical case in the previous chapter.
In this chapter, we obtain a general procedure to get an optimal bound for the quantum Fisher information based on as many expectation values of the initial state as one is ready to measure.
Two main features are worth to mention again.
First, in general this method gives us a tight bound.
Second, the bound is based on the expectation values of the initial state only, so it is not necessary to perform an evolution of the state.
This is in contrast to other approaches one can find in the literature, ours needs a significantly smaller experimental effort.

From Eq.~\eqref{eq:bg-pezze-bound}, a lower bound on the quantum Fisher information based on expectation values of the initial state is
\be
  \qfif{\rho,J_z} \geqslant \frac{\expect{J_x}^2}{\varian{J_y}}.
\ee
where the state is polarized along the $x$-axis \cite{Pezze2009}.
In the previous chapter we have also shown one of these bounds specifically designed for unpolarized Dicke states \eqref{eq:vd-unpolarized-dicke}.

The setup is described in Section~\ref{sec:bg-quantum-magnetometry} and consists of in estimating the homogeneous magnetic field strength, in this case, which points towards the $z$-axis.
Therefore, we will base our calculations on the quantum Fisher information $\qfif{\rho, J_z}$, defined in Eqs.~\eqref{eq:bg-qfi-definition-eigen-decomposition} and \eqref{eq:bg-qfi-definition-convex-roof}, see Section~\ref{sec:bg-qfi} for more information about the properties of the QFI.

\subsection{Lower bound on a convex function given some arbitrary expectation values}

Our problem could be solved with Lagrange multipliers or Legendre transforms.
We follow the latter method since obtaining a lower bound on a convex function for states with some given operator expectation values has already been studied by O. G\"uhne {\it et al.} and J. Eisert {\it et al.} in Refs.~\cite{Guehne2007, Eisert2007} respectively, mainly from the perspective of entanglement measures.
The illustrated techniques are based on the well known Legendre transform for differentiable functions, see Appendix~\ref{app:legendre-transform} for more details.
We first review in this section the state-of-the-art solution to this problem.
Later on, we extend it to the quantum Fisher information.
For simplicity in the next section, Section~\ref{sec:lt-transform-for-single-observable}, we assume that a single expectation value is given.
An extension to the case in which more expectation values are given will follow in the Section~\ref{sec:lt-transform-for-m-observables}.
Finally, we will summarize our results with an explicit formula which will be used to compute the bounds in the most general case.

\subsubsection{Lower bound based on a single observable}
\label{sec:lt-transform-for-single-observable}

When a convex function $g(\rho)$ is given together with an expectation value of some operator $w=\tr(\rho W)$, a tight lower bound, $\bound{g}(w)$, can be obtained as \cite{Rockafellar1996, Guehne2007, Eisert2007}
\be
  \label{eq:lt-lower-bound-single-parameter}
  \begin{split}
    g(\rho)\geqslant\bound{g}(w):=\,&\sup_r \{ rw - \hat{g}(rW)\}\\
    =\,& \{\inf_{\rho}g(\rho)\,|\,w = \tr(\rho W)\},
  \end{split}
\ee
where $\hat{g}(rW)$ is the Legendre transform of $g(\rho)$ and the second equality expresses the tightness of the bound.
The Legendre transform in this context is defined as
\be
  \label{eq:lt-for-convex-function-single-parameter}
  \hat{g}(rW)=\sup_{\rho}\{\expect{rW}_\rho - g(\rho)\},
\ee
where the maximization is over \emph{all} possible states.
This method to obtain the lower bound has been used to compute entanglement measures \cite{Guehne2007, Eisert2007}.

Following the theory one can find that if the convex function $g(\rho)$ is defined as a convex roof over all possible convex decompositions of the state, the optimisation of Eq.~\eqref{eq:lt-for-convex-function-single-parameter} can be reduced to an optimization over pure states only, thus simplifying the calculation \cite{Guehne2007, Eisert2007}
\be
\begin{split}
  \hat{g}(rW) & = \sup_{\rho}\{\expect{rW}_\rho - g(\rho)\} \\
  &=\sup_{\rho}\Big\{r\expect{W}_\rho - \inf_{\{p_k,\ket{\phi_k}\}}\big\{\sum_{k} p_k g(\ket{\phi_k})\big\} \Big\} \\
  &=\sup_{\{p_k, \ket{\phi_k}\}} \Big\{ \sum_k p_k \expect{rW}_{\ket{\phi_k}} - \inf_{\{p_k, \ket{\phi_k}\}} \big\{\sum_k p_k g(\ket{\phi_k}) \big\}  \Big\} \\
  &=\sup_{\{p_k, \ket{\phi_k}\}} \Big\{ \sum_k p_k \big\{ \expect{rW}_{\ket{\phi_k}} - g(\ket{\phi_k}) \big\} \Big\} \\
  &=\sup_{\ket{\psi}} \big\{ \expect{rW}_{\ket{\psi}} - g(\ket{\psi}) \big\}.
\end{split}
\ee
However, even an optimization over all pure states is feasible numerically only for small systems.
We will show later in this section how to circumvent this problem in the case of the QFI.
The convex roof construction has the following form
\be
  g(\rho) = \inf_{\{p_k,\psi_k\}} \sum_{k}p_k g(\ket{\psi_k}),
\ee
where the mixed state is decomposed into $\rho = \sum_k p_k \ketbra{\psi_k}{\psi_k}$.
Among other definitions of the QFI in the literature, there is one that defines it as the convex roof of $4\varian{J_z}$, the variance of the generator, as it has been shown in Ref.~\cite{Toth2007, Yu2013}, we can compute the Legendre transform optimizing for pure states only.
Hence, we will be able to use this simplification to apply this method to obtain the lower bound on the QFI.
Note that in this context, the QFI is the convex roof of four times the variance of the generator, Eq.~\eqref{eq:bg-qfi-definition-convex-roof}.

\subsubsection{Measuring several observables}
\label{sec:lt-transform-for-m-observables}

For some cases, it is interesting to characterize the quantum state with several measurements rather than with a single one.
For instance, we might want to use, as it is done with the spin-squeezed states the absolute polarization and the variance of one of the orthogonal components of the angular momentum to detect entanglement and metrological usefulness \cite{Pezze2009}.
So far, we studied the case in which a single measurement is used.
Its extension to several expectation values is indeed straight-forward.
We can generalize Eqs.~\eqref{eq:lt-lower-bound-single-parameter} and \eqref{eq:lt-for-convex-function-single-parameter} for several observables $\{W_i\}_{i=1}^M$ as follows \cite{Guehne2007}
\be
  \label{eq:lt-extension-bound-multiparameter}
  \bound{g}(w_1,w_2,\dots) := \sup_{\bs{r}}\big\{\bs{rw}-\sup_{\rho}\{\expect{\bs{rW}}-g(\rho)\}\big\},
\ee
where $\bs{ab} =\sum_{k=1}^M a_kb_k$, the usual notation for scalar products of two vectors.

\subsubsection[Lower bound on the QFI]{Lower bound on the quantum Fisher information}

After we have shown how to find a lower bound for a general convex function of the state based on its expectation values and how to simplify that method for the case in which the function is defined as a convex roof, now we are in the position to achieve the main goal of this chapter.
First of all, we note that for the quantum Fisher Information the inner maximization, the Legendre transform, is obtained optimizing a function quadratic in expectation values,
\be
\label{eq:lt-legendre-of-qfi}
\begin{split}
  \hat{\qfi}(rW) &= \sup_{\ket{\psi}}\big\{r\expect{W}_{\psi}-4\varian{J_z}_{\psi}\big\} \\
  &= \sup_{\ket{\psi}} \big\{ r\expect{W}_{\psi}-4\expect{J_z^2}_{\psi} + 4\expect{J_z}^2_{\psi} \big\} \\
  &= \sup_{\ket{\psi}} \big\{\expect{rW-4J_z^2}_{\psi} +
  \expect{2J_z}^2_{\psi}\big\},
\end{split}
\ee
where we have used the fact that the QFI can be expressed as a convex roof of $\varian{J_z}$ and we arrive at the problem of an optimization over a single parameter for simplicity on the following derivations.
Equation~\eqref{eq:lt-legendre-of-qfi} can be rewritten as an optimization linear in operator expectation values and over a parameter $\mu$ as
\be
  \hat{\mathcal{F}}_{\text{Q}}(rW) = \sup_{\ket{\psi},\mu}\big\{\expect{rW-4J_z^2}_{\psi}+8\mu\expect{J_z}_\psi - 4 \mu^2\mtxid\big\},
\ee
which, making use of $\max\{\expect{A}\}=\lambda_{\max}[A]$ for any observable, can be reformulated as
\be
  \label{eq:lt-legendre-for-qfi-simplified}
  \begin{split}
    \hat{\mathcal{F}}_{\text{Q}}(rW) & = \sup_{\ket{\psi}}\big\{\lambda_{\max}[rW-4J_z^2+8\mu J_z - 4 \mu^2]\big\}\\
    &=\sup_{\ket{\psi}}\big\{\lambda_{\max}[rW-4(J_z-\mu)^2]\big\},
  \end{split}
\ee
where we omitted in writing $\mtxid$ for clarity and $\lambda_{\max}[A]$ stands for the maximum eigenvalue of the operator $A$.
At the extremum, the derivative with respect to $\mu$ must be zero, hence at the optimum $\mu=\expect{J_z}_{\text{opt}}$ which represents the expectation value of $J_z$ should have considering the optimal state in Eq.~\eqref{eq:lt-legendre-of-qfi}.
This also means that we have to test $\mu$ values in the interval $-N/2\leqslant\mu\leqslant N/2$ only for spin-half systems.

The full optimization problem to be solved consists of Eqs.~\eqref{eq:lt-lower-bound-single-parameter} and~\eqref{eq:lt-legendre-for-qfi-simplified} substituting $g(\rho)$ by $\qfif{\rho,J_z}$,
\be
  \bound{\mathcal{F}}(w) = \sup_r\big\{rw-\sup_{\mu}\{\lambda_{\max}[rW-4(J_z-\mu)^2]\}\big\}.
  \label{eq:lt-bound-for-qfi}
\ee
It is crucial that the optimization over $r$ is a concave function, since the theory tells us that $\hat{\mathcal{F}}_{\text{Q}}(rW)$ is a convex function \cite{Rockafellar1996}, even when the multi-parameter case is considered.
Thus the optimum can be determined easily with simple methods, e.g., the gradient method, looking for the maximum in $r$.
Based on Eq.~\eqref{eq:lt-lower-bound-single-parameter}, we can see that even if we do not find the global optimum in $r$, we obtain a valid lower bound.
The extension of this bound to the multi-parameter case is done using the recipe given in Eq.~\eqref{eq:lt-extension-bound-multiparameter}.
On the other hand, the function to be optimized for $\mu$ does not have a single maximum in general.
Moreover, not finding the optimal $\mu$ leads to an overestimating of the bound.
Thus, a large care must be taken when optimizing over $\mu$.

We stress again the generality of these findings beyond linear interferometers covered in the following sections.
For nonlinear interferometers \cite{Luis2004, Boixo2007, Choi2008, Roy2008, Napolitano2011, Hall2012}, the phase $\theta$ must be estimated assuming unitary dynamics $U=\exp{-iG\theta}$, where $G$ is not a sum of single spin operators, hence, it is different from the angular momentum components.

\subsubsection{Exploiting the symmetries}
\label{sec:lt-symmetries}

When making calculations for quantum systems with an increasing number of qubits, we soon run into difficulties when computing the largest eigenvalue of Eq.~\eqref{eq:lt-legendre-for-qfi-simplified}.
The reason is that for $N$ qubits, we need to handle $2^N \times 2^N$ size matrices, hence we are limited to systems of 10 to 15 qubits.

We can obtain bounds for much larger particle numbers, if we restrict ourselves to the symmetric subspace \cite{Toth2007, Toth2009a}.
This approach can give optimal bounds for many systems, such as Bose-Einstein condensates of two-level atoms, which are in a symmetric multiparticle state.
The bound computed for the symmetric subspace might not be correct and generally might overestimate the real bound for general cases.

Finally, it is important to note that if the operators $W_k$ are permutationally invariant and the eigenstate with the maximal eigenvalue in Eq.~\eqref{eq:lt-legendre-for-qfi-simplified} is non-degenerate, then we can do the computations on the symmetric subspace only.
The resulting maximal eigenvalue is the maximal eigenvalue even when the whole Hilbert space is taken into account for the maximization.
Hence, the lower bound obtained in the symmetric subspace is valid even for the general case.

For completeness, we follow presenting the proof of the observation mentioned above.
Let us denote the ground state of a permutationally invariant Hamiltonian by $\ket{\Psi}.$
This is at the same time the $T=0$ thermal ground state, hence it must be a permutationally invariant pure state.
For such states $S_{kl}\ketbra{\Psi}{\Psi}S_{kl}=\ketbra{\Psi}{\Psi}$, where $S_{kl}$ is the swap operator exchanging qubits $k$ and $l$.
Based on this, follows that $S_{kl}\ket{\Psi}=c_{kl}\ket{\Psi}$, and $c_{kl}\in {-1,+1}$.
There are three possible cases to consider:
\begin{enumerate}
  \item All $c_{kl}=+1$.
  In this case, for all permutation operator $\Pi_j$ we have
  \be
    \label{eq:lt-permutating-ground-state}
    \Pi_j \ket{\Psi} = \ket{\Psi},
  \ee
  since any permutation operator $\Pi_j$ can be constructed as $\Pi_j=\prod_i S_{k_il_i}$.
  Equation~\eqref{eq:lt-permutating-ground-state} means that the state $\ket{\Psi}$ is symmetric.
  \item All $c_{kl}=-1$.
  This means that the state is anti-symmetric, however this state exists only for $N=2$ qubits.
  \item Not all $c_{kl}$ are identical to each other.
  In this case, there must be $k_+,l_+,k_-,k_-$ such that
  \be
    \label{eq:lt-different-index-pi}
    \begin{split}
      S_{k_+,l_+} \ket{\Psi} & = +\ket{\Psi},\\
      S_{k_-,l_-} \ket{\Psi} & = -\ket{\Psi}.
    \end{split}
  \ee
  Let us assume that $k_+,l_+,k_-,l_-$ are index different from each other.
  In this case, $\ket{\Psi'}=S_{k_+,k_-}S_{l_+,l_-}\ket{\Psi}$ another ground state of the Hamiltonian $H$ such that
  \be
    \label{eq:lt-different-index-pi-2}
    \begin{split}
      S_{k_+,l_+} \ket{\Psi'} & = -\ket{\Psi'},\\
      S_{k_-,l_-} \ket{\Psi'} & = +\ket{\Psi'}.
    \end{split}
  \ee
  Comparing Eqs.~\eqref{eq:lt-different-index-pi} and \eqref{eq:lt-different-index-pi-2} we can conclude that $\ket{\Psi'}\neq\ket{\Psi}$, while due to the permutational invariance of $H$ we have that $\expect{H}_{\Psi'} = \expect{H}_{\Psi}$.
  Thus, $\ket{\Psi}$ is not a non-degenerate ground state.
  The proof works in an analogous way for the only nontrivial case $k_+=k_-$, when $S_{k_+,k_-}=\mtxid$.
\end{enumerate}
Hence, if $N>2$ then only i) is possible and $\ket{\Psi}$ must be symmetric.

Next, we will demonstrate the use of our approach for several experimentally relevant situations.
In the many-particle case, often symmetric operators can be used to describe accurately the system, which makes it possible to carry out calculations for thousand of particles, as will be shown later in this chapter.

\subsection{Examples}
\label{sec:lt-examples}

In this section, we show how to obtain lower bounds based on the fidelities with respect to the GHZ state and the unpolarized Dicke state as well as with different sets of powers of collective angular momentum operators, e.g., the set $\{\expect{J_y}, \expect{J_x}, \expect{J_x^2}\}$.

\subsubsection{Fidelity measurements}
\label{sec:lt-bounds-fidelity}

Let us consider the case when $W$ is a projector onto a pure quantum state.
First, we consider GHZ states.
Hence $W$ is the projector $\ketbra{\ghz}{\ghz}$, where
\be
  \ket{\ghz} = \tfrac{1}{\sqrt{2}}(\ket{0\cdots0}+\ket{1\cdots1})
  \label{eq:lt-ghz-state}
\ee
for spin-$\frac{1}{2}$ particles, and $\expect{W}=F_{\ghz}$ is the fidelity with respect to the GHZ state.
Based on knowing $F_{\ghz}$, we would like to estimate $\qfif{\rho,J_z}$\footnote{
Not tight lower bounds on the quantum Fisher information based on the fidelity have been presented in \cite{Augusiak2016}.}.

Using Eq.~\eqref{eq:lt-bound-for-qfi}, we will obtain an analytical tight lower bound on the QFI based on the fidelity $F_{\ghz}$. The calculation that we have to carry out is computing the bound
\be
  \label{eq:lt-maximization-problem-fid-ghz}
  \bound{\mathcal{F}}(F_{\ghz}) = \sup_r \big\{r F_{\ghz} - \sup_{\mu} \{\lambda_{\max}[r\ketbra{\ghz}{\ghz} - 4 (J_z - \mu)^2]\}\big\}.
\ee
We will make our calculations in the $J_z$ orthonormal basis, which is defined with the $2^N$ basis vectors $b_0= \ket{00\dots000}$, $b_1=\ket{00\dots001}$, \dots, $b_{(2^N-2)}=\ket{11\dots110}$, and $b_{(2^N-1)}=\ket{11\dots111}$, as it can be found in Eq.~\eqref{eq:app-eigenbasis-tensor-product} for $j=\frac{1}{2}$.
It is easy to see that the matrix in the argument of $\lambda_{\max}$ in the Eq.~\eqref{eq:lt-maximization-problem-fid-ghz} is almost diagonal in the $J_z$ basis.
To be more specific, the only non-diagonal matrix block comes from $\ketbra{\ghz}{\ghz}$, which has non-trivial matrix elements only in the $\{b_0,b_{(2^N-1)}\}$ basis.
Thus, we have to diagonalize the following matrix
\be
  \label{eq:lt-expression-to-diagonalize-ghz}
  r\ketbra{\ghz}{\ghz} - 4 (J_z-\mu)^2 =
  \begin{pmatrix}
    \frac{r}{2}-4(\frac{N}{2}-\mu)^2 & \frac{r}{2}\\
    \frac{r}{2} & \frac{r}{2}-4(\frac{N}{2}+\mu)^2
  \end{pmatrix}
  \oplus D,
\ee
where $D$ is already a $(2^N-2)\times(2^N-2)$ diagonal matrix with $D_k=-4( \expect{J_z}_{b_k}-\mu)^2$ negative eigenvalues for $k=1,2,\dots, (2^N-2)$.
This means that the Eq.~\eqref{eq:lt-expression-to-diagonalize-ghz} can be diagonalized as $\text{diag}[\lambda_{+},\lambda_{-},D_1,D_2,\dots,D_{2^N-2}]$, where the two eigenvalues $\lambda_{\pm}$ are
\be
  \lambda_{\pm} = \frac{r}{2}-N^2-4 \mu^2\pm\sqrt{16\mu^2N^2+\frac{r^2}{4}}.
\ee

Next, we show a way that can simplify our calculations considerably.
As indicated in Eq.~\eqref{eq:lt-maximization-problem-fid-ghz}, we have to look for the maximal eigenvalue and then optimize it over $\mu$.
We exchange the order of the two steps, that is, we look for the maximum of each eigenvalue over $\mu$, and then find the maximal one.
The eigenvalues of $D$ are negative and for some $\mu$'s some of them can be zero.
Due to this, the problem can be simplified to the following equation
\be
  \label{eq:lt-ghz-legendre-solution}
  \begin{split}
  \sup_{\mu}\{\lambda_{\max}[r\ketbra{\ghz}{\ghz}-4(J_z-\mu)^2]\}:= & \max\{0,\sup_{\mu}(\lambda_{+})\}\\
  = & \lcor
  \begin{aligned}
    &0, && \text{ if } r<0,\\
    &\frac{r}{2}+\frac{r^2}{16N^2} && \text{if } 0\leqslant r\leqslant 4N^2,\\
    &-N^2+r && \text{if }r>4N^2,
  \end{aligned}
  \right.
  \end{split}
\ee
where we did not have to have to look for the maximum of $\lambda_{-}$ over $\mu$ since clearly $\lambda_{+}\geqslant\lambda_{-}$.
Finally, we have to substitute Eq.~\eqref{eq:lt-ghz-legendre-solution} into Eq.~\eqref{eq:lt-maximization-problem-fid-ghz}, and carry out the optimization over $r$, considering $F_{\ghz}\in[0,1]$.

\begin{figure}[htp]
  \centering
  \igwithlabel{(a)}{scale=.65}{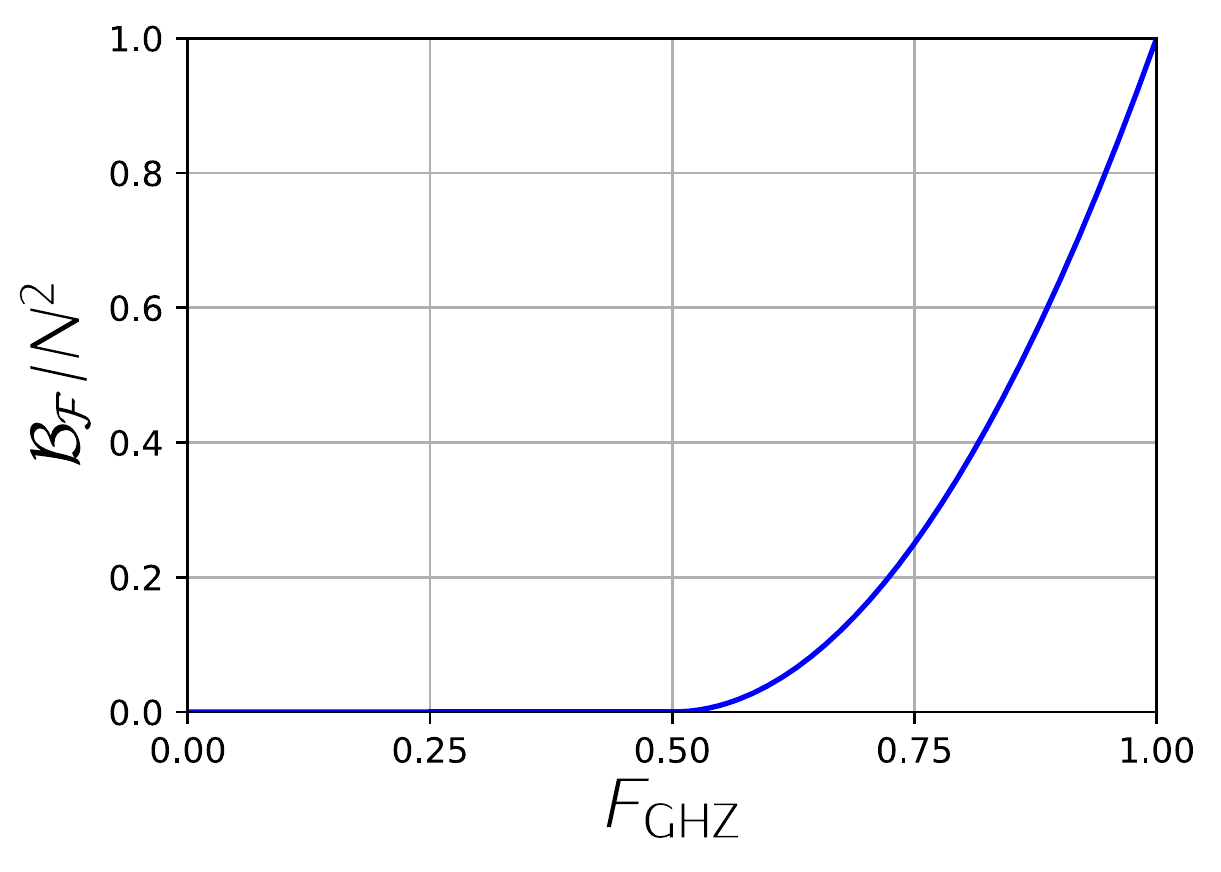}
  \igwithlabel{(b)}{scale=.65}{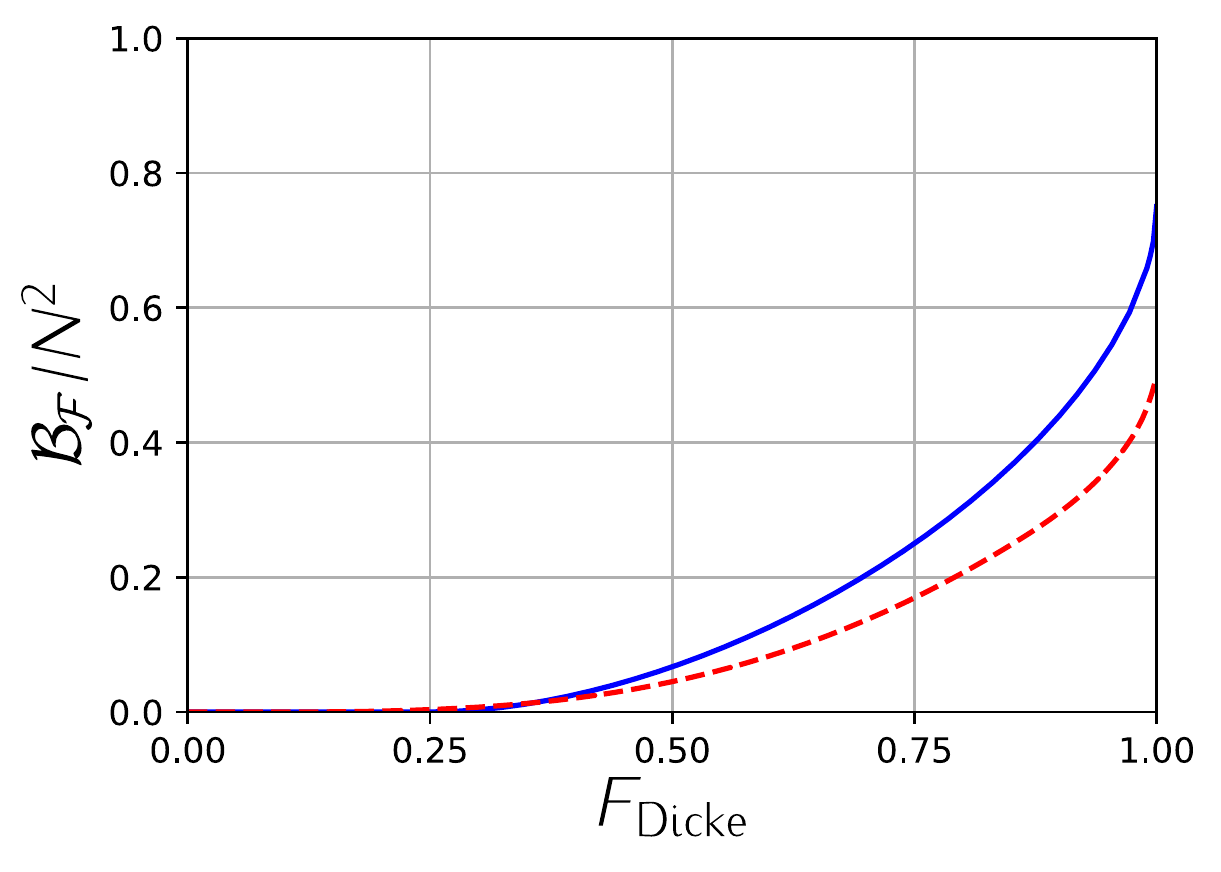}
  \caption[Lower bound for fidelities. (a) $F_{\text{GHZ}}$. (b) $F_{\text{Dicke}}$.]{
  (a) Analytical solution of the bound $\bound{\mathcal{F}}$ for different values of the fidelity with respect to the GHZ state.
  (b) Numerical results for the minimum quantum Fisher information as a function of the fidelity with respect of unpolarized Dicke states perpendicular to the magnetic field, $|\text{D}_N^0\rangle$.
  (line) For systems with 4 particles and (dashed) for systems with 40 particles.
  One may note that when the fidelity is at its maximum the bound approaches 0.5 as it is the quantum Fisher information for a large particle number.
  }
  \label{fig:lt-plots-for-fidelities}
\end{figure}

This way we arrive at a lower bound of the QFI based on the fidelity with respect to the GHZ state as
\be
  \bound{\mathcal{F}}(F_{\ghz}) = \lcor
  \begin{aligned}
    & N^2(1-F_{\ghz})^2 && \text{if } F_{\ghz} < 1/2, \\
    & 0 && \text{if } F_{\ghz}\leqslant 1/2.
  \end{aligned}
  \right.
\ee
This equation is plotted in Figure~\ref{fig:lt-plots-for-fidelities}-(a).
Note that in the figure the plot is normalized by $N^2$ and that the resulting semi-parabola is independent of the number of particles.
Moreover, the bound is zero for $F_{\ghz}\leqslant 1/2$.
This is consistent with the fact that for the product states $\rho=\ket{111\dots11}$ or $\rho=\ket{000\dots00}$ we have $F_{\ghz}=1/2$, while $\mathcal{F}_{\text{Q}}[\rho,J_z]=0$.

Next, let us consider a symmetric unpolarized Dicke state with even $N$ particles along the $x$-direction $\ket{\dicke{N}}_x$, given by Eq.~\eqref{eq:vd-unpolarized-dicke}.
This state is known to be highly entangled \cite{Toth2007, Toth2009} and allows for a Heisenberg limited interferometry \cite{Holland1993}.
In the following we may omit the subscript $x$ since this Dicke state will be always at the center of our attention, the unpolarized Dicke state perpendicular to the magnetic field in this case along the $z$-direction.
The witness operator that can be used for noisy Dicke states is $W=\ketbra{\dicke{N}}{\dicke{N}}$, hence the expectation value of the witness is just the fidelity with respect to the Dicke state, i.e., $\expect{W}=F_{\text{Dicke}}$.
In Figure~\ref{fig:lt-plots-for-fidelities}-(b), we plotted the results for symmetric Dicke states of various particle numbers.
$F_{\text{Dicke}}=1$ corresponds to $\mathcal{F}_{\text{Q}}[\rho,J_z]=N(N+2)/2$.
At this point, note that for the examples presented above, the QFI bound scales as $\mathcal{O}(N^2)$ in the asymptotic limit if the quantum state has been prepared perfectly\footnote{$\mathcal{O}(x)$ is the usual Landau notation used to describe the asymptotic behavior for large $x$ \cite{Hyllus2012, Toth2012}.}.

Note that estimating $\qfif{\rho, J_z}$ based on $F_{\text{Dicke}}$ was possible for 40 qubits for Figure~\ref{fig:lt-plots-for-fidelities}-(b), since we carried out the calculations for the symmetric subspace.
For our case, the witness operator $W$ is permutationally invariant and it has a non-degenerate eigenstate corresponding to the maximal eigenvalue.
Hence, based on the arguments of the Section~\ref{sec:lt-symmetries} the bound is valid even for the general case, i.e., non-symmetric states.

We now compute several quantities for the large $N$ case.
We show that if the fidelity with respect to the Dicke state is larger than a bound then $\bound{\mathcal{F}}>0$, where we omit the arguments for brevity.
Moreover, we have seen in Figure~\ref{fig:lt-plots-for-fidelities}-(b) that the lower bound on $\qfif{\rho,J_z}$ as a function of the fidelity $F_{\text{Dicke}}$ normalized by $N^2$ is not the same curve for all $N$.
Next, we will demonstrate by numerical evidence that the lower bound normalized by $N^2$ collapses to a nontrivial curve for large $N$.

As a first step, let us consider the state completely polarized along $z$-direction $\ket{1}_y^{\otimes N}$.
This state does not change under rotations around the $z$-axis, hence $\qfif{\rho,J_z}=0$.
Its fidelity with respect to the Dicke state $\ket{\dicke{N}}_x$ is
\be
  \label{eq:lt-fidelity-dicke-with-tp}
  F_{\text{Dicke}}(\ket{1}_y^{\otimes N}) = \frac{1}{2^N}\binom{N}{N/2}\approx \sqrt{\frac{2}{\pi N}}.
\ee
From convexity of the bound on the quantum Fisher information in $F_{\text{Dicke}}$, it immediately follows that for $F_{\text{Dicke}}$ smaller than Eq.~\eqref{eq:lt-fidelity-dicke-with-tp} the optimal bound on $\qfif{\rho,J_z}$ will give zero.

Next, we examine what happens if the fidelity is larger than Eq.~\eqref{eq:lt-fidelity-dicke-with-tp}.
For that we note first that $\qfif{\rho,J_z}$ is the convex roof of $4\varian{J_z}$ \cite{Toth2013, Yu2013}.
Hence, if we have a mixed state for which $\qfif{\rho,J_z}$ is zero, then it can always be decomposed into the mixture of pure states for which $\qfif{\ket{\Psi},J_z}$ is zero too.
As a consequence, the extremal states of the set of states for which $\qfif{\rho,J_z}=0$ are pure states, and we can restrict our search for pure states.
The optimization problem we have to solve is given as
\be
  F_{\text{opt}} = \big\{ \max_{\Psi} |\braket{\Psi}{\dicke{N}}_x|^2 \, : \, \qfif{\ket{\Psi},J_z}=0\big\}.
\ee
Hence, we have to carry out the optimization over pure states $\ket{\Psi}$ that are invariant under $U_{\theta}=\exp(-iJ_z\theta)$ for any $\theta$.
Such states are the eigenstates of $J_z$.
In order to maximize the overlap with the Dicke state $\ket{\dicke{N}}_{x}$, we have to look for symmetric eigenstates of $J_z$.
These are the symmetric Dicke states in the $z$-basis $\ket{\dicke{N,m}}_z$.
Then, using the following identity
\be
  \sum_{k=0}^q (-1)^k \binom{n}{k}\binom{n}{q-k} = \lcor
  \begin{aligned}
    &\binom{n}{q/2}(-1)^{q/2} && \text{for even }q,\\
    &0  && \text{for odd }q,
  \end{aligned}
  \right.
  \label{eq:lt-binomial-identity}
\ee
one finds that the squared overlap is given by
\be
  |\braopket{\dicke{N,m}}{_z}{\dicke{N}}_x|^2 = \lcor
  \begin{aligned}
    &\frac{\binom{N/2}{m/2}^2\binom{N}{N/2}}{2^N\binom{N}{m}} && \text{for even }m,\\
    &0 && \text{for odd }m,
  \end{aligned}
  \right.
  \label{eq:lt-dicke-overlap}
\ee
which is maximal in the case of even $N$ when $m=N$ or $m=0$, the state totally polarized along the $+z$-direction or along the $-z$-direction respectively.
We skip the case in which $N$ is odd.
For detailed calculations of Eq.~\eqref{eq:lt-dicke-overlap} see Appendix~\ref{app:calculation-dicke-overlap}.

\begin{figure}[htp]
  \centering
  \includegraphics[scale=.65]{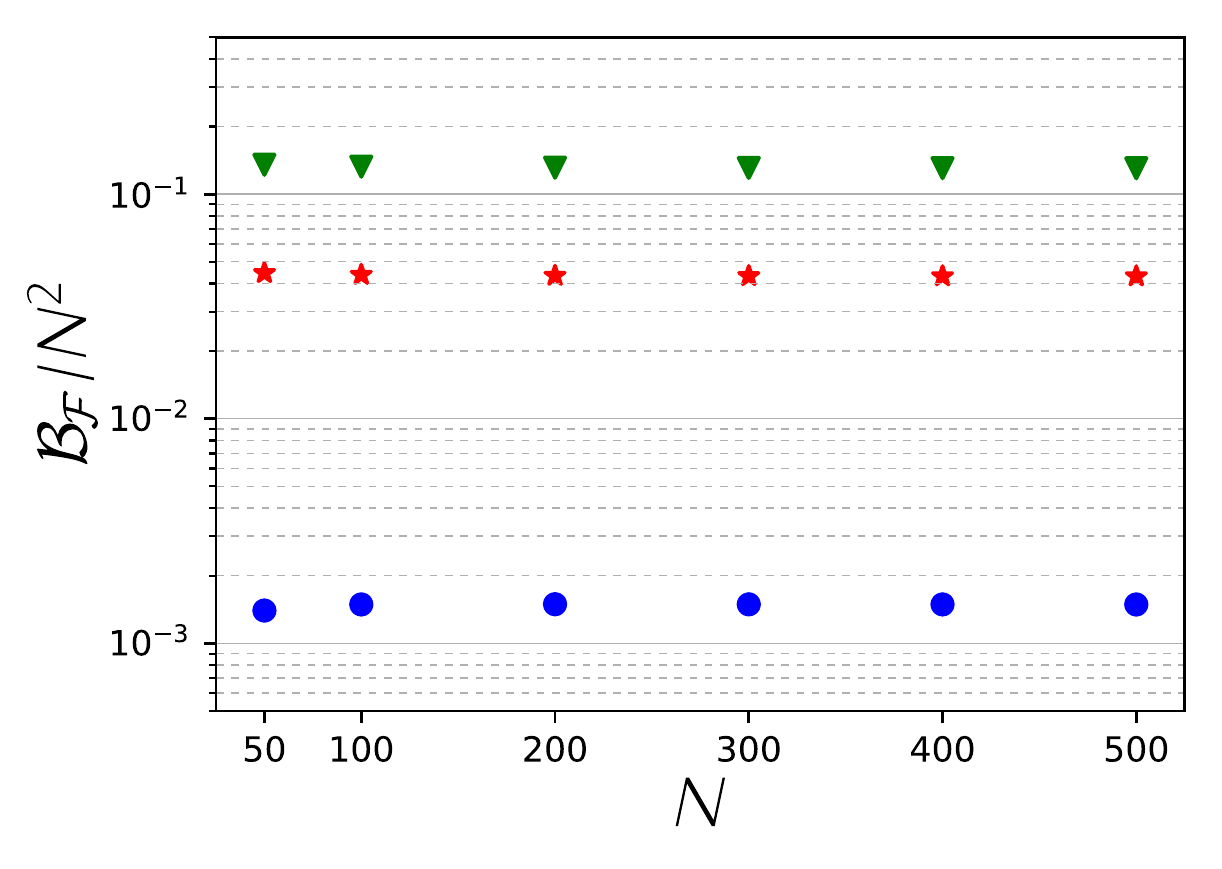}
  \caption[Lower bound QFI for $F_{\text{Dicke}}=0.2,0.3,0.7$ for different particle numbers]{
  Lower bound on QFI normalized by $N^2$ for various particle numbers $N=50,100,200,300,$ $400,500$.
  (circles) Lower bound for $F_{\text{Dicke}}=0.2$, (stars) for $F_{\text{Dicke}}=0.5$, and (triangles) for $F_{\text{Dicke}}=0.7$.
  For a better visibility we use a logarithmic scale for the $y$-axis.
  }
  \label{fig:lt-nrange-fdicke}
\end{figure}

Next, we will examine the behavior of our lower bound on $\qfif{\rho,J_z}$ based on the fidelity $F_{\text{Dicke}}$ for large $N$.
In Figure~\ref{fig:lt-nrange-fdicke}, the calculations up to $N=500$ present a strong evidence that for fidelity values $F_{\text{Dicke}}=0.2,0.5,0.7$ the lower bound on QFI has a $\mathcal{O}(N^2)$ scaling for increasing $N$.
If this is correct then reaching a fidelity larger than a certain bound for large $N$ would imply Heisenberg scaling for the bound on the quantum Fisher information.
Note that it is difficult to present a similar numerical evidence for small values of $F_{\text{Dicke}}$ since in that case the bound for QFI is nonzero only for large $N$ due to Eq.~\eqref{eq:lt-fidelity-dicke-with-tp}.

\subsubsection{Spin-squeezed states}
\label{sec:lt-bound-spsq}

In the case of spin squeezing, the quantum state has a large spin in the $y$-direction, while a decreased variance in the $x$-direction.
By measuring $\expect{J_y}$ and $\varian{J_x}$ we can estimate the lower bound on the quantum Fisher Information by Eq.~\eqref{eq:bg-pezze-bound}.
However, this formula does not necessarily give the best lower bound for all values of the collective observables.
With our approach we can find the best bound.

To give a concrete example, we choose $W_1=J_y$, $W_2=J_x^2$ and $W_3=J_x$ for the operators to be measured.
We vary $w_1$ and $w_2$ in some interval.
We also require that $w_3=0$, since we assume that the mean spin points into the $y$-direction\footnote[1]{
Due to symmetries of the problem, when minimizing $\qfif{\rho,J_z}$ with the constraints on $\expect{J_z}$ and $\expect{J_x^2}$, we do not have to add explicitly the constraint $\expect{J_x}=0$.
The optimization with only the first two constraints will give the same bound.}.
This is reasonable since in most spin-squeezing experiments we know the direction of the mean spin.

\begin{figure}[htp]
  \centering
  \includegraphics[scale=.65]{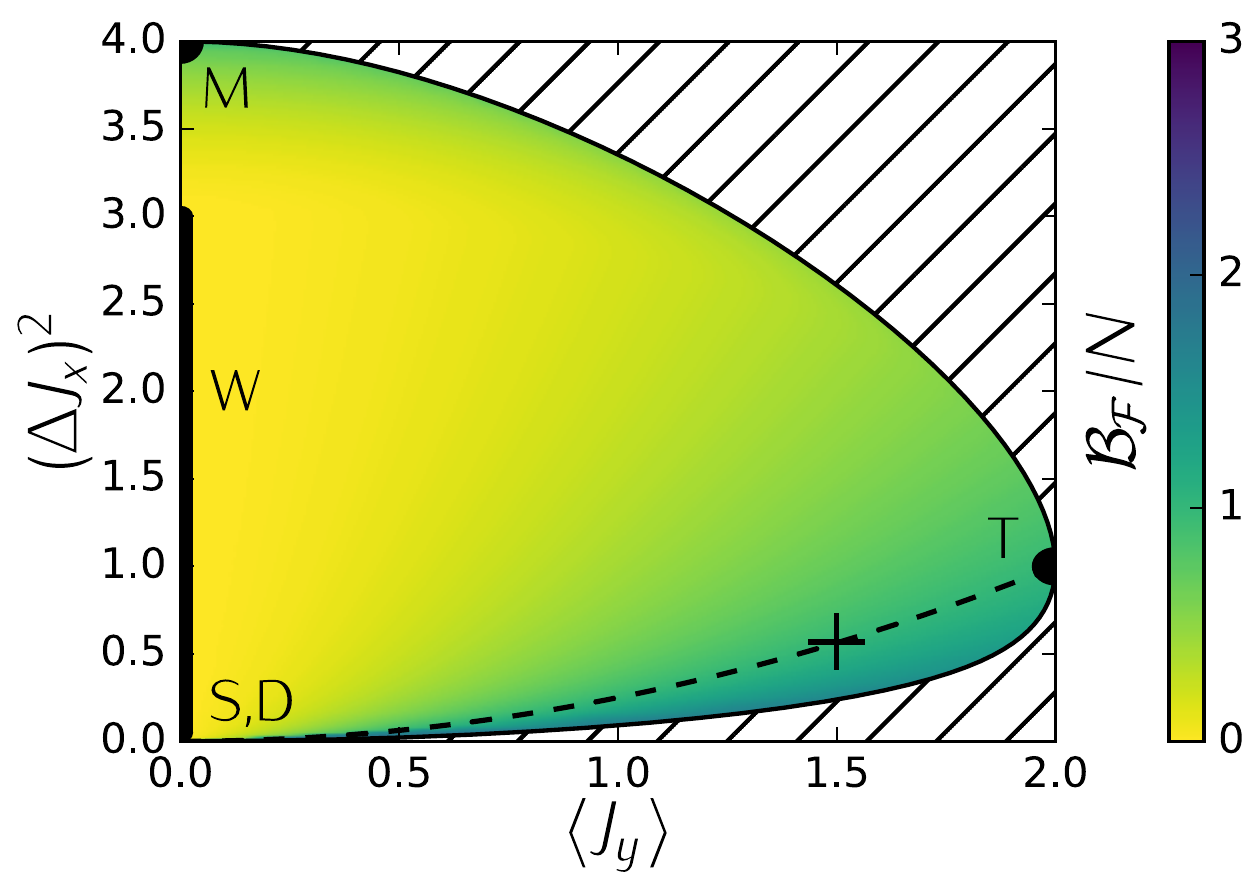}
  \caption[Precision bound for 4 particles for $\expect{J_y}$ and $\varian{J_x}$.]{
  We show as a function of the expectation value, $\expect{J_y}$, and the variance in the perpendicular direction, $\varian{J_x}$, the minimum sensitivity for a 4-qubit system.
  (hatched) The physically forbidden region is indicated.
  Interesting quantum states: (M) Mixed state defined in the text, (T) totally polarized state, (S) singlet state, and (D) Dicke state.
  (W) Any mixture of the singlet state and the completely mixed state of the symmetric subspace.
  Other states can be found on this line, for instance, the completely mixed state of the whole Hilbert space.
  (dashed) Shot-noise threshold.
  Below this line non-classical sensitivities can be achieved.
  (cross) In Figure~\ref{fig:lt-edge-diff-and-adding-jx4}, we compute the bound when an additional expectation value is measured.
  }
  \label{fig:lt-spsq2d-4}
\end{figure}

Our result can be seen in Figure~\ref{fig:lt-spsq2d-4}.
We chose $N=4$ particles since for small $N$ the main features of the plot are clearly visible.
The hatched area corresponds to non-physical combination of expectation values.
States at the boundary can be obtained as ground states of $H_{\text{bnd}}^{(\pm)}(\lambda)=\pm J_x^2 -\lambda J_y$, see Appendix~\ref{app:spin-squeezing-hamiltonian}.
In Figure~\ref{fig:lt-spsq2d-4}, the state fully polarized in the $y$-direction, and initial state for spin-squeezing experiments, corresponds to point T.
The unpolarized Dicke state along the $x$-direction Eq.~\eqref{eq:vd-unpolarized-dicke} corresponds to point D.

We add that outside the symmetric subspace, there are other states with $\expect{J_y}=\expect{J_x^2}=0$, which also correspond to the point D, e.g the singlet state labeled by the point S.
However, usual spin-squeezing procedures remain in the symmetric subspace, thus we discuss only the Dicke state.
Spin-squeezing makes $\varian{J_x}$ decrease, while $\expect{J_y}$ also decreased somewhat.
Hence, at least for small squeezing it corresponds to moving down from point T to point D following the boundary, while the metrological usefulness is increasing.
Below the dashed line $\qfif{\rho,J_z}>N$, hence the state possesses metrologically useful entanglement \cite{Pezze2009}.
The equal mixture of $\ket{000\dots00}_x$ and $\ket{111\dots11}_x$ corresponds to point M, with $\qfif{\rho,J_z} = N$.
Finally, the completely mixed state rests on the line W.
It cannot be used for metrology, hence $\qfif{\rho,J_z}=0$.

We now compare the difference between our bound and the bound of L.~Pezz\`e and A.~Smerzi Eq.~\eqref{eq:bg-pezze-bound}.
First, we consider the experimentally relevant region for which $\varian{J_x}\leqslant 1$.
We find that for points away from the physical boundary at least by 0.001 on the vertical axis, the difference between the two bounds is smaller than $2\times10^{-6}$.
Hence, Eq.~\eqref{eq:bg-pezze-bound} practically coincides with the optimal bound for $\varian{J_x}<1$.

\begin{figure}[htp]
  \centering
  \igwithlabel{(a)}{scale=.65}{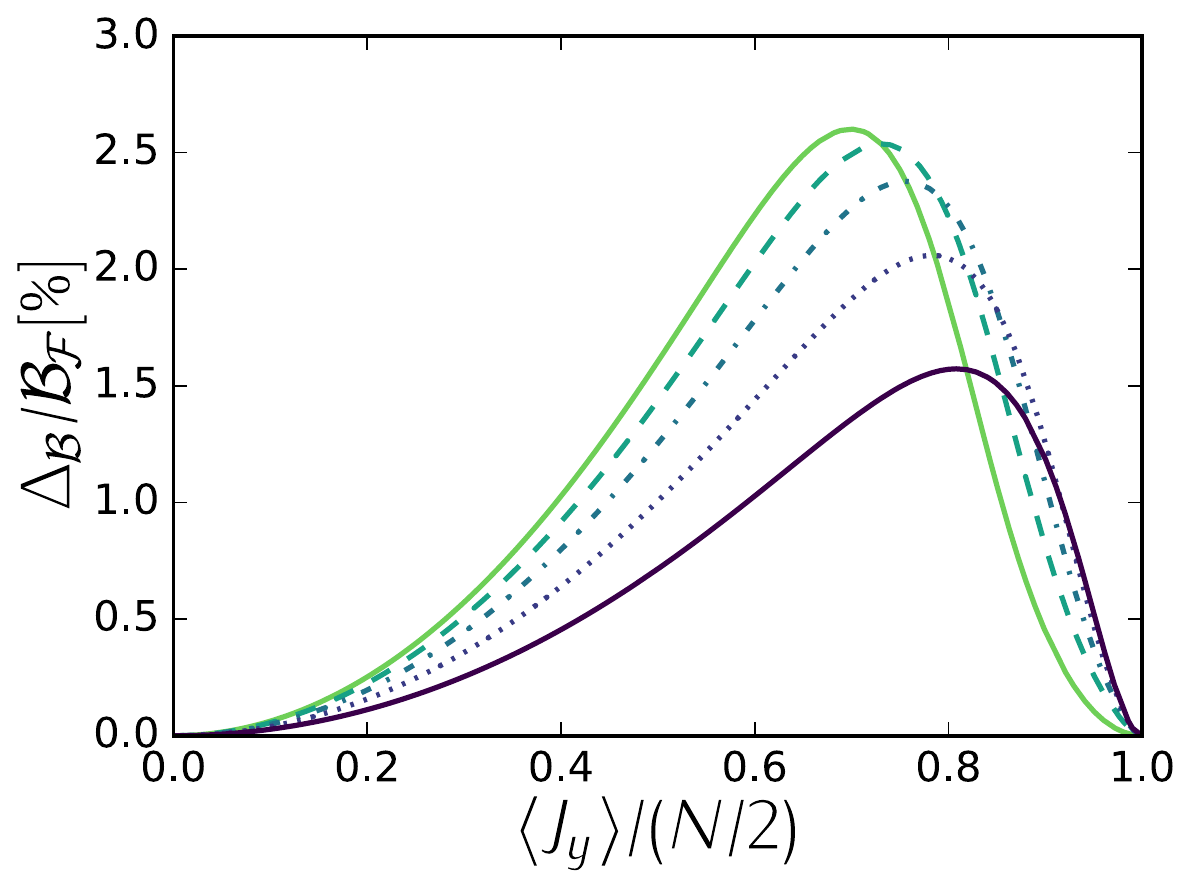}
  \igwithlabel{(b)}{scale=.65}{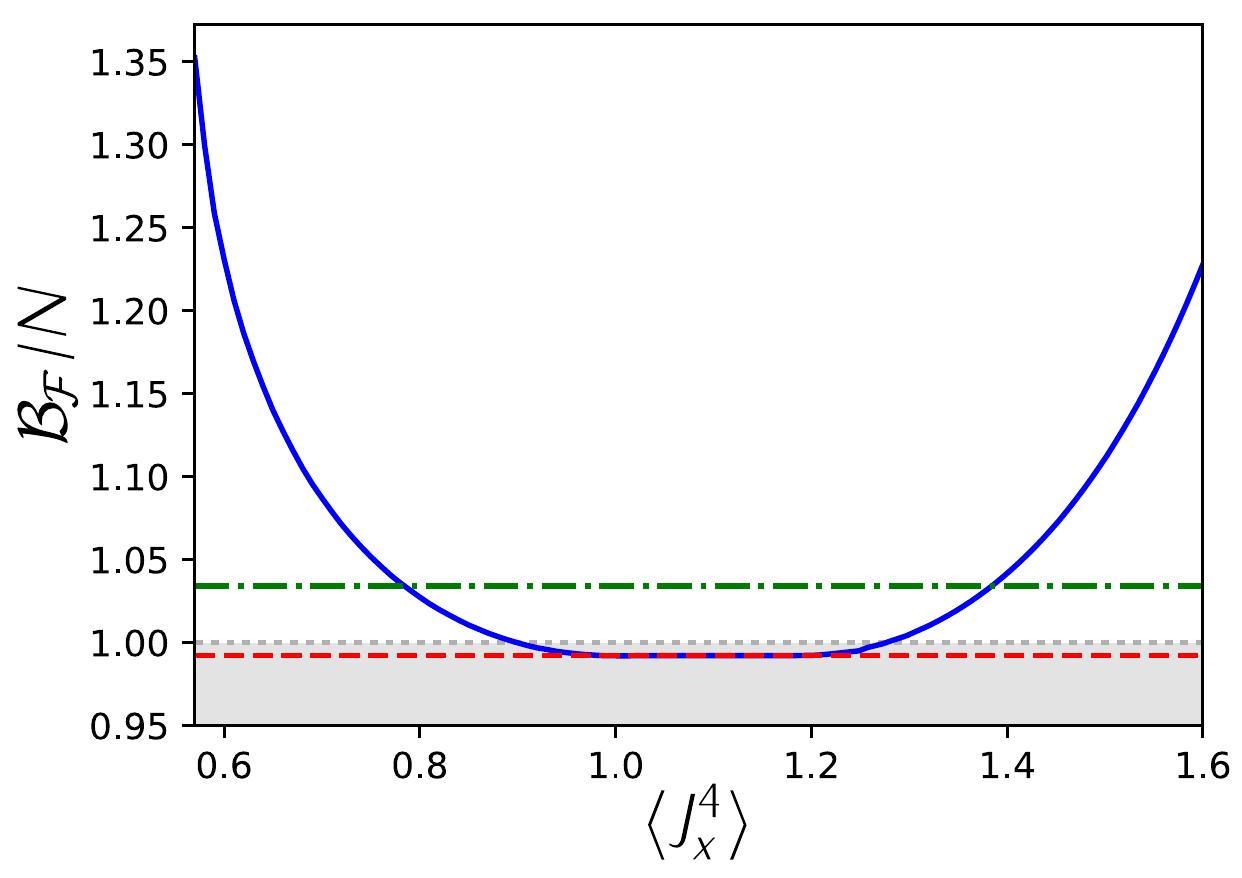}
  \caption[(a) Difference between our bound and Pezz\`e-Smerzi bound. (b) Improving the bound with a third measurement.]{
  (a) Difference between the bound of Pezz\`e-Smerzi and the optimal bound for the quantum Fisher information normalized by the value of the optimal bound itself for the bosonic ground states of $H=J_x^2-\lambda J_y$ for $\forall \lambda \in [0,\infty)$.
  From dark to lighter colors (solid-dark, dotted, dash-dotted, dashed, pointed, solid-light), results for different particle numbers, $N=4,6,10,20,1000$ respectively.
  For large particle numbers the difference is largest when the polarization is around two thirds of the maximal polarization and that this difference is less than $2.6\%$.
  (b) Lower bound on QFI for $\expect{J_y}=1.5$, $\varian{J_x}=0.567$, as a function of $\expect{J_x^4}$.
  The corresponding point in Figure~\ref{fig:lt-spsq2d-4} is denoted by a cross.
  (gray-area) Lower bound on precision below the shot-noise limit.
  (dashed) Lower bound without constraining $\expect{J_x^4}$.
  (dash-dotted) Lower bound when bosonic symmetry is considered.
  As can be seen, an additional constraint or assuming symmetry improves the bound.}
  \label{fig:lt-edge-diff-and-adding-jx4}
\end{figure}

For points at the boundary, the difference is somewhat larger, but still small, the relative difference is smaller than $2\%$ for 4 particles.
We compute the difference between the Eq~\eqref{eq:bg-pezze-bound} and our bound for different number of particles and for states at the boundary from the state totally polarized T to the unpolarized Dicke state at D, see Figure~\ref{fig:lt-edge-diff-and-adding-jx4}-(a).

We now consider regions on Figure~\ref{fig:lt-spsq2d-4} for which $\varian{J_x}>1$.
The difference between the two bounds is now larger.
It is larger at point M, for which the bound Eq.~\eqref{eq:bg-pezze-bound} is zero.
Hence for measurement values corresponding to points close to M, our method improve the formula Eq.~\eqref{eq:bg-pezze-bound}.

It is important from the point of view of applying our method to spin-squeezing experiments that the bound Eq.~\eqref{eq:bg-pezze-bound} can be substantially improved for $\varian{J_x}<1$, if we assume bosonic symmetry for the system, or we measure an additional quantity, such as $\expect{J_x^4}$ as shown in Figure~\ref{fig:lt-edge-diff-and-adding-jx4}-(b).

\subsubsection{Dicke states}
\label{sec:lt-bound-dicke-states}

In this section, we use our method to find lower bounds on the QFI for states close to the Dicke states \eqref{eq:vd-unpolarized-dicke} along the $x$-direction, based on collective measurements.
We discuss what operators have to be measured to estimate the metrological usefulness of the state.
In Section~\ref{sec:lt-many-particle-experiments}, we will test our approach for a realistic system with very many particles.

In order to estimate the metrological usefulness of states created in such experiments, we choose to measure $W_1=J_x^2$, $W_2=J_y^2$ and $W_3=J_z^2$ since the expectation values of these operators uniquely define the ideal Dicke state, and they have been already used for entanglement detection \cite{Luecke2014}.
In cold gas experiments of nowadays, the state created is invariant under transformations of the type $U_{x}(\phi)=\exp(-i J_x \phi)$ \cite{Apellaniz2015}.
For such states $\expect{J_y^2}=\expect{J_z^2}$, which we also use as a constraint in our optimization.

\begin{figure}[htp]
  \centering
  \includegraphics[scale=.65]{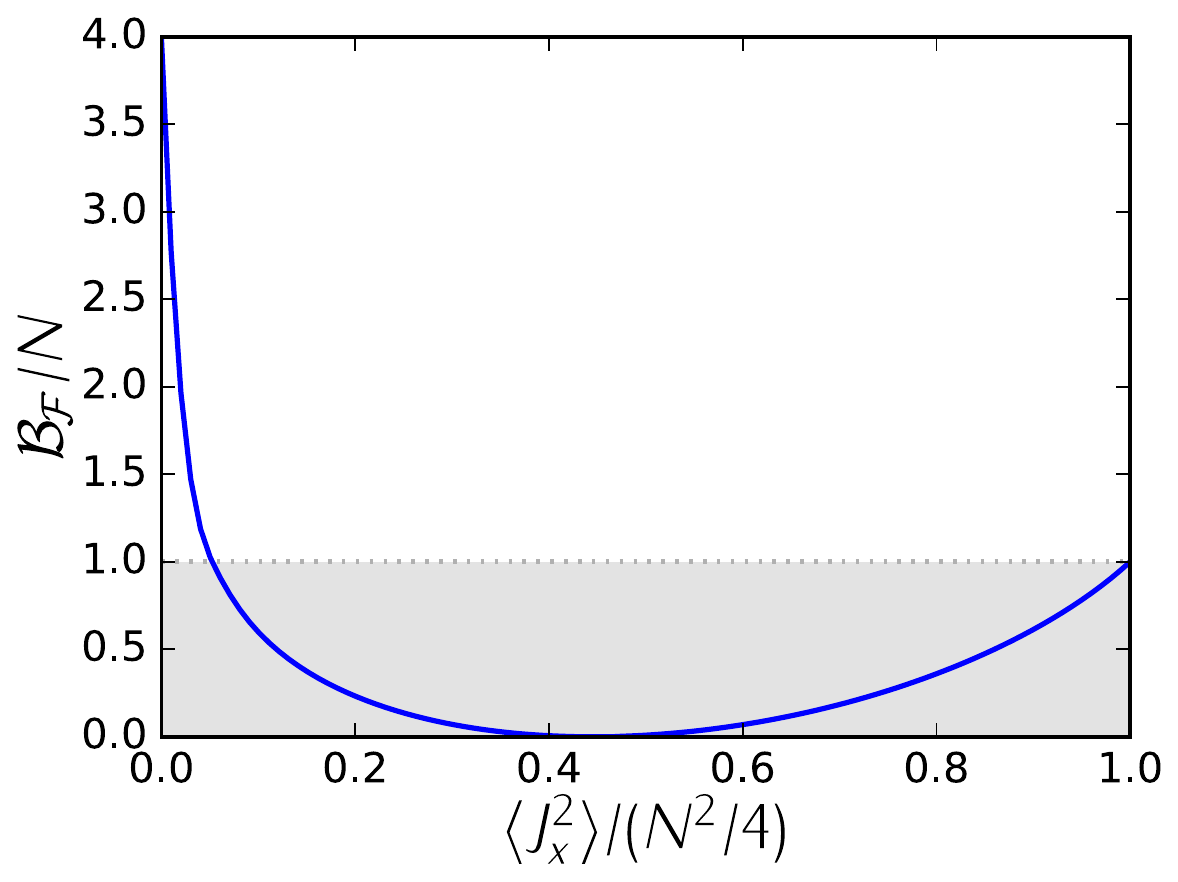}
  \caption[Bound for 6-particles with BEC symmetry for $\{\expect{J_x^2},\expect{J_y^2},\expect{J_z^2}\}$ are measured]{
  Optimal lower bound on the quantum Fisher Information for symmetric states with $\expect{J_y^2}=\expect{J_z^2}$. Even if it is metrologicaly useful for a wide range of $\expect{J_x^2}$, the numerics shows us a tiny region where the metrological gain is surpassing the shot-noise limit.}
  \label{fig:lt-symmetric-dicke-6-bound}
\end{figure}

Let us demonstrate how our method works in an example for small systems.
Figure~\ref{fig:lt-symmetric-dicke-6-bound}
shows the result for 6 qubits for symmetric states for which
\be
  \label{eq:lt-maximum-angular-momentum}
  \expect{J_x^2+J_y^2+J_z^2} = \frac{N(N+2)}{4}=:\mathcal{J}_{N/2},
\ee
which was introduced in Chapter~\ref{sec:vd}.
It can be seen that the lower bound on quantum Fisher Information is the largest for $\expect{J_x^2}=0$.
It reaches the value corresponding to the ideal Dicke state, $\qfif{\rho,J_z}/N=(N+2)/2=4$.
It is remarkable that the state is also useful for metrology if $\expect{J_x^2}$ is very large.
In this case $\expect{J_y^2}$ and $\expect{J_z^2}$ are smaller than $\expect{J_x^2}$.

\subsection{Calculations for experimental data}

In this section, we use our method to find tight lower bound on the QFI based on experimental data.
In particular, we will determine the bound for several experiments in photons and trapped ions creating GHZ states and Dicke states, in which the fidelity has been measured \cite{Krischek2011, Zhao2003, Gao2010, Leibfried2004, Sackett2000, Monz2011, Kiesel2007, Wieczorek2009, Prevedel2009, Chiuri2012}, which is much easier than obtaining the quantum Fisher Information from the density matrix \cite{Hyllus2012}, or estimation it from a metrological procedure \cite{Luecke2011}.
We will obtain a bound on the QFI for a spin-squeezing experiment with thousand of particles \cite{Gross2010}.
Based on numerical examples, we see that the bound Eq.~\eqref{eq:bg-pezze-bound} is close to the optimal even for not completely polarized states.
Assuming symmetry or knowing additional expectation values can improve the bound Eq.~\eqref{eq:bg-pezze-bound}.
Finally, we will also obtain the bound for the QFI for a recent experiment with Dicke states \cite{Luecke2014}.
The estimate of the precision based on considering the particular case when $\expect{J_x^2}$ is measured for parameter estimation \cite{Apellaniz2015} is close to the optimal bound computed by our method.

\subsubsection{Few-particle experiments}

Now, we will estimate the quantum Fisher information based on the fidelity with respect to Dicke states and GHZ states for several experiments with photons and trapped cold ions, following the ideas of Section~\ref{sec:lt-bounds-fidelity}.

Our results are summarized in Table~\ref{tab:lt-results-for-fidelities}.
The experiments in \cite{Gao2010,Chiuri2012} are with hyperentangled qubits, while in the rest of experiments a single qubit is stored in a particle.
Ref.~\cite{Monz2011} describes experiments with 2-14 ions, we presented only results of two of them.
Finally, for the experiment of Ref.~\cite{Zhao2004} we used the fidelity estimated using reasonable assumptions discussed in that paper, while the worst case fidelity is lower.
\begin{table}
  \begin{center}
    \begin{tabular}{| l | l | l | l | l |}
      \hline
      Physical&Target &\multicolumn{1}{ l|}{\multirow{2}{*}{Fidelity}}
      &\multicolumn{1}{ l|}{\multirow{2}{*}{$\bound{\mathcal{F}}/N$}}
      &\multicolumn{1}{ l|}{\multirow{2}{*}{Ref.}} \\
      system & quantum state &  &  &  \\ \hline
      \multicolumn{1}{|l|}{\multirow{6}{*}{photons}}
      & \multicolumn{1}{ l|}{\multirow{4}{*}{$\ket{\dicke{4}}$}}
      & $0.844\pm0.008$   & $1.432\pm0.044$ & \cite{Kiesel2007} \\
      & & $0.78\pm0.008$    & $1.124\pm0.236$ & \cite{Chiuri2012} \\
      & & $0.8872\pm0.0055$ & $1.680\pm0.036$ & \cite{Krischek2011} \\
      & & $0.873\pm0.005$   & $1.44\pm0.024$  & \cite{Toth2010} \\ \cline{2-5}
      & \multicolumn{1}{ l|}{\multirow{2}{*}{$\ket{\dicke{6}}$}} &
      $0.654\pm0.024$   & $0.564\pm0.076$ & \cite{Wieczorek2009} \\
      & & $0.56\pm0.02$     & $0.304\pm0.048$ & \cite{Prevedel2009} \\ \hline
      \multicolumn{1}{|l|}{\multirow{5}{*}{photons}}
      & $\ket{\ghz_{4}}$  & $0.840\pm0.007$ & $1.848\pm0.076$ & \cite{Zhao2004} \\
      & $\ket{\ghz_{5}}$  & $0.68$          & $0.65$ & \cite{Zhao2004} \\
      & $\ket{\ghz_{8}}$  & $0.59\pm0.02$   & $0.256\pm0.128$ & \cite{Huang2011} \\
      & $\ket{\ghz_{8}}$  & $0.776\pm0.06$  & $2.4376\pm0.1072$ & \cite{Gao2010} \\
      & $\ket{\ghz_{10}}$ & $0.561\pm0.019$ & $0.15\pm0.11$ & \cite{Gao2010} \\ \hline
      \multicolumn{1}{|l|}{\multirow{5}{*}{trapped-ions}}
      & $\ket{\ghz_{3}}$  & $0.89\pm0.03$   & $1.824\pm0.291$ & \cite{Leibfried2004} \\
      & $\ket{\ghz_{4}}$  & $0.57\pm0.02$   & $0.08\pm0.052$  & \cite{Sackett2000}\\
      & $\ket{\ghz_{6}}$  & $0.509\pm0.004$ & $0.0018\pm0.0018$ & \cite{Leibfried2005} \\
      & $\ket{\ghz_{8}}$  & $0.817\pm0.004$ & $3.21\pm0.08$ & \cite{Monz2011} \\
      & $\ket{\ghz_{10}}$ & $0.626\pm0.006$ & $0.64\pm0.06$ & \cite{Monz2011} \\ \hline
    \end{tabular}
  \end{center}
  \caption[Bounds on QFI for experimental data when fidelities are measured]{
  Fidelity values and the corresponding bound for the QFI for several experiments with Dicke states and GHZ states.
  Bounds normalized with $N$ are shown.
  The ones surpassing the value one in the fourth column show quantum entanglement enhanced metrological usefulness.
  For Dicke states the maximum is achieved at $(N+2)/2$, i.e., $3$ for the $\ket{\dicke{4}}$ case and $4$ for the $\ket{\dicke{6}}$ case.
  For the case in which GHZ states are used the limit for the normalized bound is $N$, the particle number.}
  \label{tab:lt-results-for-fidelities}
\end{table}

We can compare our estimate to the quantum Fisher information of the state for the experiment of Ref.~\cite{Krischek2011}, where the QFI for the density matrix was obtained as $\qfif{\rho,J_z}/N=(10.326\pm0.093)/N=(2.5816\pm0.02325)$.
As can be seen in Table~\ref{tab:lt-results-for-fidelities}, this value is larger than we obtained, however, it was calculated by knowing the entire matrixm, while our bound is obtained from the fidelity alone.

\subsubsection{Many-particle experiments}
\label{sec:lt-many-particle-experiments}

In this section, we will estimate the quantum Fisher information based on collective measurements for experiments aiming to create spin-squeezed states and Dicke states.

\subsubsubsection{Spin-squeezing experiment}

We turn our attention to a recent many-particle spin-squeezing experiment in cold gases to use our method to find lower bounds on the quantum Fisher information, following the ideas of Section~\ref{sec:lt-bound-spsq}.
With that we show that the lower bound given in Eq.~\eqref{eq:bg-pezze-bound} is close to the optimal.
We also demonstrate that we carry out calculations for real systems.

In particular, for our calculations we use the data from spin-squeezing experiments of Ref.~\cite{Gross2010}.
The particle number is $N=2300$, and the spin-squeezing parameter defined as
\be
  \label{eq:lt-spin-squeezing-parameter}
  \xi_{\textnormal{s}}^2 = N \frac{\varian{J_x}}{\expect{J_y}^2}
\ee
has the value $\xi_{\textnormal{s}}^2=-8.2\db=10^{-8.2/10}=0.1514$.
The spin length $\expect{J_y}$ has been close to maximal.
In our calculations, we choose
\be
  \expect{J_y}=\alpha \frac{N}{2},
\ee
where we will test our method with various values for $\alpha$.
For each $\alpha$ we use $\varian{J_x}$ will be given such that we get the experimentally obtained spin-squeezing parameter Eq.~\eqref{eq:lt-spin-squeezing-parameter}.
Moreover, we assume $\expect{J_x}=0$, as the $y$-direction was the direction of the mean spin in the experiment.
Based on Eq.~\eqref{eq:bg-pezze-bound}, the bound for the quantum Fisher information is obtained as
\be
  \label{eq:lt-bound-for-experiment}
  \frac{\qfif{\rho,J_z}}{N}\geqslant \frac{1}{\xi_{\textnormal{s}}^2}=6.605.
\ee

For our computations we need a tool to handle large systems.
We will carry out the calculations for symmetric states.
this way we obtain a lower bound on the QFI that we will denote by $\bound{\textnormal{sym}}$.
As already mentioned, we could obtain a bound for the QFI that is valid even for general case, not necessarily symmetric states if the matrix from which compute the maximum eigenvalue Eq.~\eqref{eq:lt-legendre-for-qfi-simplified} has a non-degenerated largest eigenvalue.
This is not the case in general for the spin-squeezing problem.
However, we still know that our bound obtained with our calculations in the symmetric subspace cannot be smaller than the optimal bound $\bound{\mathcal{F}}$, which must be larger or equal to the Eq.~\eqref{eq:bg-pezze-bound} since it cannot be smaller than the optimal bound for general states.
These relations can be summarized as
\be
  \bound{\textnormal{sym}}\geqslant \bound{\mathcal{F}}\geqslant\frac{\expect{J_y}^2}{\varian{J_x}},
\ee
where on the right-hand side we just used the bound in Eq.~\eqref{eq:bg-pezze-bound}.

Our calculations lead to
\be
  \label{eq:lt-symmetric-optimal-pezze-inequality}
  \frac{\bound{\textnormal{sym}}(\expect{J_y},\varian{J_x})}{N} = 6.605
\ee
for a wide range of values of $\alpha$.
That is, based on numerics, the left-hand side and the right-hand side of Eq.~\eqref{eq:lt-symmetric-optimal-pezze-inequality} seem to be equal.
This implies that the lower bound Eq.~\eqref{eq:bg-pezze-bound} is optimal for estimating the QFI for the system.

We follow giving the details of our calculations for $\alpha=0.5,0.85$ and we show examples in which we can improve the bound Eq.~\eqref{eq:bg-pezze-bound} with our approach, if symmetry is assumed.
We present a simple scheme that can be used to handle large systems, and make calculations for larger particle number.
Thus, we need fewer steps for the numerical optimization for large system sizes, which makes our computations faster.
Second, while we will be able to carry out the calculation for the particle number of the experiment, we will also see that we could even extrapolate the results from the results obtained for lower particle numbers.
This is useful for future application of our method for very large systems.

The basic idea is that we transform the collective quantities from $N$ to a smaller particle number using the scaling relation
\begin{align}
  \expect{J_y} & = \frac{N'}{2}\alpha,\\
  \varian{J_x} & = \xi_{\textnormal{s}}^2 \frac{N'}{4}\alpha^2.
\end{align}
We see that for the scaling we consider, for all $N'$ the bound in Eq.~\eqref{eq:bg-pezze-bound} is valid, i.e., is obtained as
\be
  \frac{\qfif{\rho_{N'},J_z}}{N'}\geqslant \frac{1}{\xi_{\textnormal{s}}^2}=6.605.
  \label{eq:lt-pezze-exp}
\ee

Let us first take $\alpha=0.85$, which is somewhat smaller than the experimental value, however, it helps us to see various characteristics of the method.
At the end of the section we will also discuss the results for other values of $\alpha$.
Based on these ideas, we compute the bound $\bound{\textnormal{sym}}$ for the quantum Fisher information for an increasing system size $N'$.

\begin{figure}[htp]
  \centering
  \igwithlabel{(a)}{scale=.65}{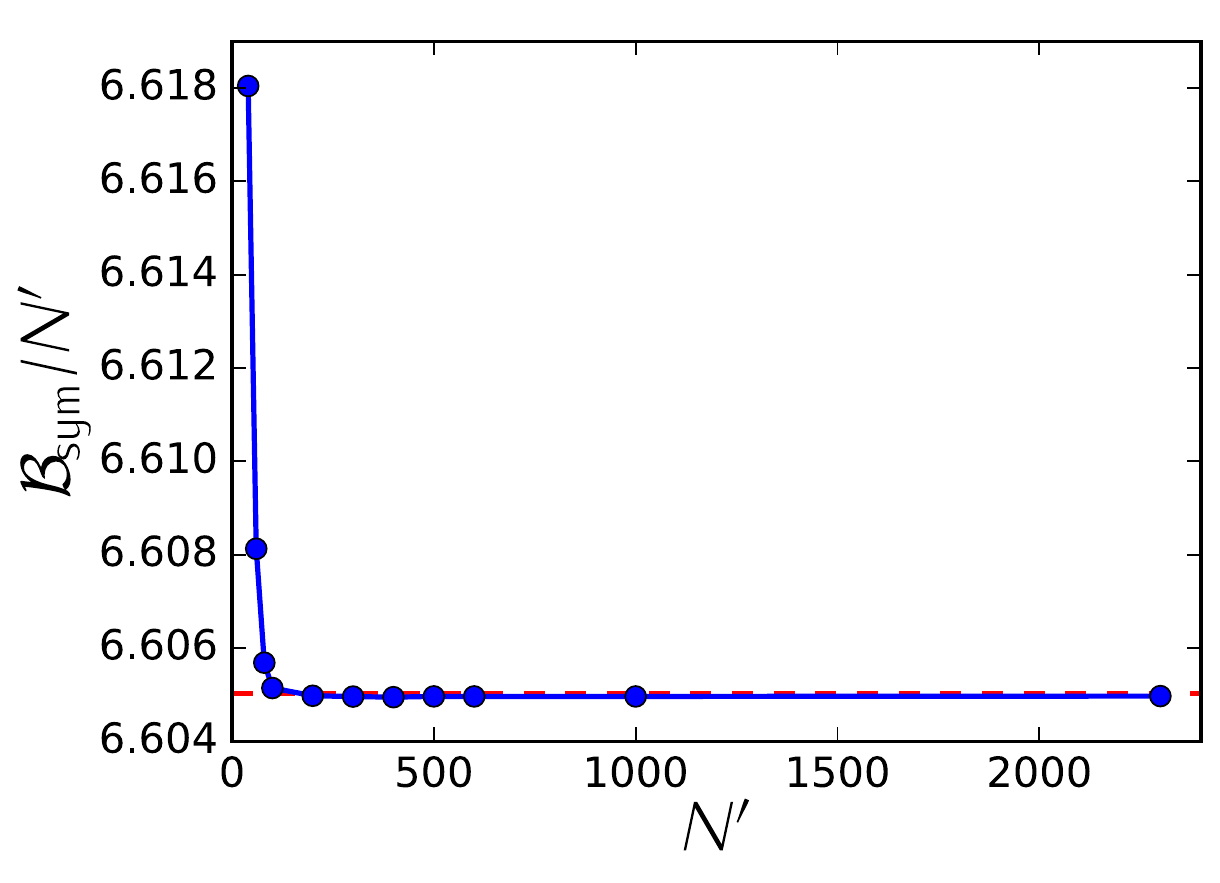}
  \igwithlabel{(b)}{scale=.65}{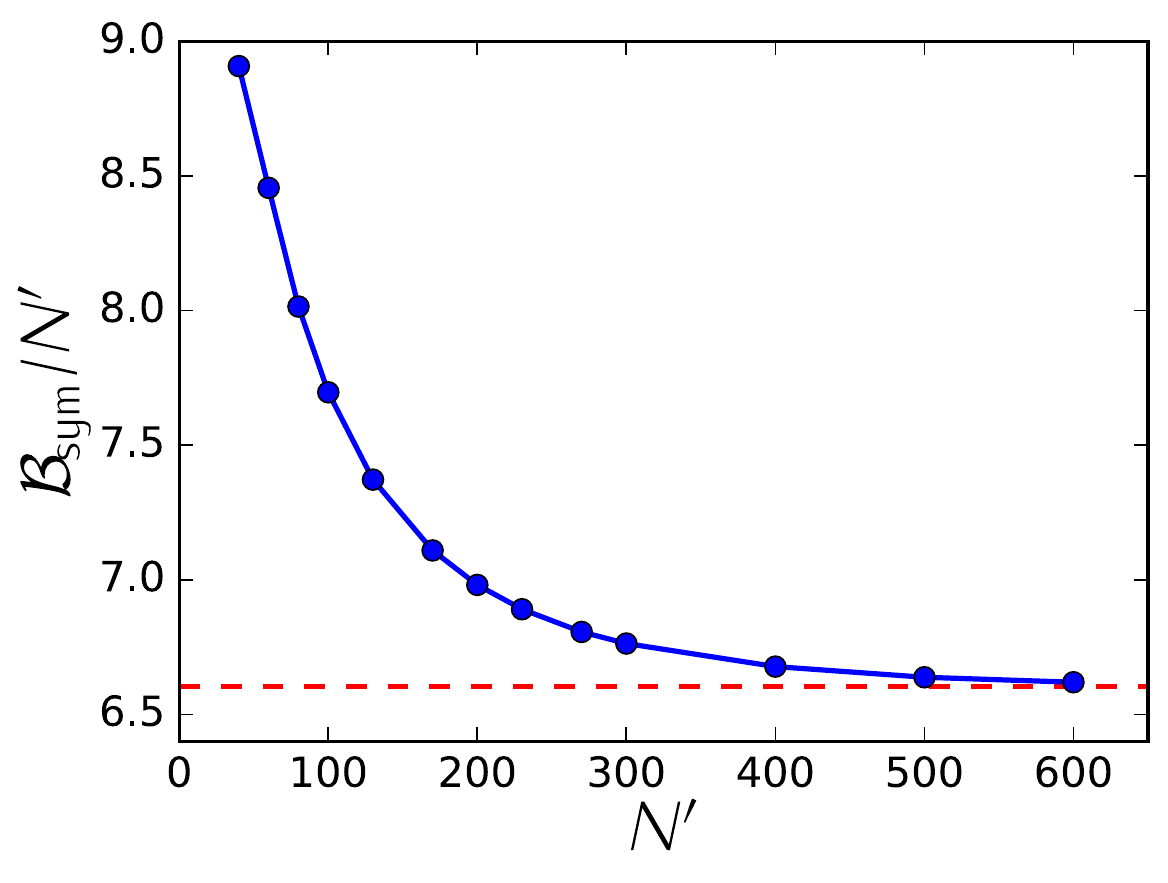}
  \caption[Asymptotic behavior of the bound for spin-squeezing experimental data]{
  (Color line) Lower bound on the QFI based on $\expect{J_y}$ and $\varian{J_x}$ obtained for the symmetric subspace for different particle numbers $N'$.
  $N{=}2300$ corresponds to the spin-squeezing experiment \cite{Gross2010}.
  (a) Almost fully polarized spin-squeezed state.
  Even for a moderate $N'$, the bound is practically identical to the right-hand side of the Eq.~\eqref{eq:lt-pezze-exp}.
  (b) Spin-squeezed state that is not fully polarized.
  For large $N'$, the bound converges to the right-hand side of the Eq.~\eqref{eq:lt-pezze-exp}, represented by a dashed line.
  (dots) Results of our calculations, which are connected by straight lines.}
  \label{fig:lt-bounds-on-symmetric-spin-squeezing}
\end{figure}

The results can be seen in Figure~\ref{fig:lt-bounds-on-symmetric-spin-squeezing}-(a).
The bound obtained this way is close to the bound in Eq.~\eqref{eq:lt-bound-for-experiment} even for small $N'$.
For larger particle number it is constant and coincides with the bound in Eq.~\eqref{eq:lt-bound-for-experiment}
This also strongly supports the idea that we could use the result for small particle numbers to extrapolate the bound for $N$.
Since for the experimental particle number we obtain that $\bound{\text{sym}}$ equals the bound in Eq.~\eqref{eq:lt-bound-for-experiment}, we find that all three lower bounds in Eq.~\eqref{eq:lt-symmetric-optimal-pezze-inequality} must be equal.
Hence, Eq.~\eqref{eq:bg-pezze-bound} is optimal for the experimental system and $\alpha$ considered before in this section.
Besides, these results present also a strong argument for the correctness of our approach.

We now give more details of the calculation. We were able to carry out the optimizations up to $N'=2300$ with a usual laptop computer using MATLAB programming language\footnote{
For MATLAB R2015a, see \url{http://www.mathworks.com}.}.
We started the calculation for each given particle number with the $r_k$ parameters obtained for the previous simulation with a smaller particle number.
This allows for faster finding of the solution than if we would start the $r_k$ parameters arbitrarily.

Let us consider a spin-squeezed state that is not fully polarized and $\alpha=0.5$.
In Figure~\ref{fig:lt-bounds-on-symmetric-spin-squeezing}-(b), we can see that for small particle numbers we have a larger bound on $\qfif{\rho, J_z}$ than the one obtained from Eq~\eqref{eq:bg-pezze-bound}.
Thus for the case in which the particle number would be smaller we could improve the bound Eq.~\eqref{eq:bg-pezze-bound} by assuming symmetry.
On the other hand, for large particle number we recover the bound Eq.~\eqref{eq:bg-pezze-bound}.

Finally, we add a note on the technical details.
We carried out our calculations with the constraints on $\varian{J_x}$ and $\expect{J_y}$, with the additional constraint $\expect{J_x}=0$.
For the experimental particle numbers, one can show that our results are valid even if we constrained only $\varian{J_x}$ and $\expect{J_y}$, and did not use the $\expect{J_x}=0$ constraint.
This way, in principle, we could only get a lower bound that is equal to the one we obtained before or lower.
However, we obtained before a value identical to the analytical bound Eq.~\eqref{eq:bg-pezze-bound}.
The optimal bound cannot be below the analytic bound, since then the analytic bound would overestimate the quantum Fisher information, and it would not be a valid bound.
Hence, even an optimization without the $\expect{J_x}=0$ constraint could not obtain a smaller value than our results.

\subsubsubsection{Experiment creating Dicke states}

In this section, we present our calculations for an experiment aiming at creating Dicke states in cold gases \cite{Luecke2014}.
The basic ideas are similar to the ones explained in Section~\ref{sec:lt-bound-dicke-states} for small systems.
The experimental data, as in previous Section~\ref{sec:vd-testing-with-experimental-data}, are $N=7900$, $\expect{J_y^2}=112\pm31$, $\expect{J_x^2}=\expect{J_z^2}=6\times10^6\pm0.6\times10^6$ \cite{Apellaniz2015}.
Applying some simple transformations, we can make calculations for a very large numbers of particles, and obtain results even for general, non-symmetric systems.

In the general, non-symmetric case, we can handle only small problems.
Thus, we have to transform the collective quantities such that the corresponding quantum state, i.e., it has to fulfill
\be
  \expect{J_x^2}_{\text{sym}} + \expect{J_x^2}_{\text{sym}} + \expect{J_x^2}_{\text{sym}} = \mathcal{J}_{N/2},
\ee
where $\mathcal{J}_{N/2}$ is defined on Eq.~\eqref{eq:lt-maximum-angular-momentum}.
A mapping of this type can be realized equally scaling all second moments of the angular momentum projections as
\be
  \label{eq:lt-expectation-values-extended-to-symmetric}
  \expect{J_l^2}_{\text{sym}, N} = \gamma \expect{J_l^2}_N,
\ee
where we now added the label $N$ to avoid confusions in upcoming equations, $l=x,y,z$ and where we used the coefficient $\gamma$ to be
\be
  \label{eq:lt-value-of-gamma}
  \gamma = \frac{\mathcal{J}_{N/2}}{\expect{J_x^2}_N +\expect{J_y^2}_N +\expect{J_z^2}_N}.
\ee
Note that $\gamma=1$ if the original state is symmetric.

Next, based on the ideas of this chapter, we calculate the lower bound on the quantum Fisher information for symmetric systems, which we denote $\bound{\text{sym},N}$.
To obtain the results for the original non-symmetric case, the convex nature of the $\bound{N}$ implies that
\be
  \bound{N}\leqslant \frac{1}{\gamma}\bound{\text{sym},N},
\ee
where $\bound{\text{sym},N}$ corresponds to the bound one would obtain in the symmetric subspace for expectation values given by the Eq.~\eqref{eq:lt-expectation-values-extended-to-symmetric}.
This can also be seen using an auxiliary state $\tilde{\rho}$ that mixes the symmetric state that has the expectation values computed with Eq.~\eqref{eq:lt-expectation-values-extended-to-symmetric} and the singlet state that has zero value for all these expectation values.
Hence, if we construct a mixture of this type as follows
\be
  \tilde{\rho}_N=(1-\gamma^{-1}) \rho_{\text{singlet},N}+\gamma^{-1}\rho_{\text{sym},N},
\ee
we have that $\tilde{\rho}_N$ has the same expectation values as the original non-symmetric case.
This way, we can relate the bound for general systems to the quantum Fisher information for symmetric cases as
\be
  \label{eq:lt-radial-linearity-dicke-bound}
  \bound{N}\leqslant \qfif{\tilde{\rho}_N,J_z}=\frac{1}{\gamma}\qfif{\rho_{\text{sym,N}},J_z}.
\ee
Here, the inequality comes due to that our bound cannot be larger than the QFI of any state having the given set of expectation values.
On the other hand, the equality holds due to the fact that both $\tilde{\rho}$ and $J_z$ can be written as block-diagonal matrix of blocks corresponding to different eigenvalues of $\bs{J}^2$.
In particular, $\rho_{\text{singlet},N}$ has non-zero elements only in the blocks for which $\expect{\bs{J}^2}=0$, while $\rho_{\text{sym},N}$ has nonzero elements only in the blocks in which $\expect{\bs{J}^2}$ is maximal.
Note that $\bs{J}^2$ is a shorthand of $J_x^2+J_y^2+J_z^2$.
Then we can use the general formula \cite{Toth2014}
\be
  \qfif{\bigoplus_k p_k\rho_k, \bigoplus_k A_k}= \sum_k p_k \qfif{\rho_k,A_k},
\ee
where $\rho_k$ are density matrices with unit trace, $\sum_k p_k=1$ and the $k$ index represent the block subspaces of the system and the operators $A_k$.

Extensive numerics for small systems show that the inequality in Eq.~\eqref{eq:lt-radial-linearity-dicke-bound} is very close to an equality within the numerical precision
\be
  \label{eq:lt-bound-extrapolation-from-symmetric-dicke}
  \bound{N}\approx\frac{1}{\gamma}\bound{\text{sym},N}.
\ee
To obtain the lower bound $\bound{N}$ we also use an increasing system size $N'$ as we have done in at the beginning of this section.
However, in this case we will not be able to do the calculation for the experimental particle number, and we will use extrapolation from the results obtained for smaller particle numbers.

First, we transform the measured second moments to values corresponding to a symmetric system using Eqs.~\eqref{eq:lt-expectation-values-extended-to-symmetric} and \eqref{eq:lt-value-of-gamma}.
For our case, $\gamma=1.301$.
This way, we obtain
\be
  \begin{split}
    \expect{J_y^2}_{\text{sym},N}&=145.69,\\
    \expect{J_x^2}_{\text{sym},N}&=\expect{J_z^2}_{\text{sym},N}=7.8\times10^6.
  \end{split}
\ee

Next, we will carry out calculations for symmetric systems.
We will consider a smaller system $N'$ that keeps expectation values such that the corresponding quantum state must be symmetric.
Hence, we will use the following relation to find the target expectation values for smaller systems
\be
  \begin{split}
    \expect{J_y^2}_{\text{sym},N'}&=\expect{J_y^2}_{\text{sym},N},\\
    \expect{J_x^2}_{\text{sym},N'}&=\expect{J_z^2}_{\text{sym},N'} =\frac{1}{2}(\mathcal{J}_{N'/2})-\expect{J_y^2}_{\text{sym},N'}),
  \end{split}
\ee
where $\mathcal{J}_{N'/2}$ is defined in Eq.~\eqref{eq:lt-maximum-angular-momentum}.
Note that with Eq.~\eqref{eq:lt-maximum-angular-momentum} holds for all $N'$, hence the state must be symmetric.
Hence, the main characteristics of the scaling relation can be summarized as follows, $\expect{J_y^2}_{\text{sym},N'}$ remains constant for all $N'$ while $\expect{J_x^2}_{\text{sym},N'}$ and $\expect{J_z^2}_{\text{sym},N'}$ are chosen such that they are equal to each other and the state is symmetric.
For large N, this implies $\expect{J_x^2}_{\text{sym},N}=\expect{J_z^2}_{\text{sym},N}\sim N(N+2)/8$.

Let us now turn to central quantities of our chapter, the lower bounds on the quantum Fisher information.
A central point in our scheme is that due to the scaling properties of the system, we can obtain the value for the particle number $N$ from the values of a smaller particle number $N'$ as \cite{Zhang2014}
\be
  \label{eq:lt-asymptotic-limit-bound-dicke-symmetric}
  \bound{\text{sym},N}\approx\frac{\mathcal{J}_{N/2}}{\mathcal{J}_{N'/2}} \bound{\text{sym},N'},
\ee
which we will verify numerically.
Note that for large $N$, we have $\mathcal{J}_{N/2}/\mathcal{J}_{N'/2}\sim N^2/(N')^2$.

\begin{figure}[htp]
  \centering
  \includegraphics[scale=.65]{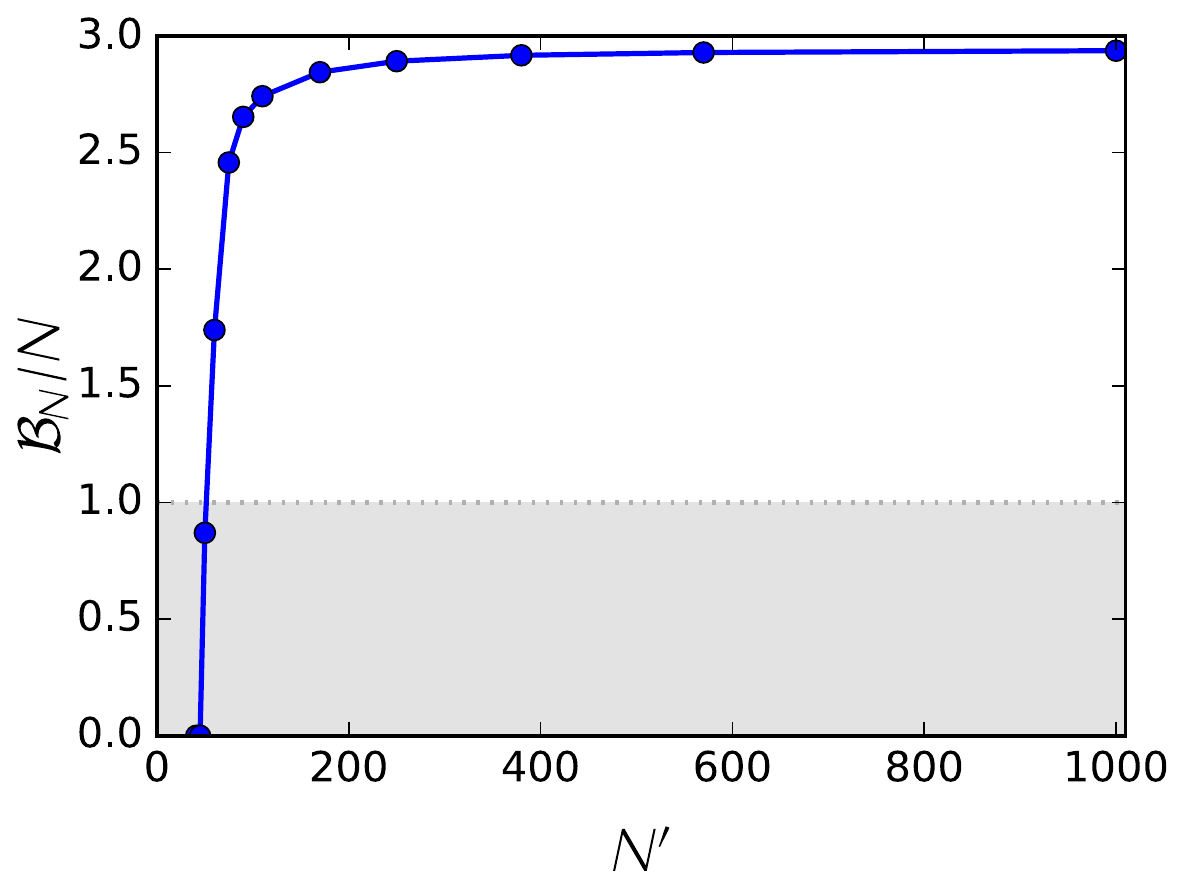}
  \caption[Asymptotic behavior of the bound for experimental data producing Dicke states]{
  Asymptotic behavior of the bound as a function of $N'$.
  The bound is first obtained for a symmetric subspace of $N'$ particles and then the bound for $N$ particles is computed using the Eq.~\eqref{eq:lt-definitive-formula-scaling-dicke}.
  The function is monotonically increasing.
  Hence, with $N'\approx200$ we already obtain a good lower bound.
  This approach does not overestimate the precision bound.
  }
  \label{fig:assimpthotic-approach-to-the-bound-from-scaled-dicke}
\end{figure}

As last step, we have to return from the symmetric system to our real system, not fully symmetric one.
Based on Eq.~\eqref{eq:lt-asymptotic-limit-bound-dicke-symmetric} and assuming Eq.~\eqref{eq:lt-bound-extrapolation-from-symmetric-dicke}, a relation for the lower bound for the original problem can be obtained from the bound on the symmetric problem with $N'$ particles as
\be
  \label{eq:lt-definitive-formula-scaling-dicke}
  \bound{N}\approx \frac{1}{\gamma}\frac{\mathcal{J}_{N/2}}{\mathcal{J}_{N'/2}} \bound{\text{sym},N'} =\frac{\expect{J_x^2}_N+\expect{J_y^2}_N+\expect{J_z^2}_N}{\mathcal{J}_{N'/2}}\bound{\text{sym},N'}.
\ee
In Figure~\ref{fig:assimpthotic-approach-to-the-bound-from-scaled-dicke},
we plotted the right-hand side of Eq.~\eqref{eq:lt-definitive-formula-scaling-dicke} as the function of $N'$ divided by $N$.
We can see that $\bound{N'}/N$ is constant or slightly increasing for $N'>400$.
This is a strong evidence that Eq.~\eqref{eq:lt-asymptotic-limit-bound-dicke-symmetric} is valid for relatively large particle numbers.
With this, we arrive at the result for the experimental system
\be
  \label{eq:lt-result-experimental-dicke}
  \frac{\bound{N}(\expect{J_y^2},\expect{J_x^2}=\expect{J_z^2})}{N}\approx 2.94.
\ee
The $\approx$ sign is used referring to the fact that we assume that the inequality in Eq.~\eqref{eq:lt-radial-linearity-dicke-bound} is close to be saturated and that we did sufficient numerics for an increasing system size $N'$ to have a good asymptotic approach to the real value Eq.~\eqref{eq:lt-asymptotic-limit-bound-dicke-symmetric}.

It is instructive to compare the value of Eq.~\eqref{eq:lt-result-experimental-dicke} to the one obtained in Section~\ref{sec:vd-testing-with-experimental-data}, where the same system was characterized base on its metrological usefulness.
Such result implies $\qfif{\rho,J_z}/N\geqslant3.3$ which is somewhat larger than our recent result as we did not use the knowledge of the fourth moments, only the second moments.
The closeness of the two results is a strong argument for the correctness of our calculations.

\subsection{Scaling of $\qfif{\rho,J_z}$ with $N$}

Recent important works examine the scaling of the quantum Fisher information with the particle number for metrology under the presence of decoherence \cite{Escher2011, Demkowicz-Dobrzanski2012}.
They consider the QFI defined now for the non-unitary, noisy evolution.
They find that for small $N$ it is close to the value obtained by considering coherent dynamics.
Hence, even the Heisenberg scaling, $\mathcal{O}(N^2)$, can be reached.
However, if $N$ is sufficiently large, then, due to the decoherence during the parameter estimation, the QFI scales as $\mathcal{O}(N)$.

We have to stress that the findings of B. M. Escher {\it et al} \cite{Escher2011} and R. Demkowicz-Dobrza{\'{n}}ski {\it et al} \cite{Demkowicz-Dobrzanski2012} are not applicable to our case.
Our methods estimate the quantum Fisher information assuming a perfect unitary dynamics.
The quantum Fisher information can be smaller that what we expect ideally only due to the imperfect preparation of the state\footnote{
This is also relevant for Ref.~\cite{Augusiak2016}, where $\qfi=\mathcal{O}(N^2)$ is reached with weakly entangled states.}.
We can even find simple conditions on the state preparation that lead to a Heisenberg scaling.
Based on Eq.~\eqref{eq:lt-ghz-legendre-solution}, if we could realize quantum states $\rho_N$ such that $F_{\text{GHZ}}(\rho_N)\geqslant0.5+\epsilon$ for $N\rightarrow\infty$ for some $\epsilon>0$, then we would reach $\bound{\mathcal{F}}(F_{\text{GHZ}}) = \mathcal{O}(N^2)$.
Strong numerical evidence suggest that a similar relation holds for fidelity $F_{\text{Dicke}}$ and $\bound{\mathcal{F}}(F_{\text{GHZ}})$, see Section~\ref{sec:lt-bound-dicke-states}.
From another point of view, our method can estimate $\qfif{\rho,J_z}$ for large particle numbers, while a direct measurement of the metrological sensitivity considerably underestimates it.

%% file: 05-gradient.tex
\section[Metrology of the gradient magnetic field]
{Precision bound for gradient field estimation with atomic ensembles}
\input{snp/doubleLineWaterMark.tex}
\label{sec:gm}

\quotes{Ronald Fisher}{To consult the statistician after an experiment is finished is often merely to ask him to conduct a post mortem examination. He can perhaps say what the experiment died of.}

\lettrine[lines=2, findent=3pt, nindent=0pt]{I}{n} this chapter, one of the most fundamental two-parameter estimation tasks in magnetometry is considered, namely gradient magnetometry.
We will add the gradient of the magnetic field as the second parameter beside the constant (homogeneous) part of the field.
While most works in magnetometry with a single ensemble focus only on the determination of the strength and direction of the magnetic field, certain measurement schemes for the gradient have already been proposed and tested experimentally.
We study gradient magnetometry with an ensemble of atoms described by a very general probability distribution function for the position of the atoms, and considering atoms with an arbitrary spin.
Some schemes use an imaging of the ensemble with a high spatial resolution, however, they do not count as single-ensemble methods in the sense we use this expression in our paper, since in this case not only collective observables are measured  \cite{Vengalattore2007,Zhou2010,Koschorreck2011}.
There is a method based on collective measurements of the spin length of a fully polarized ensemble \cite{Behbood2013}.
Finally, there is a scheme where they use as a trial state a many-body singlet states, which is described in Ref.~\cite{Urizar-Lanz2013}.

We calculate precision bounds for estimating the gradient of the magnetic field based on the quantum Fisher information.
For quantum states that are invariant under the homogeneous magnetic field, a single parameter estimation is sufficient.
In contrast to this, for states that are sensitive to the homogeneous fields, a two-parameter estimation problem must be solved to obtain the gradient parameter, since the homogeneous field must also be taken into account.
We use our method to calculate precision bounds for gradient estimation with a chain of atoms and even with two spatially separated atomic ensembles which feel different magnetic fields.
As we said, we also consider a single atomic ensemble with an arbitrary density profile, in which atoms cannot be addressed individually, which is a very relevant case for experiments.
Our model can take into account even correlations between particle positions.

The atoms will be distributed along the $x$-axis, so $y=z=0$, and in principle they will be able to feel differences in the magnetic field at different points of the axis.
The magnetic field at the atoms will be given by a linear function of the position $x$
\begin{equation}
\bs{B}(x,0,0)=\bs{B}_0 +x \bs{B}_1 + O(x^2),
\end{equation}
where we will neglect the terms of order two or higher.
We will consider the magnetic field pointing along the $z$-direction direction only, $\bs{B}_0=B_0 \bs{k}$ and $\bs{B}_1=B_1\bs{k}$, where $\bs{k}$ is the unitary vector pointing on the $z$-direction.
For this configuration, due to the Maxwell equations, with no currents or changing electric fields, we have
\begin{align}
\nabla \cdotp \bs{B}&=0, \nonumber \\
\nabla \times \bs{B}&=\bs{0},
\end{align}
where $\bs{0}\equiv (0,0,0)$ is the 3-dimensional null vector.
This implies $\sum_{l=x,y,z} \partial_l B_l=0$ and $ \partial_m B_l - \partial_l B_m =0$ for all $l\ne m$, where $\partial_m\equiv \partial/\partial_m$ stands for the partial derivative over the variable $m$.
Thus, the spatial derivatives of the field components are not independent of each other.
However, in the case of a linear arranged particle ensemble only the derivative along the $x$-axis has an influence on the quantum dynamics of the atoms.

We will determine the precision bounds for the estimation of the magnetic field gradient $B_1$ based on the quantum Fisher information \cite{Paris2009,Braunstein1994,Holevo1982,Helstrom1976,Petz2002,Petz2008}.
In this context the Heisenberg and shot-noise scaling are defined as usual.
The achievable precision in terms of the number of particles scales as $\varinv{\theta}\sim N$ and $\varinv{\theta}\sim N^2$ for shot-noise scaling and Heisenberg scaling, respectively.
We will show that with spin chains or two ensembles at different positions the Heisenberg scaling is possible.
Concerning the case of a single ensemble, we will show the following.
Since in such systems the atoms cannot be be individually addressed, we will assume that the quantum state is permutationally invariant.
We will show that for states insensitive to the homogeneous magnetic field, one can reduce the problem to a one-parameter estimation scenario.
Such states can arise in a single-ensemble scheme, however, it will be shown that the Heisenberg limit cannot be reached in this case.
When the state is sensitive to the homogeneous field, the spatial correlation between the atoms must be taken into account in order to show whether the system can overcome the shot-noise scaling and achieve the Heisenberg scaling.
Nevertheless, single-ensemble measurements have certain advantages since the spatial resolution can be higher and the experimental requirements are smaller since only a single ensemble must be prepared.

On the other hand, for states sensitive to the homogeneous field, the classical limit can be overcome only if the particle positions are highly correlated with each other.
Our calculations are generally valid for any measurement, thus they are relevant to many recent experiments \cite{Wasilewski2010,Eckert2006,Wildermuth2006, Wolfgramm2010,Koschorreck2011,Vengalattore2007,Zhou2010,Behbood2013}.
We note that in the case of the singlet, our precision bounds are saturated by the metrological scheme presented in Ref.~\cite{Urizar-Lanz2013}.

We can also connect our results to entanglement theory \cite{Werner1989,Horodecki2009,Guehne2009}.
We find that even in the case of gradient magnetometry the shot-noise scaling cannot be surpassed with separable states, while the Heisenberg scaling can be reached with entangled states.
However, in the single-ensemble scenario, the shot-noise scaling can be surpassed only if the particle positions are correlated, which is the case if the particles attract each other.
We will go into the details in Section~\ref{sec:gm-single-cloud}.

The chapter is organized as follows. In Section~\ref{sec:gm-the-setup}, we will present the setup of the system.
In Section~\ref{sec:gm-cramer-rao-bounds}, the metrological basic concepts used in the chapter are presented.
In Section~\ref{sec:gm-io-chain-and-two-ensembles}, we will show the results for the chain of ions and for when two distant ensembles are considered
In Section~\ref{sec:gm-single-cloud}, we restrict our calculations to single permutationally invariant atomic ensembles and we develop some particular cases, such as the singlet spin state or the totally polarized state.

\subsection{The setup}
\label{sec:gm-the-setup}

In this section, we will present the characteristics of our setup.
For simplicity, as well as following  recent experiments (e.g., Ref.~\cite{Koschorreck2011}), we will consider an ensemble of spin-$j$ particles placed in a one-dimensional trap or a chain.
Furthermore, we will assume that the particles are point-like and distinguishable.
On the other hand, we assume that the particles have a spin, which is a quantum degree of freedom.
Such a model is used typically to describe experiments with cold atomic ensembles.

Based on these considerations, we assume that the state is factorizable into a spatial and a spin part as
\be
\label{eq:gm-separated-internal-and-external}
\rho=\rho^{(\text{x})}\otimes\rho^{(\text{s})},
\ee
and that the spatial part can be characterized as an incoherent mixture of point-like particles that can be written as
\be
  \rho^{(\text{x})}=\int \prob(\bs{x}) \ketbra{\varphi_{\bs{x}}}{\varphi_{\bs{x}}}\,\text{d}^N\bs{x},
  \label{eq:gm-pre-thermal-state}
\ee
where $\ket{\varphi_{\bs{x}}}$ is a pure state of each particle been placed at $\bs{x}=(x_1,x_2,\dots,x_N)$, respectively.
Each part of the state acts on different Hilbert spaces denoted by $\mathcal{H}^{(\text{x})}$ and $\mathcal{H}^{(\text{s})}$, respectively.
Note that we skip to write the superscript $(\text{x})$, denoting the Hilbert space to which $\ket{\varphi_{\bs{x}}}$ belongs, for simplicity.

In order to write the operators, including the state $\rho^{(\text{x})}$, acting on the Hilbert space $\mathcal{H}^{(\text{x})}$, we will invoke the completeness relation found in the literature \citep{Sakurai2010, Cohen-Tannoudji1977} for the spatial continuous Hilbert space
\be
  \int \ketbra{\bs{x}}{\bs{x}}\,\text{d}^N\bs{x} = \mtxid,
\ee
where $\ket{\bs{x}}=\ket{x_1}^{(1,\text{x})}\otimes\ket{x_2}^{(2,\text{x})}\cdots\otimes\ket{x_N}^{(N,\text{x})}$ is the tensor product of the position eigenvectors of each particle which obey
\be
  \braket{\bs{x}}{\bs{y}} = \delta(\bs{x}-\bs{y}),
\ee
where $\delta(\bs{x}-\bs{y})$ is the Dirac delta found in the literature \citep{Sakurai2010, Cohen-Tannoudji1977}.

To see how our notation works, let us write the vector of the position operators for each particle as $\hat{\bs{x}}\equiv(x^{(1)},x^{(2)},\dots,x^{(N)})$, where we used the $\hat{\cdot}$ notation on top of $\bs{x}$ to distinguish it from the vector of position variables.
It is known that $\hat{\bs{x}}$ acting on $\ket{\bs{x}}$ will give $\bs{x}\ket{\bs{x}}$ \cite{Sakurai2010, Cohen-Tannoudji1977}.
Hence,
\be
  \hat{\bs{x}} = \hat{\bs{x}}\mtxid = \hat{\bs{x}} \int \ketbra{\bs{x}}{\bs{x}}\,\text{d}^N\bs{x} = \int \hat{\bs{x}} \ketbra{\bs{x}}{\bs{x}}\,\text{d}^N\bs{x} = \int \bs{x} \ketbra{\bs{x}}{\bs{x}}\,\text{d}^N\bs{x}.
\ee
We can follow similar arguments to rewrite the pure state $\ket{\varphi_{\bs{x}}}$.
First, note also that the expectation value of the position operator for such a state is always $\bs{x}$.
Hence, such a state must be proportional to $\ket{\bs{x}}$.
On the other hand, a pure state must be normalized to one, hence, we can construct such a state by dividing the eigenvector $\ket{\bs{x}}$ by the square-root of its norm as
\be
  \ket{\varphi_{\bs{x}}} \equiv \frac{\ket{\bs{x}}}{\sqrt{\braket{\bs{x}}{\bs{x}}}}.
  \label{eq:gm-pointlike-state}
\ee
The meaning of Eq.~\eqref{eq:gm-pointlike-state} is clear, while a rigorous form of how various limits are taken could overcomplicate our discussion. 
For illustrative purposes, we compute in this basis the expectation value of the position operator for these pure states as
\be
  \expect{\hat{\bs{x}}}_{\varphi_{\bs{x}}} = \braopket{\varphi_{\bs{x}}}{\int \bs{y} \ketbra{\bs{y}}{\bs{y}}\,\text{d}^N\bs{y}}{\varphi_{\bs{x}}} =
  \int \bs{y}
  \frac{\braket{\bs{x}}{\bs{y}}\braket{\bs{y}}{\bs{x}}}
  {\braket{\bs{x}}{\bs{x}}} \,\text{d}^N\bs{y} =
  \bs{x}\frac{\braket{\bs{x}}{\bs{x}}}{\braket{\bs{x}}{\bs{x}}} =
  \bs{x},
\ee
where in the step before the last we have used $\braket{\bs{y}}{\bs{x}}=\delta(\bs{y}-\bs{x})$ to compute the integral.
We can rewrite the spatial state Eq.~\eqref{eq:gm-pre-thermal-state} as
\be
  \rho^{(\text{x})}=\int \frac{\prob(\bs{x})}{\braket{\bs{x}}{\bs{x}}} \ketbra{\bs{x}}{\bs{x}}\,\text{d}^N\bs{x}.
  \label{eq:gm-thermal-state}
\ee
Note also that if the eigen-decomposition of the internal state is $\rho^{(\text{s})} = \sum_{\lambda} p_{\lambda}\ketbra{\lambda}{\lambda}$, then the whole state is decomposed as
\be
  \rho = \int\sum_{\lambda} \frac{\prob(\bs{x})}{\braket{\bs{x}}{\bs{x}}} p_\lambda \ketbra{\bs{x},\lambda}{\bs{x},\lambda}\,\text{d}^N\bs{x},
  \label{eq:gm-eigendecomposition-of-state}
\ee
where $\ket{\bs{x},\lambda}$ is a shorthand for $\ket{\bs{x}}^{(\text{x})}\otimes\ket{\lambda}^{(\text{s})}$, the eigenstates, where their corresponding eigenvalues are in this case $\frac{\prob(\bs{x})}{\braket{\bs{x}}{\bs{x}}} p_\lambda$.

At this point, we want to emphasize that our method could easily be extended to the case of Bose-Einstein condensates, or any other spatial configuration, not considered in this paper. In the case of BECs, the spatial state of the particles would be a pure state, and we would have $\rho^{(\text{x})}=(\ketbra{\Psi}{\Psi})^{\otimes N},$ where $\ket{\Psi}$ is a spatial single-particle state.

Although in our case the parameter to be estimated is $B_1$,
the time-evolution of the state is usually also affected by the second unknown parameter, the homogeneous field $B_0$, which means that we generally have to consider a two-parameter estimation problem.
The angular momentum of an individual atom is coupled to the magnetic field, yielding the following interaction term
\be
  h^{(n)}=\gamma B_z^{(n,\text{x})} \otimes j_z^{(n,\text{s})},
  \label{eq:gm-single-particle-hamiltonian}
\ee
where the operator $B_z^{(n)}=B_0+B_1 x^{(n)}$ acts on the spatial part of the $n^{\text{th}}$ particle Hilbert space $\mathcal{H}^{(n,\text{x})}$.

The sum  of all single-particle interactions with the magnetic field provide the total Hamiltonian
\be
\label{eq:gm-Htot}
H = \gamma \sum_{n=1}^N B_z^{(n,\text{x})} \otimes j_z^{(n,\text{s})},
\ee
which will generate the time evolution of the system.

We will calculate lower bounds on the precision of estimating $B_1$ based on measurements on the state after it passed through the unitary dynamics $U=\exp(-i\frac{H}{\hbar}t)$, where $t$ is the time spent by the system under the influence of the magnetic field.
The unitary operator can be rewritten in the following way
\be
\label{eq:gm-whole-unitary-b_i-encoded}
U=e^{-i \lpar b_0 H_0 + b_1 H_1 \rpar},
\ee
where the $b_i=\gamma B_i t/\hbar$ and therefore $b_1$ encodes the gradient of the magnetic field $B_1$.
Here, the generator describing the effect of the homogeneous field is  given as
\be
\label{eq:gm-homogeneous-h0}
H_0=\sum_{n=1}^N j_z^{(n)} = J_z,
\ee
while the generator describing the effect of the gradient is
\be
\label{eq:gm-gradient-h1}
H_1=\sum_{n=1}^N x^{(n)}j_z^{(n)}.
\ee
As in Eq.~(\ref{eq:gm-gradient-h1}), we will usually omit $\otimes$ and the superscripts $(\text{x})$ and $(\text{s})$ for simplicity, and will use it only if it is necessary to make our explanation clearer.

Note that the operators $H_{0}$ and $H_{1}$ commute with each other.
These two commuting dynamics are the two simplest in an atomic ensemble as they are based on collective operators not requiring an individual access to the particles.
This is mainly because the spatial part in Eq.~\eqref{eq:gm-single-particle-hamiltonian} is represented by a single-particle operator and not by a scalar depending on the position of the particle.
The last approach, where the position of the particle is encoded in a scalar, would require to know in advance the location of the particles to construct the Hamiltonian $H_1$ Eq.~\eqref{eq:gm-gradient-h1}, which would yield finally to a non-collective operator for $H_1$.
This approach is widely adopted by the community, since it simplifies the problem considerably \cite{Urizar-Lanz2013, Ng2014}.

Note also that it is not necessarily true that the operators we have to measure in order to estimate $b_0$ and $b_1$ must commute with each other.
On the other hand, in schemes in which the gradient is calculated based on measurements in two separate atomic ensembles or different atoms in a chain, the measuring operators can always commute with each other \cite{Wasilewski2010, Eckert2006, Zhang2014}.

\subsection{Cram\'er-Rao bound for gradient estimation}
\label{sec:gm-cramer-rao-bounds}

In this section, we show how the Cram\'er-Rao bound and the QFI help us to obtain the precision bound that is valid for any measurement scenario.
We will discuss gradient magnetometry using quantum states that are insensitive to homogeneous fields, which is a single-parameter estimation task.
Then, we discuss the case of quantum states sensitive to homogeneous fields, which is a two-parameter estimation problem.
We show that the precision bound obtained does not change under spatial translation, which is one of the main tools to derive our bounds.
For the two-parameter estimation task, we will introduce the two-parameter Cram\'er-Rao bound and the corresponding two-parameter QFI matrix, and we adapt those expressions to our problem.

For clarity, we present our main tools in subsequent paragraphs before going into details.
Here we define a functional very similar to the QFI Eq.~\eqref{eq:bg-qfi-definition-eigen-decomposition}.
This expression will be used along this chapter and it will be useful for the transition to the multi-parameter problem, i.e, it is equivalent to the QFI for the single parameter estimation problem but still it gives the chance to switch to the multi-parameter case easily.
The function is defined as follows.
For two arbitrary operators $A$ and $B$, it is written as
\be
  \label{eq:gm-FAB}
  \qfif{\rho,A,B}:=2\sum_{\lambda,\nu}
  \frac{(p_\lambda-p_\nu)^2}{p_\lambda+p_\nu}
  {A}_{\lambda,\nu}{B}_{\nu,\lambda},
\ee
where the subscript for $A$ and $B$ stand for the matrix elements on the eigenbasis of the initial state $\rho = \sum_\lambda p_\lambda\ketbra{\lambda}{\lambda}$.
If the two operators are the same, the usual form of the QFI Eq.~\eqref{eq:bg-qfi-definition-eigen-decomposition} is recovered \cite{Paris2009, Braunstein1994, Holevo1982, Helstrom1976, Petz2002, Petz2008},
\be
  \qfif{\rho,A,A}:=\qfif{\rho,A}=2\sum_{\lambda,\nu}
  \frac{(p_\lambda-p_\nu)^2}{p_\lambda+p_\nu} |{A}_{\lambda,\nu}|^2.
\ee
We mention that in our case the operators $A$ and $B$ will commute in all situations, making the computations easier.
We also make use of the fact that the QFI as written in Eq.~(\ref{eq:gm-FAB}) is linear in the second and the third arguments,
\be
  \label{eq:gm-qfi-linear-in-arguments}
  \qfif{\rho, A, \textstyle{\sum}_i b_i} = \sum_i \qfif{\rho,A,b_i}.
\ee
It also holds for commuting $A$ and $B$, that the last two arguments can be exchanged without affecting the outcome, $\qfif{\rho, A, B} = \qfif{\rho, B,A}$.

Similar to Eq.~\eqref{eq:bg-rewrite-qfi}, Eq.~(\ref{eq:gm-FAB}) can be rewritten as
\be
  \label{eq:gm-FAB-rewrite-with-trace}
  \qfif{\rho, A, B} = 4 \expect{AB}
  - 8\sum_{\lambda,\nu} \frac{(p_\lambda-p_\nu)^2}{p_\lambda+p_\nu}
  {A}_{\lambda,\nu}{B}_{\nu,\lambda},
\ee
when the operators $A$ and $B$ commute.
This form leads to simpler arguments in our derivations through the following sections.

For pure states it simplifies also to
\be
  \qfif{\ket{\psi},A,B}=4\left(\expect{AB}_{\psi}-\expect{A}_{\psi}\expect{B}_{\psi}\right).
  \label{eq:gm-FAB-pure}
\ee
Note that we recover $\qfif{\rho,A,A}=4(\Delta A)_\rho^2$ as can be found in the Eq.~\eqref{eq:bg-rewrite-qfi} for single-parameter estimation with pure states \cite{Paris2009,Toth2013}.

Another important feature of the function Eq.~\eqref{eq:gm-FAB} is that it is convex on the states.
This property is written as follows
\be
  \qfif{q\rho_1{+}(1{-}p)\rho_2}\leqslant
  p\qfif{\rho_1}{+}(1{-}p)\qfif{\rho_2},
  \label{eq:gm-FAB-convex}
\ee
where we omit in writing the last two arguments for simplicity.
Finally, it is also useful to note that additive under the tensor product as
\be
  \qfif{\rho^{(1)}\otimes \rho^{(2)}, A^{(1)}\otimes \mtxid^{(2)}+\mtxid^{(1)}\otimes A^{(2)},B^{(1)}\otimes \mtxid^{(2)}+\mtxid^{(1)}\otimes B^{(2)}} = \qfif{\rho^{(1)},A^{(1)},B^{(1)}}+ \qfif{\rho^{(2)},A^{(2)},B^{(2)}}.
  \label{eq:gm-FAB-additive-under-tensor}
\ee

In the following subsections we show the general form for the precision bounds for states insensitive to the homogeneous fields and for states sensitive to them. We also show that both bounds are invariant under spatial translation of the system which makes the computing for particular cases much easier.

\subsubsection{States insensitive to the homogeneous field:
Single-parameter estimation}

Let us consider quantum states that are  insensitive to the homogeneous field.
For these states, $[\rho, H_0]=0$ and hence the evolved state is a function of a single unknown parameter, $b_1$.
For the unitary dynamics we consider, the QFI for single-parameter estimation problem can be expressed in terms of the eigenstates and eigenvalues of the density matrix as \cite{Paris2009, Braunstein1994, Holevo1982, Helstrom1976, Petz2002, Petz2008}
\be
  \label{eq:gm-general one parameter quantum fisher information}
  \qfif{\rho,H_1}=2\sum_{\lambda,\nu} \frac{(p_\lambda-p_\nu)^2}{p_\lambda+p_\nu} |\braopket{\lambda}{H_1}{\nu}|^2.
\ee
Note that here the eigenstates $\ket{\lambda}$ and $\ket{\nu}$ live on both the external and internal Hilbert spaces.
Due to the Cram\'er-Rao formula, the QFI gives us an upper
bound for the precision
\be
  \label{eq:gm-one parameter precision bound}
  (\Delta b_1)^{-2}|_{\max} = \qfif{\rho,H_1}.
\ee
Note that it is \emph{always} possible to find a measurement that saturates the precision bound above.
Hence, we denote it using the "$|_{\max} = $" notation.
Here, $\qfif{\rho,H_1}$ denotes the QFI that depends, in the case of unitary transformation of the form Eq.~\eqref{eq:gm-whole-unitary-b_i-encoded}, on the state $\rho$ and on the generator of the evolution $H_1$.

For the particular case in which the state has the form of Eqs.~\eqref{eq:gm-separated-internal-and-external} and \eqref{eq:gm-thermal-state}, Eq.~\eqref{eq:gm-general one parameter quantum fisher information} can be simplified in the following way.
Note that we have to compute the matrix elements of $H_1$ which is already diagonal in the spatial subspace.
Therefore, the following holds for the matrix elements of $H_1$
\be
  \begin{split}
    (H_1)_{\bs{x},\lambda;\bs{y},\nu}
    &=\bra{\bs{x},\lambda}{H_1}\ket{\bs{y},\nu}\\
    &=\bra{\bs{x},\lambda}{\sum_{n=1}^N x^{(n)}j^{(n)}}\ket{\bs{y},\nu}\\
    &=\delta(\bs{x}-\bs{y})\sum_{n=1}^N x_n\bra{\lambda}{j^{(n)}}\ket{\mu},
  \end{split}
  \label{eq:gm-simplify-h1-matrix-els}
\ee
where $\ket{\lambda}$ and $\ket{\mu}$ refer now to eigenstates of the internal state $\rho^{(\text{s})}$ and we use $\bra{\bs{x}}x^{(n)}\ket{\bs{y}}=\delta(\bs{x}-\bs{y})x_n$.
We will use the Dirac delta function appearing in Eq.~\eqref{eq:gm-simplify-h1-matrix-els} to further simplify the Eq.~\eqref{eq:gm-general one parameter quantum fisher information}.

We show now that spatial translations does not change the sensitivity of gradient estimation.
The translation operator $U_d$ moves the state to a new position at a distance $d$ from its previous location, and it is written as
\be
  U_d = e^{-idP_x/\hbar},
\ee
where $P_x$ is the sum of all single-particle linear momentum operators $p_x^{(n)}$ in the $x$-direction and it only acts on the external degrees of freedom of the state, i.e., the external Hilbert space $\mathcal{H}^{(\text{x})}$.
To show that the precision is unchanged, we use the Heisenberg picture in which the operators are transformed instead of the states.
Thus, we compute the transformation of $H_1$ as
\be
\begin{split}
\label{eq:gm-shifted h1 generator}
U_d:H_1 \rightarrow H_1(d)
&=  U_d^{\dagger}H_1U^{\phantom\dagger}_d\\
&=\sum_{n=1}^N U_d^{\dagger}x^{(n)}U^{\phantom\dagger}_d \otimes j_z^{(n)}\\
&=\sum_{n=1}^N (x^{(n)}-d)j_z^{(n)}\\
&=H_1-dH_0.
\end{split}
\ee
Hence, the new unitary evolution operator to represent the translated system, instead of Eq.~\eqref{eq:gm-whole-unitary-b_i-encoded}, is
\be
  U=e^{-i(b_0H_0+b_1 H_1(d))}=e^{-i((b_0-b_1d)H_0+b_1H_1)}.
  \label{eq:gm-shifted-unitary}
\ee
Eq.~\eqref{eq:gm-shifted-unitary} is equivalent to Eq.~\eqref{eq:gm-whole-unitary-b_i-encoded} for states insensitive to the homogeneous fields, since in this case $[\rho, H_0]=0$.

To compute the QFI, we used the Dirac delta function appearing in Eq.~\eqref{eq:gm-simplify-h1-matrix-els}, and the state defined by Eq.~\eqref{eq:gm-eigendecomposition-of-state}.
See Appendix~\ref{app:matrix-elements-of-QFI} for details.

The following bound in the precision of the estimation of the gradient parameter $b_1$ holds for states insensitive to the homogeneous magnetic fields
\be
  \label{eq:gm-bound-for-insensitive-and-thermal-state}
  (\Delta b_1)^{-2}|_{\max} = \sum_{n,m}^N \int \prob(\bs{x})x_n x_m \,\text{d}^N\bs{x}\,
  \qfif{\rho^{(\text{s})}, j_z^{(n)}, j_z^{(m)}},
\ee
where the integral has the form of a two-point correlation function of the spatial state.

\subsubsection{States sensitive to the homogeneous field:
Two-parameter dependence}

In order to obtain the precision bound for states sensitive to the
homogeneous field, one has to consider the effect on the state of a second
unknown parameter, in this case $b_0$, which represents the homogeneous magnetic field.
The homogeneous field will rotate all the spins in the same way,
while the field gradient rotates the spins
differently depending on the position of the particles.
Now, instead to the Cram\'er-Rao bound Eq.~\eqref{eq:gm-one parameter precision bound},
we have a matrix inequality \cite{Paris2009}
\be
  \text{Cov}(b_0,b_1) \geqslant \mathbfcal{F}^{-1},
  \label{eq:gm-matrix-inequality}
\ee
where $\text{Cov}(b_0,b_1)$ is the covariance matrix for $b_0$ and $b_1$.

For the matrix inequality Eq.~\eqref{eq:gm-matrix-inequality}, we have the inverse of QFI matrix $\mathbfcal{F}$ on one hand, which depends on $\rho$ and the two generators $H_0$ and $H_1$, and the covariance matrix on the other hand.
In this section, we are only interested in the variance of the gradient parameter, $\varian{b_1}$.
Since we have to compute the inverse of the QFI matrix and then look at the element corresponding to the $\varian{b_1}$, the determinant of $\mathbfcal{F}$ cannot be zero.
$H_0$ and $H_1$ are Hermitian operators and commute with each other.
For unitary dynamics of the type Eq.~\eqref{eq:gm-whole-unitary-b_i-encoded}, the QFI matrix elements are computed as $\mathbfcal{F}_{ij}\equiv \qfif{\rho, H_i, H_j},$ following the definition given in Eq.~\eqref{eq:gm-FAB}.

In the two-parameter estimation problem, $\mathbfcal{F}$ is a $2 \times 2$ matrix and the precision bound for the estimation of the gradient is
\be
\label{eq:gm-bound-for-b1-with-qfi-els}
  \varinv{b_1}\leqslant \mathbfcal{F}_{11}-\frac{\mathbfcal{F}_{01}\mathbfcal{F}_{10}}{\mathbfcal{F}_{00}},
\ee
where the inequality is saturated only if there exists a set of compatible measurements to determine both parameters $b_0$ and $b_1$, which is not true in general and must be studied for each particular case \cite{Paris2009, Ragy2016}.
We distinguish this case from the Eq.~\eqref{eq:gm-one parameter precision bound}, in which the bound is surely saturated by some measurement, using an inequality "$\leqslant$" instead of "$|_{\max}=$".

To compute the bound Eq.~\eqref{eq:gm-bound-for-b1-with-qfi-els}, we will need to simplify the matrix elements of $H_0$ and $H_1$ written in the eigenbasis of the state Eq.~\eqref{eq:gm-eigendecomposition-of-state}, see Eq.~\eqref{eq:gm-FAB}.
Note that the matrix elements for $H_1$ were already computed in Eq.~\eqref{eq:gm-simplify-h1-matrix-els}.
Hence, we now calculate $(H_0)_{\bs{x},\lambda;\bs{y},\nu}$ in a similar way as we did for Eq.~\eqref{eq:gm-simplify-h1-matrix-els}
\be
\begin{split}
  (H_0)_{\bs{x},\lambda,\bs{y},\nu}
  &=\bra{\bs{x},\lambda}{H_0}\ket{\bs{y},\nu}\\
  &=\bra{\bs{x},\lambda}{J_z}\ket{\bs{y},\nu}\\
  &=\delta(\bs{x}-\bs{y}) \braopket{\lambda}{J_z}{\nu}.
\end{split}
\label{eq:gm-simplify-h0-matrix-els}
\ee
With this we are now in the position to compute the missing matrix elements of $\mathbfcal{F}$.
One can find most of the computations of the matrix elements of $\mathbfcal{F}$ in the Appendix~\ref{app:matrix-elements-of-QFI}.
First of all, we compute $\mathbfcal{F}_{11}=\qfif{\rho, H_1, H_1}$ which turns to be equal to Eq.~\eqref{eq:gm-bound-for-insensitive-and-thermal-state} for obvious reasons,
\be
  \mathbfcal{F}_{11}=\sum_{n,m}^N \int x_n x_m \prob(\bs{x}) \,\text{d}^N\bs{x}\, \qfif{\rho^{(\text{s})}, j_z^{(n)}, j_z^{(m)}}.
  \label{eq:gm-f11-thermal}
\ee
Second, the most trivial matrix element is $\mathbfcal{F}_{00}$ which turns to depend only on the internal state $\rho^{(\text{s})}$,
\be
  \mathbfcal{F}_{00} = \qfif{\rho^{(\text{s})}, J_z}.
\ee
Finally, we compute both $\mathbfcal{F}_{01}$ and $\mathbfcal{F}_{10}$.
To compute them, note that the two matrix elements are equal $\mathbfcal{F}_{01}=\mathbfcal{F}_{01}$, due to the properties of Eq.~\eqref{eq:gm-FAB} for commuting $H_0$ and $H_1$.
Therefore, we have to compute only one of them
\be
  \mathbfcal{F}_{01} = \sum_{n=1}^N \int x_n \prob(\bs{x})  \,\text{d}^N\bs{x}\,
  \qfif{\rho^{(\text{s})},j_z^{(n)},J_z}.
\ee

With these results the bound for the precision for states sensitive to the homogeneous field which have the form of Eq.~\eqref{eq:gm-eigendecomposition-of-state} is
\be
\label{eq:gm-bound-for-sensitive-and-thermal-state}
\begin{split}
  \varinv{b_1}\leqslant&
  \sum_{n,m}^N \int x_n x_m \prob(\bs{x}) \,\text{d}^N\bs{x}\,
  \qfif{\rho^{(\text{s})}, j_z^{(n)}, j_z^{(m)}}
  - \frac{\left(\sum_{n=1}^N\int x_n \prob(\bs{x}) \,\text{d}^N\bs{x}\, \qfif{\rho^{(\text{s})}, j_z^{(n)},J_z}\right)^2}{\qfif{\rho^{(\text{s})}, J_z}}.
\end{split}
\ee

To simplify our calculations, we show that the bound Eq.~\eqref{eq:gm-bound-for-sensitive-and-thermal-state} is invariant under displacements of the system.
We use the linearity of the last two arguments of $\qfif{\rho, A, B}$, Eq.~\eqref{eq:gm-qfi-linear-in-arguments}, the fact that $H_0$ remains unchanged in the Heisenberg picture, and we also use the shifted $H_1(d)$ operator computed in Eq.~\eqref{eq:gm-shifted h1 generator}.
Hence, the translated QFI matrix elements are written as
\begin{subequations}
  \begin{align}
  \mathbfcal{F}_{00}(d)&=\qfif{\rho, H_0(d)} = \qfif{\rho, H_0},\\
  \mathbfcal{F}_{01}(d)&=\qfif{\rho, H_0(d), H_1(d)}\nonumber\\
    &= \qfif{\rho, H_0,H_1-dH_0} = \mathbfcal{F}_{01}-d\mathbfcal{F}_{00},\\
  \mathbfcal{F}_{11}(d)&=\qfif{\rho, H_1(d), H_1(d)}\nonumber\\
    &=\qfif{\rho, H_1-dH_0,H_1-dH_0}\nonumber\\
    &=\mathbfcal{F}_{11}-2d\mathbfcal{F}_{01}+d^2\mathbfcal{F}_{00}.
  \end{align}
\end{subequations}
Simple algebra shows that the bound for the precision of the estimation of the gradient remains constant under spatial translations,
\be
\begin{split}
  \varinv{b_1}_d\leqslant\,&\mathbfcal{F}_{11}(d)-\frac{(\mathbfcal{F}_{01}(d))^2}{\mathbfcal{F}_{00}(d)}\\
  =\,& \mathbfcal{F}_{11}-2d\mathbfcal{F}_{01}+d^2\mathbfcal{F}_{00}\\
  &-\frac{\mathbfcal{F}_{01}^2-2d \mathbfcal{F}_{01}\mathbfcal{F}_{00}+d^2\mathbfcal{F}_{00}^2}{\mathbfcal{F}_{00}}\\
  =\,& \mathbfcal{F}_{11}-\frac{\mathbfcal{F}_{01}^2}{\mathbfcal{F}_{00}}.
\end{split}
\ee

These observations make the computations of the precision bounds in the next sections easier, since now we can place the system arbitrarily wherever we choose.
It also allows us, for example, to place the origin of our coordinate system as well as the system itself where the magnetic field is zero.
So, we can write the linear magnetic field simply as $\bs{B}(x)= xB_1\bs{k}$ where $\bs{k}$ is the unitary vector pointing on the $z$-direction perpendicular to $x$- and $y$-axis.
The discourse we have had on the preceding section has a vital importance to understand properly our results.

\subsection{Testing our approach}
\label{sec:gm-io-chain-and-two-ensembles}

Despite the generality of the tools we developed in Section~\ref{sec:gm-cramer-rao-bounds}, it is always useful to start with simple but concise examples.
For this, we consider two of the most relevant cases for the external state $\rho^{(\text{x})}$ which we know that behave well for estimating the gradient parameter.
We will study the chain of atoms, where the atoms are placed one-by-one in a $1$-dimensional lattice with a constant separation $a$, and the two-ensembles of atoms, where half of the atoms are in $x=-a$ and the rest in $x=+a$.

\subsubsection{Distinguishable atoms in a 1D lattice}

\begin{figure}[htp]
  \begin{center}
    \igwithlabel{(a)}{scale=1.15}{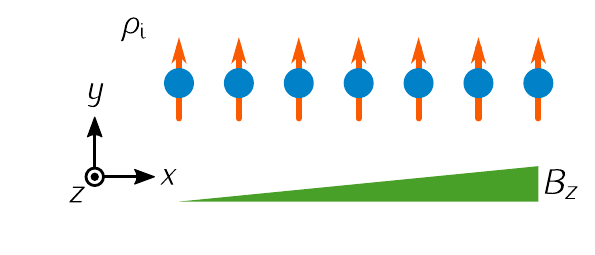}
    \igwithlabel{(b)}{scale=1.15}{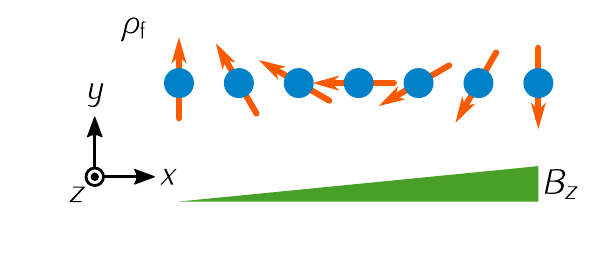}
    \caption[1-D chain of atoms polarized along $y$-axis under a gradient magnetic field]{
    (blue-circles) A system of $N$-atoms of spin-$j$ trapped in a chain configuration.
    (green-area) Magnetic field gradient.
    Note that the field is pointing outwards of the figures.
    (red-arrows) Spins of the particles initially all aligned.
    (a) The ensemble is initially totally polarized along a
    perpendicular direction, in this case $y$-direction, of the magnetic field $B_z$.
    The internal state can be written as $\ket{+j}_{y}^{\otimes N}$, the number represents $m_y$ the eigenvalue of the one particle operator $j_y^{(n)}$.
    (b) One can see how the gradient field affects with a varying field strength the different spins when they are placed in different positions. }
    \label{fig:ionchain-evolution}
  \end{center}
\end{figure}

As we have said, the first spatial state we consider will be given by $N$ particles all placed equidistantly from each other in a one-dimensional spin chain, i.e., a chain of atoms in a $1$-dimensional lattice
\cite{Altenburg2016},
see Figure~\ref{fig:ionchain-evolution}.
The probability density function describing the system is
\be
  \prob(\bs{x})=\prod_{n=1}^N \delta(x_n-na).
\ee

With this at hand we compute the single-point averages and the two-point averages corresponding to the ion-chain.
For the single-point average, one of the integrals appearing in Eq.~\eqref{eq:gm-bound-for-sensitive-and-thermal-state}, we have that
\be
  \int  x_n\prob(\bs{x})\,\text{d}^N\bs{x} = na,
\ee
and for the two-point correlation, which appears in Eqs.~\eqref{eq:gm-bound-for-insensitive-and-thermal-state} and \eqref{eq:gm-bound-for-sensitive-and-thermal-state}, we have the following for the case of the chain of atoms,
\be
  \int x_n x_m\prob(\bs{x})\,\text{d}^N\bs{x} = nma^2.
\ee

On the other hand, we use first a totally polarized state in the $y$-direction for the internal degrees of freedom, $\rho^{(\text{s})}=(\ketopbra{+j}{_y}{+j}_y)^{\otimes N}$ appeared in Eq.~\eqref{eq:bg-totally-polarized}.
This state is sensitive to the homogeneous field, so using that the state is pure and separable
\be
  \qfif{\ket{+j}_y^{\otimes}, j_z^{(n)},j_z^{(m)}}=4(\expect{j_z^{(n)},j_z^{(m)}}-\expect{j_z^{(n)}}\expect{j_z^{(m)}})=2j\delta_{n,m}.
\ee
Since this function is linear in the second and third arguments, we have that $\qfif{\ket{+j}_y^{\otimes}, j_z^{(n)},J_z}=2j$ and $\qfif{\ket{+j}_y^{\otimes}, J_z}=2jN$.
With this, we can now write the precision bound for a chain of atoms when the internal state is totally polarized along the $y$-axis as
\be
\begin{split}
  \varinv{b_1}_{\text{ch,tp}} & \leqslant a^2 \left\{ \sum_{n=1}^N n^2 2j - \frac{(\sum_{n=1}^N n2j)^2}{N2j}\right\}\\
  &=a^2\frac{N^2 -1}{12} 2jN .
\end{split}
\label{eq:gm-result-for-chain-without-sigma}
\ee

Despite that Eq.~(\ref{eq:gm-result-for-chain-without-sigma}) is a third order function of the particle number $N$ and that it seems to overcome the ultimate scaling Heisenberg scaling, note that the length of the chain increases as we introduce more particles into the system.
Hence, we have to normalize the bound with the effective size of the system, such that separable states would scale as shot-noise scaling.
This restores the ultimate threshold of the precision to $\sim N^2$ as usual.

In this section, we will use the standard deviation of the averaged particle position as the length measure of the system.
We also include in our next definitions the averaged correlation of two different particle positions, since it will appear in the following sections and for completeness.
They are computed as
\begin{subequations}
  \begin{align}
    \mu &= \int \frac{\sum_{n=1}^N x_n}{N} \prob(\bs{x}) \, \text{d}^N\bs{x},
    \label{eq:gm-mean} \\
    \sigma^2 &= \int \frac{\sum_{n=1}^N x_n^2}{N} \prob(\bs{x}) \, \text{d}^N\bs{x}  - \mu^2,
    \label{eq:gm-variance} \\
    \eta & =  \int \frac{\sum_{n\neq m} x_n x_m}{N(N-1)}\prob(\bs{x}) \, \text{d}^N\bs{x} - \mu^2,
    \label{eq:gm-correlation}
  \end{align}
\end{subequations}
where $\mu$ denotes the averaged position of the particles, $\sigma^2$ the variance of position of the particles and $\eta$ the position correlation between different particles.
So, the system effective width for the chain of atoms, computed by the variance Eq.~\eqref{eq:gm-variance}, is given as
\be
  \sigma_{\text{ch}}^2 = a^2\frac{N^2-1}{12}.
  \label{eq:gm-variance-chain}
\ee
It turns out that Eq.~\eqref{eq:gm-variance-chain} exactly coincides with one of the factors we have in Eq.~\eqref{eq:gm-result-for-chain-without-sigma}.
Substituting this into the Eq.~\eqref{eq:gm-result-for-chain-without-sigma} we have that,
for ion-chains where the particles are separated by a constant distance and where the spin-state $\rho^{(\text{s})}$ is the totally polarized state along the $y$-direction, the precision bound is given by
\be
  \varinv{b_1}_{\text{ch,tp}} \leqslant \sigma_{\text{ch}}^2 2jN,
\ee
in terms of $\sigma_{\text{ch}}$, the spin of each particle $j$ and the particle number $N$.

\subsubsection{Differential magnetometry with two ensembles}

We now turn our attention to the case of two  ensembles of distinguishable atoms.
Two ensembles of spin-$j$ atoms spatially separated from each other have been realized in cold gases (e.g., Ref.~\cite{Julsgaard2001}), and can be used for differential interferometry \cite{Eckert2006, Landini2014}.
We will also use an internal state such the maximal QFI is achieved so the reader gets familiar with our approach and sees how the best state to measure the gradient parameter looks like in our framework.

The spatial part is described by the following probability density function, where for an even number of particles half of the particles are at one position and the rest at another, both places at a distance of $a$ from the origin
\be
  \prob(\bs{x})=\prod_{n=1}^{N/2} \delta(x_n+ a)\prod_{n=N/2+1}^{N} \delta(x_n-a).
  \label{eq:gm-double-well-spatial-pdf}
\ee
This could be realized in a double-well trap, where the width of the wells is negligible compared to the distance of the wells.
Note that a state defined by Eq.~\eqref{eq:gm-double-well-spatial-pdf} and Eq.~\eqref{eq:gm-thermal-state} is a pure state in the position Hilbert space.
To distinguish the two wells we will use the labels "L" and "R" for the left and right wells respectively, which is a shorthand which collects the first half of the particles and the last half into a single index, respectively.
With this we are able to compute the single-point and two-point correlation functions as
\begin{subequations}
  \begin{align}
    \int  x_n\prob(\bs{x})\,\text{d}^N\bs{x} &= \lcor
    \begin{aligned}
      & {-a} && \text{if }  n\in \text{L},\\
      &  a   && \text{if }  n\in \text{R},
    \end{aligned}\right.
    \label{eq:single-point-function-double-well}\\
    \int x_n x_m\prob(\bs{x})\,\text{d}^N\bs{x} &= \lcor
    \begin{aligned}
      &{-a^2} && \text{if } (n,m)\in \text{(L,L) or (R,R),}\\
      & a^2   && \text{if } (n,m)\in \text{(L,R) or (L,R).}
    \end{aligned}
    \right.
    \label{eq:two-point-function-double-well}
  \end{align}
\end{subequations}

\begin{figure}[htp]
  \begin{center}
    \includegraphics[scale=1.15]{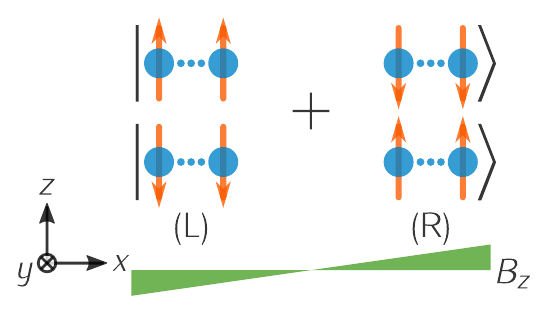}
    \caption[Best entangled state for the spatial two-ensembles]{
    (blue-circle) Atoms located at (L) or (R).
    (red-arrow) Spin state of each of the atoms.
    (green-area) Linear magnetic field $B_z$.
    Note that the $z$-axis is to represent the direction of the spins.
    On the other hand, the state is a linear superposition of the upper state and the lower state, represented by $\ket{\cdot}$ and $+$ sign.
    Note that all particles at (L) or (R) are assumed to be in the same spatial spot.}
    \label{fig:gm-double-well}
  \end{center}
\end{figure}

We will try states insensitive to the homogeneous fields.
Since the spatial part is a pure state in the position subspace, we will try pure states in the spin subspace due to the fact that the QFI is maximized by pure states as it was shown in Section~\ref{sec:bg-quantum-metrology}.
For pure states we have that the QFI is computed directly as four times the variance of the gradient Hamiltonian, $\qfif{\rho, H_1}=4\varian{H_1}$.
So, we just choose a spin-state that maximizes the variance of $H_1$ when the spatial part is Eq.~\eqref{eq:gm-double-well-spatial-pdf}.
Hence, the best state in this case is
\be
  \ket{\Psi} = \frac{1}{\sqrt{2}}(\ket{{+j}\cdots {+j}}^{(\text{L})}\ket{{-j}\cdots{-j}}^{(\text{R})}+\ket{{-j}\cdots {-j}}^{(\text{L})}\ket{{+j}\cdots{+j}}^{(\text{R})}),
  \label{eq:best-state}
\ee
where it can be seen as the superposition of the state with the smallest energy and the state with the greatest energy when the Hamiltonian is $H_1$, see Figure~\ref{fig:gm-double-well}.
The state $\ket{\Psi}$ is indeed insensitive to the homogeneous field since both states of the superposition in Eq.~\eqref{eq:best-state} are eigenstates of $H_0$ with the same eigenvalue.
Moreover, the state is also maximally entangled.

We compute first the $\qfif{\rho^{(\text{s})},j_z^{(n)},j_z^{(m)}}$ for $\ket{\Psi}$ as
\be
  \qfif{\ket{\Psi},j_z^{(n)},j_z^{(m)}}=\lcor
  \begin{aligned}
    & 4j^2  && \text{if } (n,m)\in\text{(L,L) or (R,R)},\\
    & {-4j^2} && \text{if } (n,m)\in\text{(L,R) or (R,L)}.
  \end{aligned}
  \right.
\ee
Now, if we separate the terms of the sum in Eq.~\eqref{eq:gm-bound-for-insensitive-and-thermal-state} into two groups, such that one of the sums is for indexes $(n,m)\in\text{(L,L) or (R,R)}$, and the other is for indexes $(n,m)\in\text{(L,R) or (R,L)}$, we can compute the bound for the best state for the two ensemble case as
\be
\begin{split}
  \varian{b_1}|_{\max} &=
  \sum_{\ver{(n,m)\in}{ \text{(L,L) or (R,R)}}} a^2 4j^2 + \sum_{\ver{(n,m)\in}{ \text{(L,R) or (R,L)}}} -a^2 (-4j^2)\\
  &= 4 a^2 N^2j^2.
\end{split}
\ee

On the other hand, if we compute now the standard deviation Eq.~\eqref{eq:gm-variance}, as we did before for the case of the chain Eq.~\eqref{eq:gm-variance-chain},
we have that for the two ensembles case $\mu=0$ and the standard deviation for the spatial state is
\be
  \sigma_{\text{te}}^2 = a^2,
  \label{eq:gm-variance-two-ensembles}
\ee
with which
\be
  \varian{b_1}_{\text{HL}}|_{\max} = 4 \sigma_{\text{te}}^2 N^2j^2.
  \label{eq:gm-hl-twoens}
\ee
For a general $\sigma^2$, the Eq.~\eqref{eq:gm-hl-twoens} can be seen as the Heisenberg limit for gradient metrology.
Note that the state is insensitive to the homogeneous field, hence this bound is saturable by some measurement, and that the state $\ket{\Psi}$ maximizes the variance of $H_1$ for any given $\sigma^2$.
Before concluding, we want to show another more usual approach to the same problem.

Knowing that the QFI is convex in the state and considering the spatial state to be Eq.~\eqref{eq:gm-double-well-spatial-pdf}, we reduce our problem to the internal subspace in which the state that maximizes the QFI is the one that maximizes $(\Delta H_1)^2$.
In this case, taking into account the particle locations are given and that we have zero magnetic field at the origin, we obtain
\be
  H_{1,\text{eff}}^{(\text{s})} = a(\mtxid^{(\text{L})}\otimes J_{z}^{(\text{R})}-J_{z}^{(\text{L})}\otimes \mtxid^{(\text{R})}),
  \label{eq:gm-effective-hamiltonian-double-well}
\ee
where we write the effective Hamiltonian that the particles on the left and right feel.
This proves that we have used the right state, since it maximizes the variance $\varian{H_{1,\text{eff}}^{(\text{s})}}$ \cite{Landini2014}.

\subsubsubsection{States separable into $\ket{\psi}^{({\rm L})}\otimes\ket{\psi}^{({\rm R})}$}

Now that we have already introduced the reader to the case of the two ensembles, we take the opportunity to show some more important results for states of the form of $\ket{\psi}^{(\text{L})} \otimes \ket{\psi}^{(\text{R})}$.
These states can reach the Heisenberg limit, while they are easier to realize experimentally than states in which the particles on the left and particles on the right are entangled with each other.

First, we will compute the bound for states insensitive to the homogeneous field.
For such states we only have to compute the QFI for $H_{1,\text{eff}}$ Eq.~\eqref{eq:gm-effective-hamiltonian-double-well} such that
\be
  \varinv{b_1}|_{\max} = \qfif{\ket{\psi}^{(\text{L})}\otimes\ket{\psi}^{(\text{R})},a( \mtxid^{(\text{L})}\otimes J_{z}^{(\text{R})}-J_{z}^{(\text{R})}\otimes \mtxid^{(\text{L})})}= 2a^2\qfif{\ket{\Psi}^{(\text{L})},J_z^{(\text{L})}},
  \label{eq:gm-bound-insensitive-twoens-simplified}
\ee
where we used the general rule Eq.~\eqref{eq:bg-qfi-additive-for-tensor-product} and that any scalar multiplying the second argument of the QFI can be extracted outside the function squared.

Now, we analyze how the bound would look like for states sensitive to the homogeneous field.
Note that the single-point correlation function for particles at "(L)" and "(R)" is $a$ and $-a$ respectively, and the two-point correlation function is $a^2$ for both. Thus, in the case of computing the bound for the states sensitive to the homogeneous fields, we have that $F_{01}^{(\text{L})} = -F_{01}^{(\text{R})}$, which we used the superscript to indicate in this case over which subspace is computed the QFI matrix element, whereas the other two matrix elements we have to compute are equal for both subspaces "(L)" and "(R)".
The precision bound for states sensitive to the homogeneous fields is obtained as
\be
\begin{split}
\varinv{b_1} \leqslant &  F_{11}+\frac{(F_{01})^2}{F_{00}}\\
 =& F_{11}^{(\text{L})}+F_{11}^{(\text{R})} +\frac{(F_{01}^{(\text{L})}+F_{01}^{(\text{R})})^2} {F_{00}^{(\text{L})}+F_{00}^{(\text{R})}}\\
 =& 2F_{11}^{(\text{L})} +\frac{(F_{01}^{(\text{L})}-F_{01}^{(\text{L})})^2} {2F_{00}^{(\text{L})}}\\
 =& 2F_{11}^{(\text{L})}\\
 =& 2 a^2 \qfif{\rho^{(L)}, J_z^{(L)}},
\end{split}
\label{eq:gm-bound-sensitive-twoens-simplified}
\ee
where we use in the first line the identities for additions under tensor products Eqs.~\eqref{eq:bg-qfi-additive-for-tensor-product} and \eqref{eq:gm-FAB-additive-under-tensor}, and in the last line we extract the common factor $a^2$ and we use the linearity on the arguments the QFI.
Note that this is the same precision bound we will obtain for states insensitive to the homogeneous fields.
Note that this bound relates how good the state on the "(L)" or "(R)" subsystem is in sensing the homogeneous field with the precision achievable for the gradient parameter.
This is reasonable because the state in "(L)" is not interacting neither correlated with "(R)".
Hence, after the homogeneous field is estimated for "(L)" and "(R)" independently,
the gradient can also be estimated as the difference between the two estimates divided by the square distance $a^2$.

In the literature one can find several states that can be used to measure a homogeneous field, such as the GHZ states \cite{Greenberger1989}, unpolarized Dicke states, and spin-squeezed states.
Note that if $\ket{\Psi}$ is separable, then based in Eq.~\eqref{eq:bg-shot-noise-limit} and for any of the two bounds Eqs.~\eqref{eq:gm-bound-insensitive-twoens-simplified} and \eqref{eq:gm-bound-sensitive-twoens-simplified}, we have
\be
  \qfif{\ket{\Psi}_{\text{sep}}^{(\text{L})}\ket{\Psi}_{\text{sep}}^{(\text{R})},\,a(J_z^{(\text{L})} \mtxid^{(\text{R})}-\mtxid^{(\text{L})} J_z^{(\text{R})})}
  = 2a^2\qfif{\ket{\Psi}_{\text{sep}}^{(\text{L})}, J_z^{(\text{L})}}
  = 2a^24\frac{N}{2}j^2
  \leqslant 4a^2Nj^2.
  \label{eq:gm-snl-two-ensembles}
\ee
Note that each of the ensembles has half of the total particle number $N$.
Eq.~\eqref{eq:gm-snl-two-ensembles} can be seen as the shot-noise limit when two ensembles are used for gradient metrology.
In Table~\ref{tab:result-states-two-ensembles}, we summarized the precision bounds for states of type $\ket{\Psi}^{(\text{L})}\otimes \ket{\Psi}^{(\text{R})}$ for the two-ensemble case.
\begin{table}
  \begin{center}
    \begin{tabular}{|l|c|c|}
    \hline
    States in (L) and (R) & $\qfif{\rho,J_z}$ for $N/2$ & $(\Delta b_1)^{-2}\leqslant$ \\
    \hline
    $\ket{+j}_l^{\otimes N/2} \otimes \ket{+j}_l^{\otimes N/2} $ & $Nj$ & $ 2a^2Nj$ \\
    \hline
    $\ket{\ghz}\otimes\ket{\ghz}$ & $N^2/4$ & $ a^2N^2/2$\\
    \hline
    $\ket{\dicke{{N/2}}}_{x}\otimes \ket{\dicke{{N/2}}}_{x}$ & $N(N+4)/8$ & $ a^2N(N+4)/4$\\
      \hline
    $\ket{\Psi}_{\text{sep}}\otimes\ket{\Psi}_{\text{sep}}$ & $2Nj^2$  & $ 4a^2Nj^2$ \\
    \hline
    \end{tabular}
  \end{center}
\caption[Bounds on the precision for different states $\ket{\psi}^{(\text{L})}\otimes\ket{\psi}^{(\text{R})}$]{
(first column) The complete state as a tensor product of the state in (L) and the state in (R). Note that for the GHZ state and the unpolarized Dicke state, the spin is $j=\frac{1}{2}$.
The last state $\ket{\Psi}$ is the best separable state for the estimation of the homogeneous field.
Hence, the bound for $\ket{\Psi}$ coincides with the shot-noise limit for gradient metrology with two ensembles.
(second column) Precision of the estimation of the homogeneous magnetic field in one of the ensembles.
(third column) From the second column and based on Eqs.~\eqref{eq:gm-bound-insensitive-twoens-simplified} and \eqref{eq:gm-bound-sensitive-twoens-simplified}, we compute the precision for differential magnetometry for various product quantum states in two ensembles spatially separated from each other by a distance $a$.
Note that all states are sensitive to the homogeneous field so the saturability of of the bound is not ensured.
This is the reason we use "$\leqslant$" instead of "$|_{\max}$".
}
\label{tab:result-states-two-ensembles}
\end{table}

In this section we have shown to the reader how one should handle the spatial width of the system for classifying it for gradient metrology as well as a state-of-the-art system in which the Heisenberg limit is achieved.
Moreover, we have shown how to use the tools developed in the previous section to compute simple bounds.
In the next section we will focus on single cold-atom ensembles since they play an important role in quantum technology, and many groups are trying to realize them whit great success but with few theoretical support.

\subsection{Magnetometry with a single atomic ensemble}
\label{sec:gm-single-cloud}

\begin{figure}[htp]
  \begin{center}
  \includegraphics[scale=1.15]{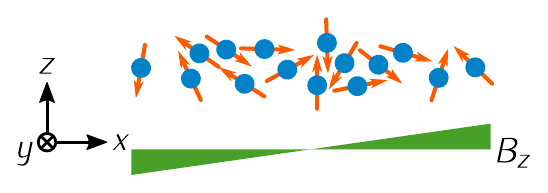}
  \caption[Single atomic ensemble for gradient magnetometry]{
  (blue-point) Atomic ensemble with their spins (red-arrow) pointing randomly in any direction coupled with a linear magnetic field $B_z$ (green).
  The spatial state $\rho^{(\text{x})}$ is assumed to be permutationally invariant.
  The ensemble is centered around the place at which the magnetic field is zero due to the invariance of the precision bound under translations of the system.}
  \label{fig:gm-single-ensemble-in-gradient}
  \end{center}
\end{figure}

In this section, we discuss magnetometry with a single
atomic ensemble in more detail.
We consider a one-dimensional ensemble of spin-$j$ atoms
placed in a one dimensional trap, which is elongated
in the  $x$-direction.
The magnetic field points in the $z$-direction,
and has a constant gradient along the $x$-direction.
The setup is depicted
in Fig.~\ref{fig:gm-single-ensemble-in-gradient}.
In the last part of this section, we calculate precision bounds for the
gradient estimation for some important multi-particle quantum states,
for instance, Dicke states or GHZ states.
Note that all these states are permutationally invariant, since we
assume that a permutationally invariant procedure prepared the states.

\subsubsection{Precision bound}

In an atomic ensemble of very many atoms, typically
the atoms cannot be individually addressed.
This can be taken into account, if we consider
quantum states that are permutationally invariant.
Hence, we will consider states for which both the internal state $\rho^{(\text{s})}$ and the probability distribution function
$\prob(\bs{x})$, appearing in Eq.~(\ref{eq:gm-thermal-state}),
are permutationally invariant.
The permutational invariance of $\prob(\bs{x})$ implies that
\be
  \label{eq:pi-for-pdf}
  \prob(\bs{x})=\tfrac{1}{N!}\sum_{k}\mathcal{P}_k [\prob(\bs{x})],
\ee
where the summation is over all the possible permutations of the variables $x_n$ denoted by $\mathcal{P}_k$.

As we have shown in Section~\ref{sec:gm-cramer-rao-bounds}, the precision bound is invariant under spatial translations.
This allows us to place the "center of mass" of the system at the origin of the coordinate system.
With this simplifying assumption and based on Eqs.~\eqref{eq:gm-mean}, \eqref{eq:gm-variance} and \eqref{eq:gm-correlation}, the single-point average appearing in Eq.~\eqref{eq:gm-bound-for-sensitive-and-thermal-state} is
\be
  \int x_n \prob(\bs{x}) \, \text{d}^N\bs{x} = \int \frac{\sum_{n=1}^N x_n}{N} \prob(\bs{x}) \, \text{d}^N\bs{x} = \mu = 0,
  \label{eq:gm-single-point-average-sinens}
\ee
where we used the permutational invariance of $\prob(\bs{x})$ substitute $x_n$ by the sum appearing in Eq.~\eqref{eq:gm-mean}.
In a similar way we obtain
\be
  \int x_n x_m \prob(\bs{x}) \, \text{d}^N\bs{x} = \lcor
  \begin{aligned}
    &\sigma^2         &&\text{for } n=m,\\
    &\frac{\eta}{N-1} &&\text{for } n\neq m,
  \end{aligned}\right.
  \label{eq:gm-two-point-correlation-sinens}
\ee
where we used that the system is placed at the origin $\mu=0$.
An interesting property of the covariance of this type is that its a value is bounded from below and from above by the variance itself and the particle number $N$ in the following way,
\be
  \frac{-\sigma^2}{N-1}\leq \eta\leq \sigma^2.
\ee
Note that in the first sum in Eq.~\eqref{eq:gm-bound-for-sensitive-and-thermal-state} there are in total $N(N-1)$ terms proportional to $\eta/(N-1)$ and $N$ terms proportional to $\sigma^2$.

From the linearity in the second and third arguments of $\qfif{\rho, A, B}$ Eq.~\eqref{eq:gm-FAB} and for states insensitive to the homogeneous field, where $\qfif{\rho, J_z}=0$, we have that
\be
  \sum_{n=1}^N \qfif{\rho, j_z^{(n)}} = -\sum_{n\neq m}^N \qfif{\rho,j_z^{(n)},j_z^{(m)}}.
  \label{eq:gm-qfi-identity-insensitive}
\ee

From the definition of the QFI for states insensitive to the homogeneous field, Eq.~\eqref{eq:gm-bound-for-insensitive-and-thermal-state}, we compute the bound for single ensembles as
\be
\begin{split}
  (\Delta b_1)^{-2}|_{\max} &= \sum_{n,m}^N \int x_n x_m \prob(\bs{x}) \, \text{d}^N\bs{x} \qfif{\rho^{(\text{s})}, j_z^{(n)}, j_z^{(m)}}\\
  &=\sum_{n=1}^N \sigma^2 \qfif{\rho,j_z^{(n)}} + \sum_{n\neq m}^N \eta \qfif{\rho,j_z^{(n)},j_z^{(m)}}.
\end{split}
\ee
Together with Eq.~\eqref{eq:gm-qfi-identity-insensitive} we write the precision bound for states insensitive to the homogeneous fields as
\be
(\Delta b_1)^{-2}|_{\max} = (\sigma^2-\eta) \sum_{n=1}^{N} \qfif{\rho^{(\text{s})},j_z^{(n)}}.
\label{eq:gm-bound-insensitive-sinens}
\ee
Note that the bound in Eq.~\eqref{eq:gm-bound-insensitive-sinens}
can be saturated by an optimal measurement.
Nevertheless, it cannot surpass the
shot-noise scaling, $\sim N$, because $\qfif{\rho^{(\text{s})},j_z^{(n)}}$, the QFI for the single-particle operator $j_z^{(n)}$, cannot be larger than $j^2$.

To compute the bound for states sensitive to the homogeneous field, note that in the second term appearing in Eq.~\eqref{eq:gm-bound-for-sensitive-and-thermal-state} is proportional to the single-point average Eq.~\eqref{eq:gm-single-point-average-sinens} which was chosen to be equal to zero.
Hence, we only have to compute the first term of the Eq.~\eqref{eq:gm-bound-for-sensitive-and-thermal-state} as
\be
\begin{split}
  \varinv{b_1}\leqslant& \sum_{n,m}^N \int x_n x_m \prob(\bs{x}) \, \text{d}^N\bs{x} \qfif{\rho^{(\text{s})}, j_z^{(n)}, j_z^{(m)}}\\
  =& \sum_{n=1}^N \sigma^2 \qfif{\rho,j_z^{(n)}} + \sum_{n\neq m}^N \eta \qfif{\rho,j_z^{(n)},j_z^{(m)}}\\
  =&(\sigma^2- \eta)\sum_{n=1}^N  \qfif{\rho,j_z^{(n)}} + \eta\sum_{n, m}^N  \qfif{\rho,j_z^{(n)},j_z^{(m)}},
\end{split}
\ee
where in the second line we compute the diagonal and the off-diagonal terms of the sum separately and in the last line we add $\eta\sum_{n=1}^N \qfif{\rho,j_z^{(n)}}$ to the last term and subtract it from the first term to make the expression more similar to Eq.~\eqref{eq:gm-bound-insensitive-sinens}.

Hence, for states sensitive to homogeneous fields,
the precision of estimating the gradient is bounded from above as
\be
\label{eq:gm-bound-sensitive-sinens}
\varinv{b_1} \leqslant (\sigma^2-\eta) \sum_{n=1}^N \qfif{\rho^{(\text{s})},j_z^{(n)}} + \eta \qfif{\rho^{(\text{s})},J_z}.
\ee
The second term on the right-hand side of Eq.~\eqref{eq:gm-bound-sensitive-sinens} is new in the sense that it did not appear in the bound for states insensitive to homogeneous fields.
Note that the bound in Eq.~(\ref{eq:gm-bound-sensitive-sinens}) is not necessarily saturable if the optimal measurements to estimate the gradient parameter and the homogeneous parameter do not commute with each other.
Note also that even if the first term cannot overcome the shot-noise scaling, in the second term the covariance is multiplied by QFI for estimating the homogeneous field and therefore this concrete term can make the bound, for extremely correlated particle positions, to scale as Heisenberg scaling.

\subsubsection{Precision bounds for different spin-states}

In this section, we present the precision limits for different classes of important quantum states such as the totally polarized state, the state having the largest precision among separable states, or the singlet state.
We will calculate the precision bounds presented before, Eqs.~\eqref{eq:gm-bound-insensitive-sinens} and \eqref{eq:gm-bound-sensitive-sinens}, for these systems.
We show first the results for singlets that are insensitive to homogeneous
fields.
In this case, the bounds can be achieved by choosing
the appropriate magnitude to measure.
The rest of the results are for states sensitive to homogeneous
fields which in general are not necessarily achievable bounds.

Before going into the details of our computations we present a summary of the results obtained in this section.
The summary for different states can be found in the Table~\ref{tab:compare all the states}.
\begin{table}
  \begin{center}
  \begin{tabular}{|l|c|}
    \hline
    States & $\varinv{b_1}$ \\
    \hline
    permutationally invariant singlet states & $ |_{\max} = (\sigma^2-\eta) \frac{4Nj(j+1)}{3}$ \\
    \hline
    $\ket{+j}_{y}^{\otimes N}$ & $\leqslant \sigma^2 2Nj$ \\
    \hline
    Best separable state $\ket{\Psi}$ & $\leqslant \sigma^2 4N j^2$ \\
    \hline
    $\ket{\dicke{N}}_{z}$ & $ |_{\max}=(\sigma^2-\eta) N$\\
    \hline
    $\ket{\dicke{N}}_{x}$ & $ \leqslant (\sigma^2 -\eta) N + \eta \frac{N(N+2)}{2}$ \\
    \hline
    $\ket{\ghz}$ & $\leqslant (\sigma^2 - \eta) N  + \eta N^2$ \\
    \hline
  \end{tabular}
  \end{center}
\caption[Bounds on the precision for different states for a single atomic ensemble.]{
Precision bounds for  differential magnetometry for various quantum states.
For the definition of the states, see the text.
If the bound are proved to be saturable then
the "$|_{\max}=$" subscript is used instead of an inequality.}
\label{tab:compare all the states}
\end{table}

\subsubsubsection{Permutationally invariant singlet states}

We consider now the singlet state, which is invariant under the influence
of a homogeneous field along any direction.
So, we have to compute the formula for the bound of the precision Eq.~(\ref{eq:gm-bound-insensitive-sinens}).
A pure singlet state is an eigenstate of the collective $J_z$ and $J^2$
operators, with an eigenvalue zero in both cases.
There are many different singlet states for an ensemble of $N$
spin-$j$ particles, which some of them are permutationally invariant.
Surprisingly the precision bound we compute is the same
for any permutationally invariant singlet.
Atomic ensembles in a singlet state have been experimentally
created with cold gases \cite{Toth2010, Behbood2014}.

In an $N$-particle system, there are several singlets pairwise orthogonal to each other.
The number of such singlets, $D_0$, depends on the particle spin $j$ and the number of particles $N$.

The most general singlet state can be written in the total angular momentum basis, using $D$ to label the degenerate states, see Appendix~\ref{app:angular-subspaces}.
In its eigenbasis the singlet is written as
\be
\rho^{(\text{s})}=\sum_{D=1}^{D_0}\lambda_D\ketbra{0,0,D}{0,0,D},
\label{eq:gm-general-singlet}
\ee
where $\sum_D \lambda_D=1$.
In its complete form the eigenvalues of the spin density matrix are $\lambda_{J,M_z,D}=\delta_{0,J}\lambda_D$.

Looking at Eq.~\eqref{eq:gm-bound-insensitive-sinens},
we must compute the QFI for the one-particle operator $j_z^{(n)}$ in order to compute the precision bound for permutationally invariant singlet states.
For that purpose we use the fact that when $j_z^{(n)}$ acts on a singlet state, it produces a state outside of the singlet subspace.
This can be proved by noting that
\be
  e^{i \pi J_x} j_z^{(n)} e^{-i \pi J_x}=-j_z^{(n)}
\ee
and that $e^{-i\pi J_x} \ket{0,0,D} = \ket{0,0,D}$
holds for any pure singlet state.
Hence, we can arbitrarily flip the sign of $j_z^{(n)}$ so
\be
  \bra{0,0,D}{j_z^{(n)}}\ket{0,0,D'}
  =-\bra{0,0,D}{j_z^{(n)}}\ket{0,0,D'},
\ee
which implies
\be
  \bra{0,0,D}{j_z^{(n)}}\ket{0,0,D'}=0,
  \label{eq:gm-expectation-jzn-for-singlets}
\ee
for any pair of pure singlet singlet states.

In order to compute the QFI for the singlet state we use
Eq.~\eqref{eq:gm-FAB-rewrite-with-trace}.
Hence, we can write the following for the second term of Eq.~(\ref{eq:gm-FAB-rewrite-with-trace}),
\be
  8\sum_{D,D'}
  \tfrac{\lambda_D\lambda_{D'}}
  {\lambda_D+\lambda_{D'}}
  |\bra{0,0,D}{j_z^{(n)}}\ket{0,0,D'}|^2=0.
\ee
It follows that the QFI of $j_z^{(n)}$ for any singlet is indeed simply
\be
  \label{eq:gm-qfi-as-trace-singlet}
  \qfif{\rho^{(\text{s})}, j_z^{(n)}}
  =4\tr({\rho^{(\text{s})} (j_z^{(n)})^2}).
\ee

Finally, we must compute the expectation value of the operator $(j_z^{(n)})^2$.
For that we have that
\be
  \tr(\rho^{(\text{s})}(j_k^{(n)})^2)
  =\tr(\rho^{(\text{s})}(j_l^{(n)})^2),
\ee
for any pair $k,l\in x,y,z$ due to the rotational invariance of the singlet, i.e, all the singlets remain invariant under a $SU(2)$ transformation of the kind $U=e^{i\phi J_{\bs{n}}}$, where $\bs{n}$ is an unitary vector belonging to the positional space.
Then we can write that
\be
\expect{(j_x^{(n)})^2+(j_y^{(n)})^2+(j_z^{(n)})^2}=j(j+1),
\ee
for any state, since it represents the spin number of the particle, which is fixed.
Hence, the expectation value of $(j_z^{(n)})^2$ on the singlet is
\be
  \label{eq:gm-expectation-jzn2-for-singlets}
  \tr(\rho^{(\text{s})}(j_z^{(n)})^2)=\frac{j(j+1)}{3},
\ee
for all the singlets.
Inserting this into Eq.~\eqref{eq:gm-qfi-as-trace-singlet} and using Eq.~\eqref{eq:gm-bound-insensitive-sinens}, we
obtain
\be
  \label{eq:gm-precision-singlet}
  \varinv{b_1}_{\text{s}}|_{\max} =\lpar\sigma^2-\eta\rpar \frac{4Nj(j+1)}{3}.
\ee

To conclude, singlet states are insensitive to homogeneous magnetic fields, hence determining the gradient leads to a single-parameter estimation problem.
This implies that there is an optimal operator that saturates the precision bound given by Eq.~\eqref{eq:gm-precision-singlet}.
However, it is usually very hard to find this optimal measurement,
although a formal procedure for this exists \cite{Paris2009, Ragy2016}.
In Ref. \cite{Urizar-Lanz2013}, a particular set-up for determining the magnetic gradient with permutationally invariant singlet states was suggested by the measurement of the $J_x^2$ collective operator.
For this scenario the precision is given by the error propagation formula as
\be
\label{eq:gm- Jx2_acc}
(\Delta b_1)^{-2}
= \frac{|\partial_{b_1}\expect{J_x^2}|^2}{\expect{J_x^4}-\expect{J_x^2}^2}.
\ee

\subsubsubsection{Totally polarized state}

The totally polarized state can easily be prepared experimentally.
It has already been used for gradient magnetometry with a single atomic ensemble \cite{Koschorreck2011,Vengalattore2007}.
For the gradient measurement as for the measurement of the homogeneous field, the polarization must be perpendicular to the field we would like to measure in order to take advantage of the interaction between the particles and the field.
Here we chose as before the totally polarized state along the $y$-axis which is written as $\ket{j}_y^{\otimes N}$.
Note that this state is sensitive to the homogeneous field, hence, we must use the Eq.~(\ref{eq:gm-bound-sensitive-sinens}) to compute the bound.

For the pure states we have that $\qfif{\ket{\psi}, A} = 4(\Delta A)^2$.
Together with,
$(\Delta j_z^{(n)})^2=j/2$ and $(\Delta J_z)^2=Nj/2$, when the polarization is
perpendicular to the $z$-direction, the precision will be computed straightforward from Eq.~\eqref{eq:gm-bound-sensitive-sinens}.

Therefore, the Cram\'er-Rao bound fixes the highest value for the precision
of the totally polarized state as
\be
  \varinv{b_1}_{\text{TP}}\leqslant \sigma^2 2Nj  .
\ee
Note that the precision bound for the totally polarized state
is smaller than that of the optimal separable state we present later on.
We can see clearly that the precision scales as $\mathcal{O}(N)$.

Let us now see, which quantities have to be measured to estimate the field gradient with a totally polarized state.
The homogeneous field rotates all spins by the same angle, while the gradient rotates the spins at different positions by a different angle.
Due to that, the homogeneous field rotates the collective spin, but does not change its absolute value.
On the other hand, the field gradient decreases the absolute value of the spin, since it has been prepared to be maximal, which has been used in Ref.~\cite{Behbood2013} for gradient magnetometry, see Figure~\ref{fig:ionchain-evolution}.

\subsubsubsection{The best separable state}

We will now turn our attention to the precision bound for all separable spin states.
It is useful to obtain this value so we have a direct comparison on what the best classically achievable precision is.
It will turn out that for $j>\frac{1}{2},$ it is possible
to achieve a precision higher than with the fully polarized state.
One has to take into account that if the state is insensitive to the homogeneous field the bound can be saturated for sure, and if the state is sensitive to homogeneous fields, it would depend on the measurements compatibility and on the system as we discussed before.
From another point of view and instead of using the Eqs.~\eqref{eq:gm-bound-insensitive-sinens} and \eqref{eq:gm-bound-sensitive-sinens}, what we have is that the bound is the same $\qfif{\rho, H_1}$ for both cases.
Note that we can place the system at the point in which the magnetic field is zero without changing the result.
Thus, it is easy to argue that the precision bound itself is a convex function of the states.
Moreover, it is also a convex function of the states when the external state $\rho^{(\text{x})}$ is fixed and only the internal $\rho^{(\text{s})}$ is considered.

In the single ensemble configuration, Eq.~\eqref{eq:gm-f11-thermal} must be computed only, where the two-point correlation function returns $\sigma^2$ or $\eta$ based on Eq.~\eqref{eq:gm-two-point-correlation-sinens}.
On the other hand, for pure states we have that $\qfif{\rho^{(\text{s})}, j_z^{(n)}, j_z^{(m)}}$ is four times the correlation $\expect{j_z^{(n)}j_z^{(m)}}-\expect{j_z^{(n)}}\expect{j_z^{(m)}}$.
If the state is a product state, then we have  $\expect{j_z^{(n)}j_z^{(m)}}-\expect{j_z^{(n)}}\expect{j_z^{(m)}}=0$ for all $n\neq m$.
Hence, $\eta$ does not play any role in the precision.
Finally, we have to maximize only the variance of each of the single-particle operators $4(\Delta j_z^{(n)})^2$.
From the definition of the variance,
\be
  \varian{j_z^{(n)}} = \expect{(j_z^{(n)})^2}-\expect{j_z^{(n)}}^2.
\ee
Hence, We try a state that approaches to zero its polarization on the $z$-direction and maximizes \expect{(j_z^{(n)})^2}.

We have that  $\ket{\Psi}=(\ket{+j}+\ket{-j})/\sqrt{2}$ is ideal for this, for any $j$.
Hence, we write the entire internal state as $\rho^{(\text{s})} =(\ketbra{\Psi}{\Psi})^{\otimes N}$.
This state gives $(\Delta j_z^{(n)})^2=j^2$ which can be used in Eq.~\eqref{eq:gm-f11-thermal} after multiplying by four.
Note that this state is permutationally invariant, hence we have finished the search for the best separable permutationally invariant state.
Moreover, the state is sensitive to the homogeneous field.

Finally, the best achievable precision for separable states is written as
\be
  \varinv{b_1}_{\text{SNL}} \leqslant  \sigma^2 4 N j^2,
  \label{eq:gm-best_separable}
\ee
where the state itself is sensitive to homogeneous fields and the shot-noise limit is achieved.
Note that in the two ensembles case $\sigma^2_{\text{te}}=a^2$ which tells us that both bounds Eqs. \eqref{eq:gm-snl-two-ensembles} and \eqref{eq:gm-best_separable} are equal.
This bound coincides with the totally polarized state studied before when the spin number $j=\frac{1}{2}$.

In the following we try to find a better precision bound making use of the presumably better entangled states.
Note that the bound for the singlet state, even if it is entangled, is above the bound for the totally polarized state but below of the bound defined for the best separable state.
Nevertheless, when the singlet state is used the effect of the homogeneous magnetic field has not to be compensated since the state is insensitive to it and thus the bound can be saturated with an optimal estimator for the gradient field.
\pagebreak

\subsubsubsection{The unpolarized Dicke states $\ket{\dicke{N}}_z$ and $\ket{\dicke{N}}_x$}

Unpolarized Dicke states play an important role in quantum optics and quantum information science.
The unpolarized Dicke state $\ket{\dicke{N}}_l$ with a maximal $\expect{J_x^2+J_y^2+J_z^2}=\mathcal{J}_{N/2}$, defined in Eq.~\eqref{eq:app-maximum-total-angular-momentum}, and $\langle J_l\rangle=0$ for any $l\in x,y,z$ is particularly interesting due to its entanglement properties and its metrological usefulness.
This state has been created in photonic experiments \cite{Kiesel2007,Wieczorek2009,Chiuri2012} and in cold atoms \cite{Luecke2011,Hamley2012}, while a Dicke state with $\langle J_z\rangle>0$ has been created with cold trapped ions \cite{haeffner2005}.

The Dicke state $\ket{\dicke{N}}_z$ is an eigenstate of $J_z$ so it is insensitive to homogeneous magnetic field pointing into the $z$-direction, thus the precision can be saturated by some measurement.
In the following, $\ket{\dicke{N}}_z$ without the subscript $z$ refers to $\ket{\dicke{N}}_z$.
On the other hand, the Dicke state $\ket{\dicke{N}}_x$ is sensitive to the homogeneous field.
Moreover it is very useful for homogeneous magnetometry as it has been shown in Ref.~\cite{Holland1993}.
Here we consider large particle numbers, to make the results simpler.

Since both Dicke states are pure, and following the procedure we used in previous sections, we have that to compute all the $\qfif{\rho^{(s)},j_z^{(n)}}=4(\expect{j_z^{(n)}j_z^{(m)}}-\expect{j_z^{(n)}}\expect{j_z^{(m)}})$ and $\qfif{\rho^{(s)},J_z}$ appearing in Eqs.~\eqref{eq:gm-bound-insensitive-sinens} and \eqref{eq:gm-bound-sensitive-sinens}.
Since both states are unpolarized and permutationally invariant, we have that $\expect{J_z}=0$ and $\expect{j_z^{(n)}}=0$ for both cases.
Therefore, we only need to compute the second moments to compute the needed variances.

To distinguish between the to cases, $\ket{\dicke{N}}$ and $\ket{\dicke{N}}_x$, we will denote their expectation values by $\expect{\cdot}_{\dicke{}}$ and $\expect{\cdot}_{\dicke{},x}$, respectively.

First of all, from the definition of the Dicke states we have that
\be
  \expect{J_x^2+J_y^2+J_z^2} = \mathcal{J}_{N/2} = \frac{N}{2}\lpar\frac{N}{2}+1\rpar,
  \label{eq:gm-square-total-angular-momentum-dicke}
\ee
for both cases.
Moreover, $\expect{J_l^2}=0$ holds for $\ket{\dicke{N}}_l$.
The other two second moments of Eq.~\eqref{eq:gm-square-total-angular-momentum-dicke} are equal to the invariance of the states under rotations around the $l$-axis.
Hence, we can write that
\begin{subequations}
  \begin{align}
    \expect{J_z^2}_{\dicke{}} &= 0,\\
    \expect{J_z^2}_{\dicke{},x} &= \frac{\mathcal{J}_{N/2}}{2}.
  \end{align}
  \label{eq:gm-expect-jz2-both-dicke}
\end{subequations}

For the single spin components
\be
  \expect{(j_x^{(n)})^2+(j_y^{(n)})^2+(j_z^{(n)})^2}=\mathcal{J}_{1/2}
\ee
holds.
Invoking the rotational symmetry again and that $\expect{J_l^2}=\sum_{n,m}^N \expect{j_l^{(n)}j_l^{(m)}}$, we arrive at
\begin{subequations}
  \begin{align}
    \expect{(j_z^{(n)})^2}_{\dicke{}}   &= \frac{1}{4},\\
    \expect{(j_z^{(n)})^2}_{\dicke{},x} &= \frac{1}{4},
  \end{align}
  \label{eq:gm-expect-jzn2-both-dicke}
\end{subequations}
after solving a system of linear equations.

Substituting Eqs.~\eqref{eq:gm-expect-jz2-both-dicke} and \eqref{eq:gm-expect-jzn2-both-dicke} into Eqs.~\eqref{eq:gm-bound-insensitive-sinens} and \eqref{eq:gm-bound-sensitive-sinens}, the bounds for unpolarized Dicke states insensitive to the homogeneous field and sensitive to the homogeneous field are
\begin{subequations}
  \begin{align}
    \varinv{b_1}_{\dicke{}}|_{\max} &= (\sigma^2 -\eta)N,
    \label{eq:gm-bound-dicke-insensitive} \\
    \varinv{b_1}_{\dicke{},x} &\leqslant (\sigma^2 -\eta)N + \eta \frac{N(N+2)}{2},
    \label{eq:gm-bound-dicke-sensitive}
  \end{align}
\end{subequations}
where Eq.~\eqref{eq:gm-bound-dicke-sensitive} shows in principal a Heisenberg scaling behavior in the second term, whenever the particles are very correlated among each other in the position subspace.
This is due to the metrological enhancement of sensing the homogeneous field.
In the next section, we will see another example of a state useful for homogeneous field estimation that is also useful for gradient magnetometry.

\subsubsubsection{The GHZ state}

The Greenberger-Horne-Zeilinger (GHZ) states are also highly entangled and play an important role in quantum information theory \cite{Greenberger1990}.
They have been created experimentally in photonic systems \cite{Pan2000,Yao2012,Lu2007} and trapped ions \cite{Sackett2000,Monz2011}.

We invoke the definition of the GHZ states Eq.~\eqref{eq:lt-ghz-state} given as
\be
  \ket{\ghz} = \tfrac{1}{\sqrt{2}}(\ket{0\cdots0}+\ket{1\cdots1}),
  \label{eq:gm-ghz-state}
\ee
where $\ket{0}$ and $\ket{1}$ stands for particles with eigenvalue $-1/2$ and $+1/2$ respectively for the one-particle $j_z^{(n)}$ operator.
The state Eq.~\eqref{eq:gm-ghz-state} is very sensitive to the homogeneous field.

In order to calculate the bound explicitly, let us recall that for pure states the QFI is simplified to $\qfif{\rho,A}=4\varian{A}$ Eq.~\eqref{eq:bg-qfi-for-pure-states}.
Following the Eq.~\eqref{eq:gm-bound-sensitive-sinens}, for the GHZ state the expectation values of $j_z^{(n)}$ and $J_z$ are equal to zero, and $\expect{(j_z^{(n)})^2}=\frac{1}{4}$ and $\expect{J_z^2} = \frac{N^2}{4}$.
Hence, the variances of $j_z^{(n)}$ and $J_z$ can be computed.
Finally, we obtain the precision bound for gradient magnetometry for the GHZ state as
\be
\label{eq:gm-precision bound for ghz}
\varinv{b_1}_{\ghz} \leqslant (\sigma^2 - \eta) N  + \eta N^2.
\ee
This means that we can reach the Heisenberg-limit with such states, but only in
cases where $\eta$ is positive, i.e, that the particles stay spatially correlated.

In summary, we have considered the experimentally most relevant spatial distributions of particles, which could be used for gradient metrology.
We have have also applied our methods to calculate the quantum Fisher information for various spin states.
As  we have seen, in some cases the system overcomes the shot-noise limit, even when the spatial state is a single ensemble of atoms, which opens up the possibility of ultra-precise gradient magnetometry with a smaller experimental effort.

%% file: 06-conclusions.tex
\section{Conclusions}

\input{snp/singleLineWaterMark.tex}
\lettrine[lines=2, findent=3pt, nindent=0pt]{I}{n} this thesis we have presented some aspects of quantum metrology from three different perspectives.
Besides the introductory part, Chapters~\ref{sec:in} and \ref{sec:bg}, our main results can be found in Chapters~\ref{sec:vd}, \ref{sec:lt} and \ref{sec:gm}.
In Chapter~\ref{sec:vd}, we have developed the theory of quantum metrology for metrology with noisy Dicke states.
In Chapter~\ref{sec:lt}, we have presented a method for witnessing the QFI with expectation values of some general observables.
Finally in Chapter~\ref{sec:gm}, we have computed precision bounds for gradient magnetometry.

In Chapters~\ref{sec:vd} and \ref{sec:lt}, we were constructing bounds on the quantum Fisher information based on the expectation values of some observables of the initial state.
It turns out that to compute the quantum Fisher information is not a trivial task and there is not a measurement scheme to obtain it from the initial state apart from a complete tomography, which is very demanding for large system sizes.
Hence, some shortcuts to compute the bound of the QFI are necessary.

In Chapter~\ref{sec:vd}, we computed the precision bound for noisy unpolarized Dicke states based on some initial expectation values.
Moreover, we first reduced the number of expectation values needed to four.
More explicitly, we have to measure only the second and the fourth moments of the $y$-component and the $x$-component of the collective angular momentum in order to estimate the metrological usefulness of the system.
In practice, the fourth-order moments can also be well approximated by the second-order moments.

In Chapter~\ref{sec:lt}, we developed a method based on the Legendre transform.
Based on this method, we are able to obtain a tight lower bound on the quantum Fisher information as a function of a set of expectation values of the initial state.
Furthermore, we tested our approach on extensive experimental data of photonic and cold gas experiments, and demonstrated that it works even for the case of thousands particles.
In the future, it would be interesting to use our method to test the optimality of various recent formulas giving a lower bound on the quantum Fisher information \cite{Zhang2014, Oudot2015}, as well as to improve the lower bounds for spin-squeezed states and Dicke states allowing for the measurement of more observables than the ones used in this publication.

On the other hand, in Chapter~\ref{sec:gm}, we have investigated the precision limits for measuring the gradient of a magnetic field with atomic ensembles arranged in different geometries and initialized in different states.
In particular, we studied spin-chain configurations as well as the case of two atomic ensembles localized at two different positions, and also the experimentally relevant set-up of a single atomic ensemble with an arbitrary density profile of the atoms was considered.
We discussed the usefulness of various quantum states for measuring the field strength and the gradient.
Some quantum states, such as singlet states, are insensitive to the homogeneous field.
Using these states, it is possible to estimate the gradient and saturate be Cramér-Rao bound, while for states that are sensitive to the homogeneous magnetic field, compatible measurements are needed for this task.
For spin chains and the two-ensemble case, the precision of the estimation of the gradient can reach the Heisenberg limit.
For the single ensemble case, only if strong correlation between the particles is allowed can the shot-noise limit be surpassed and even the Heisenberg limit be achieved.
However, even if the Heisenberg limit is not reached, single-ensemble methods can have a huge practical advantage compared to methods based on two or more atomic ensembles, since using a single ensemble makes the experiment simpler and can also result in a better spatial resolution.

%% file: appendix.tex
\renewcommand\thesubsection{\Alph{subsection}}
\numberwithin{equation}{subsection}
\numberwithin{figure}{subsection}
\afterpage{\phantomsection}
\section*{Appendices}
\addcontentsline{toc}{section}{Appendices}

\subsection{Multiparticle angular momentum subspaces}
\label{app:angular-subspaces}

As we mentioned in the Section~\ref{sec:bg-the-quantu-state}, when dealing with many particle systems, the Hilbert space is represented by tensor a product of the subspaces with a fixed spin.
So, the final dimension is the product of all single-particle dimensions which lead to an exponentially large Hilbert space.
In order to simplify our calculations, it is worth to note that some interesting structures arise from this kind of tensor product construction.

Let us name some basic assumptions with which the problem of adding angular momentum subspaces can be simplified.
First of all, the single-particle Hilbert space must be discrete and finite, hence it can be represented by a $d$-level system or qudit, where $d$ is the dimension of the single-particle system.
When $d$ equals two, we have the well known 2-level system or qubit.
The basis of such systems is composed of $d$ different eigenstates of the spin operator $j_z^{(n)}$, where $n$ denotes the Hilbert space in which the operator is defined.
We also assume that all parties have the same dimension $d$, so the total dimension is $d^N$.
The spin $j$ is defined as $j:=(d-1)/2$.
It is usual to use the single-particle angular momentum projector operator in the $z$-direction to completely characterize the basis
\be
  j_z^{(n)}\ket{m} = m \ket{m},
\ee
where $m = -j,-j+1,\dots,+j-1,+j$ and this way the necessary $d$ different pure states are defined.

In quantum information, the two eigenstates $\ket{-1/2}$ and $\ket{+1/2}$ of the 2-level systems, or qubits, are usually identified with $\ket{0}$ and $\ket{1}$ respectively, since the qubit case is the most studied case in which the dichotomized representation of classical \emph{bit}s, ones and zeros, is directly related with.
For qudits, particles whit $d$ levels, one can map directly the $m=-j,-j+1,\dots,+j-1,+j$ to $\tilde{m}=0,1,\dots,d-1$ in a similar way as for qubits, where $\tilde{m}$ is usual label used in the quantum information framework, whereas the $m$ gives directly the eigenvalue of the state when $j_z^{(n)}$ is applied.

A usual approach is using the tensor products of the single-party basis states given as
\be
  \begin{split}
    & \ket{-j,-j,\dots,-j,-j},\\
    & \ket{-j,-j,\dots,-j,-j+1},\\
    & \vdots \\
    & \ket{-j,-j,\dots,-j+1,-j},\\
    & \ket{-j,-j,\dots,-j+1,-j+1},\\
    & \vdots \\
    & \ket{+j,+j,\dots,+j,+j},
  \end{split}
  \label{eq:app-eigenbasis-tensor-product}
\ee
as a basis for the whole Hilbert space.
Here we have used the notation $\ket{m_1,m_2,\dots,m_{N-1},m_N}\equiv \ket{m_1}\otimes\ket{m_2}\dots\otimes\ket{m_N}$.
This basis is yet an eigenbasis of $J_z = \sum_{n} j_z^{(n)}$.
On the other hand, it is not an eigenbasis of the total angular momentum $\bs{J}^2 = J_x^2+J_y^2+J_z^2$, neither all the eigenstates are permutationally invariant, which would be useful for dealing with symmetric subspaces.

We will explain shortly how to write a basis in which all basis states are eigenstates of the total angular momentum $\bs{J}^2$ as well as of the $J_z$ operator.
This is a usual procedure when adding angular momentum operators, see Ref.~\cite{Cohen-Tannoudji1977, Sakurai2010} for more details.
For that, we have the ladder operators $J_{\pm} := J_x \pm i J_y$  which increase or decrease the eigenvalue of the state for $J_z$ without changing the eigenvalue for $\bs{J}^2$.
Therefore, if we start from $\ket{-j,-j,\dots,-j}$, which is an eigenstate of $\bs{J}^2$ with the maximal eigenvalue $Nj(Nj+1)$, and we apply $J_+$, we obtain all the states belonging to that subspace in which $\bs{J}^2$ is maximal.
We use the following notation for the maximal eigenvalue of the $\bs{J}^2$ operator
\be
  \mathcal{J}_{Nj} \equiv Nj(Nj+1),
  \label{eq:app-maximum-total-angular-momentum}
\ee
since it appears many times throughout the thesis.
Then, we use orthogonal states and we keep doing this until we have all the subspaces characterized.

Hence, the eigenstates are characterized with only three simple numbers $\ket{J,M,D}$ instead of Eq.~\eqref{eq:app-eigenbasis-tensor-product}.
First of all, we have the total angular momentum number $J$, where $J=0,1,\dots,Nj$ for this particular case in which we are adding together $N$ spin-$j$ particles, and define the eigenvalue of the $\bs{J}$ operator as $J(J+1)$.
Then, we have the quantum number of the angular momentum projection into the $z$-direction, $M=-J,-J+1,\dots,+J-1,+J$, which corresponds to the eigenvalue of the $J_z$ operator.
And finally, the degeneracy number of the $J$ subspaces, $D=1,2,\dots,D_J$, that is always one for the $J=Nj$ subspace and for the rest it depends in general in the number of particles as well as in the spin-number $j$.

We now show the definition of some of the states most used in this thesis.
For the spin-$\frac{1}{2}$ particles, i.e., qubits, we can mention several important quantum states.
For instance, the symmetric states, i.e., the states that after interchanging any pair of particles remain the same, are all confined into the subspace where $J^2$ is maximal, and equals $Nj(Nj+1)$.
They can be constructed taking $N-n$ particles in the $\ket{-1/2}$ or, using another notation, in the $\ket{0}$ state, and $n$ in the $\ket{+1/2}$ or $\ket{1}$ state and symmetrizing them as
\be
  \label{eq:ap-symmetric-states-qubits}
  \ket{J=N/2, M} \equiv \binom{N}{N/2+M}^{-\frac{1}{2}} \sum_k \mathcal{P}_k(\ket{0}^{\otimes N-n}\otimes\ket{1}^{\otimes{n}}),
\ee
where the sum is over all possible different permutations of the state denoted by $\mathcal{P}_k$ and $M=N/2+n$.
Note that the Dicke states, are named after R. H. Dicke, who used them to explain the coherent superradiance effect \cite{Dicke1954}.
If we consider two modes corresponding to the two states, then Eq.~\eqref{eq:ap-symmetric-states-qubits} is equivalent to the twin Fock state for a constant particle number.
Therefore apart from Eq.~\eqref{eq:ap-symmetric-states-qubits}, we use the following notation for these states
\be
  \ket{\dicke{N, n}} \equiv \ket{J=N/2, M=N/2-n}=\binom{N}{n}^{-\frac{1}{2}} \sum_k \mathcal{P}_k(\ket{0}^{\otimes N-n}\otimes\ket{1}^{\otimes{n}}),
  \label{eq:app-definition-of-dicke-states}
\ee
where $n$ gives the number of particles that are in $\ket{0}$.
One particularly interesting case of these states is the unpolarized Dicke state
\be
  \ket{\dicke{N}}\equiv \ket{\dicke{N,N/2}},
\ee
since it appears many times in this thesis, we skip writing the second subscript for simplicity.

Finally, we present another interesting state.
It is the permutationally invariant singlet state for spin-$\frac{1}{2}$ particles.
This is a uniquely defined state that can be constructed in several different ways.
One can start by the product state of pairwise singlets $\ket{\Psi^{-}} = \frac{1}{\sqrt{2}}(\ket{01}-{10})$ and later impose the permutational invariance for the density matrix.
Or one can find the thermal ground states of the Hamiltonian $H=\bs{J}^2$.
Finally, it can be constructed as the completely mixed state of the subspace where $J=0$.
All these alternative constructions are collected in the following equation
\be
  \lpar\frac{N!}{2^{N/2}(N/2)!}\rpar^{-\frac{1}{2}} \sum_{k\in \sigma_{\text{s}}}\mathcal{P}_k(\ketbra{\Psi^-}{\Psi^-}^{\otimes{N/2}})
  \equiv
  \lim_{\beta\rightarrow\infty}\frac{\exp(-\bs{J}^2\beta)}{\tr(\exp(-\bs{J}^2\beta))}
  \equiv
  \frac{1}{D_0}\sum_{D=1}^{D_0}\ketbra{0,0,D}{0,0,D},
\ee
where $\beta$ is the inverse of the Boltzman constant $k_\text{B}$ times the temperature $T$, and $\sigma_{\text{s}}$ is the set of all possible unique permutations, in this case $\frac{N!}{2^{N/2}(N/2)!}$.

\subsection{Husimi $\mathcal{Q}$-representation and the Bloch sphere}
\label{app:husimi-representation}

To graphically represent states with an angular momentum larger than $J=\frac{1}{2}$, it is convenient to use the so-called Husimi $\mathcal{Q}$-representation on the Bloch-sphere.
In fact it is straightforward to represent in a 3D-sphere all the possible states as it is done for qubits.

The Husimi $\mathcal{Q}$-representation must be normalized to 1.
Hence,
\be
  \label{eq:ap2-husimi-integral-to-one}
  \int \mathcal{Q}_\rho(\Omega)\,\text{d}\Omega = 1,
\ee
where $\Omega$ represents the solid angle of the sphere, i.e., the function is a function of $\varphi$ and $\theta$, the azimuth angle and the polar angle respectivelly, and $\text{d}\Omega= \sin(\theta)\text{d}\varphi\text{d}\theta$.
We will use it to describe states belonging to the symmetric subspace or for states belonging to the maximum angular momentum subspace.
Therefore, the $\mathcal{Q}_\rho(\Omega)$ function will be proportional to the fidelities of totally polarized states that point to different directions represented by $\Omega$.

In the case of many qubits such totally polarized states can be written as
\be
   \ket{\Omega} \equiv \ket{N/2,N/2}_{\Omega},
\ee
where can be reformulated as the eigenstate with the maximum eigenvalue for $J_{\Omega}=\cos(\varphi)\sin(\theta) J_x + \sin(\varphi)\sin(\theta)\, J_y + \cos(\theta) J_z$ operator.
An alternative way to obtain such totally polarized states $\ket{\Omega}$ is to rotate a totally polarized state along the $z$-direction by $\theta$ angle along the $y$-axis and then applying a rotation of $\varphi$ angle along the $z$-axis.
Hence,
\be
  \begin{split}
    \ket{\Omega} & = e^{-i\varphi J_z} e^{-i\theta J_y} \ket{11\dots 1},\\
    J_{\Omega}\ket{\Omega} & = \frac{N}{2} \ket{\Omega}.
  \end{split}
\ee
We write the quasi-probability $\mathcal{Q}(\Omega)$ proportional to the fidelity with respect to $\ket{\Omega}$ of the state as
\be
  \mathcal{Q}_\rho(\Omega)\propto \tr(\rho \ketbra{\Omega}{\Omega}).
\ee
The normalization constant comes from Eq.~\eqref{eq:ap2-husimi-integral-to-one}.
To obtain is the totally mixed state in the symmetric subspace can be used for which $\tr(\frac{\mtxid}{N+1} \ketbra{\Omega}{\Omega})=\frac{1}{N+1}$.
Integrating Eq.~\eqref{eq:ap2-husimi-integral-to-one} we obtain the proportionality factor shown in the following equation
\be
  Q_\rho(\Omega)=\frac{1}{4\pi(N+1)}\tr(\rho \ketbra{\Omega}{\Omega}),
\ee
which must be true for $N$ qubits in the symmetric subspace.
Similar definitions could be obtained for different subspaces or even for different spin number of the constituents.

\subsection{\texorpdfstring{Calculation of $\expect{\{J_x^2,J_y^2\}+\{J_x,J_y\}^2}$}{Calculation of Eq.~\eqref{eq:vd-result-before-simp}}}
\label{app:simplification-of-4th-moments}

The expectation value appearing in Eq.~\eqref{eq:vd-result-before-simp} which we want to simplify has 6 different terms, all with two $J_x$ and another two $J_y$,
\be
  \expect{J_x^2J_y^2} + \expect{J_xJ_yJ_xJ_y} + \expect{J_xJ_y^2J_x}
  + \expect{J_yJ_x^2J_y} + \expect{J_yJ_xJ_yJ_x} + \expect{J_y^2J_x^2}.
\ee
We can immediately see that the third term is zero, since  $J_x\ket{\dicke{N,N/2}}_x=0$.

We use the commutation relations of the angular momentum operators $[J_k,J_l]=\epsilon_{klm} iJ_m$, where $\epsilon_{klm}$ is the Levi-Civita symbol, to rearrange all operators,
\begin{subequations}
\begin{align}
  \expect{J_x^2J_y^2} & = i\expect{J_xJ_zJ_y}+\expect{J_xJ_yJ_xJ_y},
  \label{eq:ap-simplification-1} \\
  \expect{J_xJ_yJ_xJ_y} & = i\expect{J_xJ_yJ_z}+\expect{J_xJ_y^2J_x},
  \label{eq:ap-simplification-2} \\
  \expect{J_xJ_y^2J_x} & = \expect{J_xJ_y^2J_x},
  \label{eq:ap-simplification-3} \\
  \expect{J_yJ_x^2J_y} & = i\expect{J_yJ_xJ_z} + \expect{J_yJ_xJ_yJ_x},
  \label{eq:ap-simplification-4} \\
  \expect{J_yJ_xJ_yJ_x} & = -i\expect{J_zJ_yJ_x} + \expect{J_xJ_y^2J_x},
  \label{eq:ap-simplification-5} \\
  \expect{J_y^2J_x^2} & = -i\expect{J_yJ_zJ_x} + \expect{J_yJ_xJ_yJ_x}.
  \label{eq:ap-simplification-6}
\end{align}
\end{subequations}
One may note that with those relations is enough to see that we have six $\expect{J_xJ_y^2J_x}$, for instance, Eq.~\eqref{eq:ap-simplification-1} is $i\expect{J_xJ_zJ_y}$ plus Eq.~\eqref{eq:ap-simplification-2}, which at the same time is $i\expect{J_xJ_yJ_z}$ plus Eq.~\eqref{eq:ap-simplification-3}.
So each equation has at the end one $\expect{J_xJ_y^2J_x}$ plus or minus some expectation value of the product of three operators.

For the three terms operators and again using the commutation relations we can further simplify this expression.
We transform each term such that a  $\expect{J_xJ_yJ_z}$ term appears in the expressions obtained, and we arrive at the following,
\begin{subequations}
\begin{align}
  i\expect{J_xJ_zJ_y} & = \expect{J_x^2}+i\expect{J_xJ_yJ_z},
  \label{eq:ap-last-terms-1} \\
  2i\expect{J_xJ_yJ_z} & = 2i\expect{J_xJ_yJ_z},
  \label{eq:ap-last-term-2} \\
  i\expect{J_yJ_xJ_z} & = \expect{J_z^2}+i\expect{J_xJ_yJ_z},\\
  -i\expect{J_yJ_zJ_x} & = \expect{J_y^2} - \expect{J_z} - i\expect{J_xJ_yJ_z}, \\
\begin{split}
  -3i\expect{J_zJ_yJ_x} & = -3\expect{J_x^2} - 3i \expect{J_yJ_zJ_x} \\
  & = -3\expect{J_x^2} + 3 \expect{J_y^2} - 3i\expect{J_yJ_xJ_z} \\
  & = -3\expect{J_x^2} + 3 \expect{J_y^2} - 3\expect{J_z^2}
   - 3i \expect{J_xJ_yJ_z}.
\end{split}
\end{align}
\end{subequations}
We now sum all the terms.
Several terms cancel each other and the six $\expect{J_xJ_y^2J_x}$ terms add up.
The resulting expression is
\be
   4\expect{J_y^2} - 3 \expect{J_z^2} - 2\expect{J_x^2} + 6\expect{J_xJ_y^2J_x}.
\ee

\subsection{Properties of the spin-squeezing Hamiltonian}
\label{app:spin-squeezing-hamiltonian}

In this section, we discuss how to obtain spin-squeezed states as the ground states of the spin-squeezing Hamiltonian \cite{Sorensen2001a}, which is defined as
\be
  H_{\lambda} = J_x^2 + \lambda J_y,
  \label{eq:app-spsq-h}
\ee
where $\lambda$ is a real number.

Since $H$ is permutationally invariant, if the ground state $\ket{\text{GS}_{\lambda}}$ is non-degenerate, then $\ket{\text{GS}_{\lambda}}$ is a symmetric state.
We can prove that by using the permutation operator $\Pi$ which permutes any two particles when acting on the state.
With the permutation operator, we can write that
\be
\begin{split}
  [H_\lambda, \Pi] \ket{\text{GS}_{\lambda}} &= (H_\lambda \Pi - \Pi H_\lambda ) \ket{\text{GS}_{\lambda}}, \\
  &= H_\lambda \Pi \ket{\text{GS}_{\lambda}} - \Pi E_{0,\lambda}, \ket{\text{GS}_{\lambda}} \\
  &= (H_\lambda - E_{0,\lambda})\Pi\ket{\text{GS}_{\lambda}},
  \label{eq:app-commutator-spsq-h}
\end{split}
\ee
where $E_{0,\lambda}$ is eigenvalue corresponding to the state $\ket{\text{GS}_{\lambda}}$.
Eq.~\eqref{eq:app-commutator-spsq-h} must be zero since $H_\lambda$ is permutationally invariant and we have that $[H_\lambda, \Pi]=0$.
Hence, $\Pi\ket{\text{GS}_{\lambda}}$ must be an eigenstate of $H_{\lambda}$ with eigenvalue $E_{0,\lambda}$.
It turns out that, $\Pi\ket{\text{GS}_{\lambda}}$ and $\ket{\text{GS}_{\lambda}^{\pm}}$ must be the same state for any permutation $\Pi$.
Hence, the ground states of $H_\lambda$ must be symmetric if they are non-degenerate.

We have to show now that for a given expectation value of $\expect{J_y}$ of the ground state $\ket{\text{GS}_{\lambda}}$, these states are the ones that minimize $\expect{J_x^2}$.
Hence, using the linearity of the expectation values, we have that for any state
\be
\begin{split}
  \expect{H_\lambda} &= \expect{J_x^2} + \lambda \expect{J_y},\\
  \expect{J_x^2} &= \expect{H_\lambda} - \lambda \expect{J_y}.
\end{split}
\ee
Then for a particular $\expect{J_y}$, if we want to minimize the value of $\expect{J_x^2}$, then we have to minimize $\expect{H_{\lambda}}$, which by definition is done by the ground states $\ket{\text{GS}_{\lambda}}$.
Hence, these states sit at the lower boundary of the set of the expectation values $(\expect{J_x^2},\expect{J_y})$, which appears in Figure~\ref{fig:lt-spsq2d-4}.
Based on similar arguments, states at the upper boundary are ground states of the Hamiltonian
\be
  H_{\lambda}^{-} = -J_x^2 + \lambda J_y.
  \label{eq:app-spsq-h-minus}
\ee

\subsection{Legendre transform in 1-dimension}
\label{app:legendre-transform}

\begin{figure}[htp]
  \centering
  \includegraphics[scale=.65]{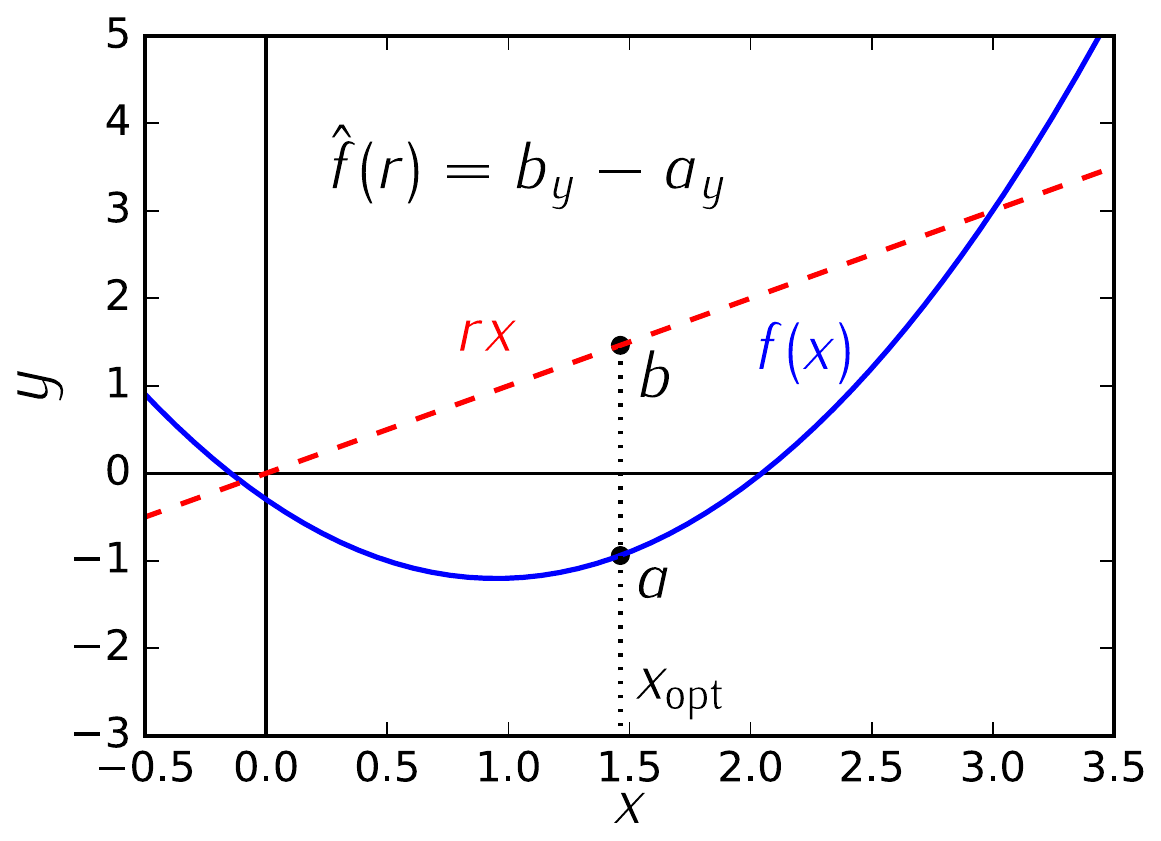}
  \caption[Graphical representation of the Legendre transform.]{
  Graphical representation of the Legendre transform.
(blue-line) Convex function, $f(x)=x^2-1.9x-0.3$, to be transformed.
(red-dashed) Line passing by the coordinate system origin, $rx$.
The Legendre transform is the maximal difference between $rx$ and $f(x)$ at the same $x$.
In this case, the vertical distance between $a$ and $b$.}
  \label{fig:lt-geometric-legendre}
\end{figure}

The Legendre transform of a convex function $f(x)$ is defined as the maximum distance between the line $y=rx$ and the function $f(x)$ for any $x$.
It can be written as follows,
\be
  \hat{f}(r):=\max_{x}\{rx-f(x)\},
\ee
where $\hat{f}(r)$ denotes the transformed function~\cite{Rockafellar1996}.
A geometric representation of the transform is given in Figure~\ref{fig:lt-geometric-legendre}.

The inverse transformation is simply obtained by applying again the same technique.
One fully recovers the
\be
  f(x) = \max_{r}\{rx-\hat{f}(r)\}.
\ee

Let us calculate the concrete example shown in the Figure~\ref{fig:lt-geometric-legendre}, where the function is $f(x)=x^2-1.9x-0.3$.
In this case the problem is well defined on the complete real axis.
Now, one has to find the maximum of $g(r,x)=rx-f(x)$ for all $r$.
The maximum is easily obtained in this particular case with usual techniques.
One has to solve for x the following equation $\partial_x g(r,x) = 0$. Thus, the maximum is at $x_{\text{opt}} = \frac{r+1.9}{2}$ and hence, the Legendre transform is
\be
  \hat{f}(r) = \frac{r^2}{4}+0.95r+1.2025.
\ee
If one applies again the transformation the resulting function is again the original one.

\subsection{\texorpdfstring{Calculation of $|\braopket{\dicke{N,m}}{_z}{\dicke{N,N/2}}_x|^2$}{Calculation of Eq.~\eqref{eq:lt-dicke-overlap}}}
\label{app:calculation-dicke-overlap}

To compute the Eq.~\eqref{eq:lt-dicke-overlap}, we use the Dicke states $\ket{\dicke{N,m}}_l$ defined as Eq.~\eqref{eq:app-definition-of-dicke-states}, see Appendix~\ref{app:angular-subspaces} for more details about Dicke states.
The Dicke states are multi-qubit states, hence we use $\ket{0}_l^{(n)}$'s and $\ket{1}_l^{(n)}$'s to describe them, where the superscript denotes to which particle the state belongs.

The scalar product of $\ket{0}_l^{(n)}$'s and $\ket{1}_{l'}^{(n')}$ is zero if $n\ne n'$.
For $n=n'$ and $l=l'$, these states form an orthonormal basis, i.e., $\braket{a}{b}=\delta_{a,b}$.
The rest of the possibilities are
\be
  \braopket{0}{_z}{1}_x = \braopket{1}{_z}{0}_x = \braopket{1}{_z}{1}_x = \braopket{1}{_z}{0}_y = \braopket{1}{_z}{1}_y = \frac{1}{\sqrt{2}},
  \label{eq:app-single-scalar-prod-xyz-j2-trivial}
\ee
and
\be
  \braopket{0}{_z}{0}_x = \frac{-1}{\sqrt{2}}, \quad
  \braopket{0}{_z}{0}_y = \frac{-i}{\sqrt{2}}, \quad
  \braopket{0}{_z}{1}_y = \frac{+i}{\sqrt{2}},
  \label{eq:app-single-scalar-prod-xyz-j2-special}
\ee
and all their complex conjugates.
On the other hand using the tensor product properties we have that
\be
  \braket{a_1,a_2,\dots,a_N}{b_1,b_2,\dots,b_N} = \braket{a_1}{b_1} \braket{a_2}{b_2}\dots \braket{a_N}{b_N}.
\ee

For the overlap of the different Dicke states, in this case for different $m$'s of $|\braopket{\dicke{N,m}}{_z}{\dicke{N,N/2}}_x|^2$, we have first that
\be
  \braopket{\dicke{N,m}}{_z}{\dicke{N,N/2}}_x =
  \binom{N}{m}^{-\frac{1}{2}}\binom{N}{N/2}^{-\frac{1}{2}}
  \sum_{k,k'\in \sigma_s}
  \mathcal{P}_k ( \bra{0}_z^{\otimes N-m}\bra{1}_z^{\otimes m} ) \,
  \mathcal{P}_{k'} ( \ket{0}_x^{\otimes N/2}\ket{1}_x^{\otimes N/2} ),
\ee
where $\mathcal{P}_k$ denotes one of the unique permutations of its argument.

From Eqs.~\eqref{eq:app-single-scalar-prod-xyz-j2-trivial} and \eqref{eq:app-single-scalar-prod-xyz-j2-special}, we see that all of the scalar products will be either $\frac{1}{\sqrt{2}}$ or $\frac{-1}{\sqrt{2}}$.
This way, we can take $(\frac{1}{\sqrt{2}})^N$ out of the sum, and use redefine states such that $\brawopketw{0}{_z}{0}{_x}=-1$ and $\brawopketw{0}{_z}{1}{_x}=\brawopketw{1}{_z}{0}{_x}=\brawopketw{1}{_z}{1}{_x}=+1$.
Moreover, note that for each permutation on $k'$, $k$ already permutes all of the possible pairings.
We can drop that sum then and substitute it by a constant which is all the possible distinguishable permutations of $N$ objects when $N/2$ are of one kind and $N/2$ of another, which turns to be $\binom{N}{N/2}$.

Hence, we get
\be
  \braopket{\dicke{N,m}}{_z}{\dicke{N,N/2}}_x =
  \frac{\binom{N}{m}^{-\frac{1}{2}}\binom{N}{N/2}^{\frac{1}{2}}}{\sqrt{2^N}}
  \sum_{k\in \sigma_s}
  \mathcal{P}_k ( \braw{0}_z^{\otimes N-m}\braw{1}_z^{\otimes m} ) \,
  \ketw{0}_x^{\otimes N/2}\ketw{1}_x^{\otimes N/2}.
  \label{eq:app-dicke-overlap-primitive}
\ee
The terms in the sum are either $+1$ or $-1$.
At this point, we have to solve this by using notions in permutation sets.
We have to assign $-1$ to the term in the sum whenever there are an odd number of $\braw{0}_z$'s in the first half particles of $\mathcal{P}_k (\braw{0}_z^{\otimes N-m}\braw{1}_z^{\otimes m})$.

For $0\leqslant m\leqslant N/2$, we can start having all the zeros in the first half of the particles, which are $\binom{N/2}{m}$ different possibilities with the same sign.
Then, putting one $\braw{0}$'s to the second half of the particle numbers, we have $\binom{N/2}{m-1}\binom{N/2}{1}$ possibilities where the sign of the terms flipped.
We have to do so until there are no more $\braw{0}$'s left in the first half of the particles, which clearly returns $+1$ for the summing term.
For the sum then, this yields
\be
  \sum_{k\in \sigma_s}
  \mathcal{P}_k ( \braw{0}_z^{\otimes N-m}\braw{1}_z^{\otimes m} ) \,
  \ketw{0}_x^{\otimes N/2}\ketw{1}_x^{\otimes N/2} = \sum_{i=0}^m (-1)^{i} \binom{N/2}{m-i}\binom{N/2}{i}.
  \label{eq:app-sum-of-dicke-overlap-m-small}
\ee
The key point now is to note that for $N/2 < m \leqslant N$ we can swap all the $\braw{0}_z$'s with $\braw{1}_z$'s, which yields the same combinations, but where we have to count the number of $\braw{1}_z$'s in the first half to determine the sign.
Hence, the following holds
\be
  \sum_{k\in \sigma_s}
  \mathcal{P}_k ( \braw{0}_z^{\otimes N-m}\braw{1}_z^{\otimes m} ) \,
  \ketw{0}_x^{\otimes N/2}\ketw{1}_x^{\otimes N/2} = \sum_{i=0}^{N-m} (-1)^{N/2-i} \binom{N/2}{N-m-i}\binom{N/2}{i}.
  \label{eq:app-sum-of-dicke-overlap-m-small-inverse}
\ee
Note that if $N/2$ is even, then Eq.~\eqref{eq:app-sum-of-dicke-overlap-m-small} is equal to Eq.~\eqref{eq:app-sum-of-dicke-overlap-m-small-inverse}, whereas if $N/2$ is odd, then Eq.~\eqref{eq:app-sum-of-dicke-overlap-m-small} is the negative of Eq.~\eqref{eq:app-sum-of-dicke-overlap-m-small-inverse}.
Since we are interested in the square of $\braopket{\dicke{N,m}}{_z}{\dicke{N,N/2}}_x$, we can compute it for $0\leqslant m \leqslant N/2$ at the same time we obtain the results for $N/2<m\leqslant N$.
This is represented by
\be
  |\braopket{\dicke{N,m}}{_z}{\dicke{N,N/2}}_x|^2 = |\braopket{\dicke{N,(N-m)}}{_z}{\dicke{N,N/2}}_x|^2.
\ee

Finally, using the binomial identity of Eq.~\eqref{eq:lt-binomial-identity} we have that Eq.~\eqref{eq:app-sum-of-dicke-overlap-m-small} yields
\be
\begin{split}
  \sum_{i=0}^m (-1)^{i} \binom{N/2}{m-i}\binom{N/2}{i} = \lcor
  \begin{aligned}
    &\binom{N/2}{m/2}(-1)^{m/2}&& \text{for even }m,\\
    &0 && \text{for odd }m.
  \end{aligned}
  \right.\\
\end{split}
\ee
Hence, for even $m$ the overlapping coefficient is
\be
  |\braopket{\dicke{N,m}}{_z}{\dicke{N,N/2}}_x|^2 = \frac{\binom{N/2}{m/2}^2\binom{N}{N/2}}{2^N\binom{N}{m}}.
  \label{eq:app-dicke-overlap}
\ee
If we substitute $m$ by $N-1$, using some binomial identities the formula remains the same, hence, the Eq.~\eqref{eq:app-dicke-overlap} is valid for any $m$.
Note that it also coincides with Eq.~\eqref{eq:lt-dicke-overlap}.

\subsection{Calculation of the QFI matrix elements}
\label{app:matrix-elements-of-QFI}

We start with states insensitive to the homogeneous fields.
Later, we will discuss states sensitive to it.
Next, we must compute the matrix elements of the quantum Fisher information defined for the generators $H_0$ and $H_1$.
We use the functional defined in Eq.~\eqref{eq:gm-FAB} to compute the matrix elements.
We also use thermal states with respect to the spatial degrees of freedom Eq.~\eqref{eq:gm-eigendecomposition-of-state}, since it is one of the most common situations in the experiments.
Moreover, we consider the eigen-decomposition of the state appearing in Eq.~\eqref{eq:gm-eigendecomposition-of-state}.

First of all, we compute the $\mathbfcal{F}_{11}\equiv\qfif{\rho, H_1, H_1}\equiv\qfif{\rho,H_1}$, since it is valid for states that are insensitive to the homogeneous field as well as for states that are sensitive to it.
We have the following for the QFI
\be
\begin{split}
  \qfif{\rho,H_1} &= 2\iint \sum_{\lambda,\nu}\frac{1}{\braket{\bs{x}}{\bs{x}}}
  \frac{(\prob(\bs{x})p_\lambda - \prob(\bs{y})p_\nu)^2}
  {\prob(\bs{x})p_\lambda + \prob(\bs{y})p_\nu}
  |(H_1)_{\bs{x},\lambda;\bs{y},\nu}|^2
  \,\text{d}^N\bs{x} \text{d}^N\bs{y}\\
  &= 2\iint \sum_{\lambda,\nu}\frac{1}{\braket{\bs{x}}{\bs{x}}}
  \frac{(\prob(\bs{x})p_\lambda - \prob(\bs{y})p_\nu)^2}
  {\prob(\bs{x})p_\lambda + \prob(\bs{y})p_\nu}
  (\delta(\bs{x}-\bs{y}))^2\sum_{n,m}^N x_n x_m
  \braopket{\lambda}{j_z^{(n)}}{\nu}\braopket{\nu}{j_z^{(m)}}{\lambda}
  \,\text{d}^N\bs{x} \text{d}^N\bs{y},
\end{split}
\ee
where we use $\braket{\bs{x}}{\bs{x}}=\braket{\bs{y}}{\bs{y}}$.
From the definition of the Dirac delta, we arrive at
\be
\begin{split}
  \qfif{\rho,H_1} &= 2\int \sum_{\lambda,\nu}\frac{\prob(\bs{x})}{\braket{\bs{x}}{\bs{x}}}
  \frac{(p_\lambda - p_\nu)^2}
  {p_\lambda + p_\nu}
  \delta(\bs{x}-\bs{x})\sum_{n,m}^N x_n x_m
  \braopket{\lambda}{j_z^{(n)}}{\nu}\braopket{\nu}{j_z^{(m)}}{\lambda}
  \,\text{d}^N\bs{x}\\
  &= 2\int \prob(\bs{x}) \sum_{\lambda,\nu}
  \frac{(p_\lambda - p_\nu)^2}{p_\lambda + p_\nu}
  \sum_{n,m}^N x_n x_m
  \braopket{\lambda}{j_z^{(n)}}{\nu}\braopket{\nu}{j_z^{(m)}}{\lambda}
  \,\text{d}^N\bs{x}\\
  &= \sum_{n,m}^N \int x_n x_m \prob(\bs{x}) \,\text{d}^N\bs{x}\,
  \qfif{\rho_{\text{s}},j_z^{(n)},j_z^{(m)}},
\end{split}
\label{eq:app-computing-f11}
\ee
where we used the Eq.~\eqref{eq:gm-simplify-h1-matrix-els} for the simplification of the matrix elements of $H_1$, where we simplified $\braket{\bs{x}}{\bs{x}}$ with $\delta(\bs{x}-\bs{x})$, and where in the third line we reconstructed $\qfif{\rho_{\text{s}},j_z^{(n)},j_z^{(m)}}$ using the factor $2$, the sum over $\lambda$ and $\nu$, and the matrix elements of $j_z^{(n)}$ and $j_z^{(m)}$ appearing in the third line.
We finally reordered all the terms in order to group what has to be integrated together between "$\int$" and "$\text{d}^N\bs{x}$", which in this case represents a two-point correlation function of $x_n$ and $x_m$ over the PDF $\prob(\bs{x})$.

We carry out a similar calculation for $\mathbfcal{F}_{01}$ and $\mathbfcal{F}_{00}$.
We use again a state of the form Eq.~\eqref{eq:gm-eigendecomposition-of-state} to compute these matrix elements of the QFI.
We also use the simplified expression for the matrix elements of $H_0$ for this case Eq.~\eqref{eq:gm-simplify-h0-matrix-els}.
For $\mathbfcal{F}_{01}$, the computation looks like
\be
\begin{split}
  \mathbfcal{F}_{01} &= 2\iint \sum_{\lambda,\nu}\frac{1}{\braket{\bs{x}}{\bs{x}}}
  \frac{(\prob(\bs{x})p_\lambda - \prob(\bs{y})p_\nu)^2}
  {\prob(\bs{x})p_\lambda + \prob(\bs{y})p_\nu}
  (H_0)_{\bs{x},\lambda;\bs{y},\nu}(H_1)_{\bs{y},\nu;\bs{x},\lambda}
  \,\text{d}^N\bs{x} \text{d}^N\bs{y}\\
  &= 2\iint \sum_{\lambda,\nu}\frac{1}{\braket{\bs{x}}{\bs{x}}}
  \frac{(\prob(\bs{x})p_\lambda - \prob(\bs{y})p_\nu)^2}
  {\prob(\bs{x})p_\lambda + \prob(\bs{y})p_\nu}
  (\delta(\bs{x}-\bs{y}))^2\sum_{n,m}^N x_m
  \braopket{\lambda}{j_z^{(n)}}{\nu}\braopket{\nu}{j_z^{(m)}}{\lambda}
  \,\text{d}^N\bs{x} \text{d}^N\bs{y}.
\end{split}
\ee
Using the definition of the Dirac delta inside one of the integrals we have that
\be
\begin{split}
  \mathbfcal{F}_{01} &= 2\int \sum_{\lambda,\nu}\frac{\prob(\bs{x})}{\braket{\bs{x}}{\bs{x}}}
  \frac{(p_\lambda - p_\nu)^2}
  {p_\lambda + p_\nu}
  \delta(\bs{x}-\bs{x})\sum_{n,m}^N x_m
  \braopket{\lambda}{j_z^{(n)}}{\nu}\braopket{\nu}{j_z^{(m)}}{\lambda}
  \,\text{d}^N\bs{x}\\
  &= 2\int \prob(\bs{x}) \sum_{\lambda,\nu}
  \frac{(p_\lambda - p_\nu)^2}{p_\lambda + p_\nu}
  \sum_{n,m}^N x_m
  \braopket{\lambda}{j_z^{(n)}}{\nu}\braopket{\nu}{j_z^{(m)}}{\lambda}
  \,\text{d}^N\bs{x},
\end{split}
\ee
which follows from the definition of Eq.~\eqref{eq:gm-FAB}
\be
\begin{split}
  \mathbfcal{F}_{01} &= \sum_{n,m}^N \int x_m \prob(\bs{x}) \,\text{d}^N\bs{x}\,
  \qfif{\rho_{\text{s}},j_z^{(n)},j_z^{(m)}}\\
  &= \sum_{n=1}^N \int x_n \prob(\bs{x}) \,\text{d}^N\bs{x}\,
  \qfif{\rho_{\text{s}},j_z^{(n)},J_z},
\end{split}
\label{eq:app-computing-f01}
\ee
where we have used similar arguments as when computing Eq.~\eqref{eq:app-computing-f11}.
We also use the linearity on the second and third arguments of the functional Eq.~\eqref{eq:gm-FAB} in the last line to remove the one of the summation indexes.
Note that the main difference with respect to $\mathbfcal{F}_{11}$ is that in this case the integral represents a single-point average instead of a two-point correlation function.
Finally, for the matrix element $\mathbfcal{F}_{00}$ we have that
\be
\begin{split}
  \mathbfcal{F}_{00} &= 2\iint \sum_{\lambda,\nu}\frac{1}{\braket{\bs{x}}{\bs{x}}}
  \frac{(\prob(\bs{x})p_\lambda - \prob(\bs{y})p_\nu)^2}
  {\prob(\bs{x})p_\lambda + \prob(\bs{y})p_\nu}
  |(H_0)_{\bs{x},\lambda;\bs{y},\nu}|^2
  \,\text{d}^N\bs{x} \text{d}^N\bs{y}\\
  &= 2\iint \sum_{\lambda,\nu}\frac{1}{\braket{\bs{x}}{\bs{x}}}
  \frac{(\prob(\bs{x})p_\lambda - \prob(\bs{y})p_\nu)^2}
  {\prob(\bs{x})p_\lambda + \prob(\bs{y})p_\nu}
  (\delta(\bs{x}-\bs{y}))^2\sum_{n,m}^N
  \braopket{\lambda}{j_z^{(n)}}{\nu}\braopket{\nu}{j_z^{(m)}}{\lambda}
  \,\text{d}^N\bs{x} \text{d}^N\bs{y}.
\end{split}
\ee
We use now the definition of the Dirac delta such that
\be
\begin{split}
  \mathbfcal{F}_{00} &= 2\int \sum_{\lambda,\nu}\frac{\prob(\bs{x})}{\braket{\bs{x}}{\bs{x}}}
  \frac{(p_\lambda - p_\nu)^2}
  {p_\lambda + p_\nu}
  \delta(\bs{x}-\bs{x})\sum_{n,m}^N
  \braopket{\lambda}{j_z^{(n)}}{\nu}\braopket{\nu}{j_z^{(m)}}{\lambda}
  \,\text{d}^N\bs{x}\\
  &= 2\int \prob(\bs{x}) \sum_{\lambda,\nu}
  \frac{(p_\lambda - p_\nu)^2}{p_\lambda + p_\nu}
  \sum_{n,m}^N
  \braopket{\lambda}{j_z^{(n)}}{\nu}\braopket{\nu}{j_z^{(m)}}{\lambda}
  \,\text{d}^N\bs{x}.
\end{split}
\ee
We apply the definition of Eq.~\eqref{eq:gm-FAB} and we reorder terms as
\be
\begin{split}
  \mathbfcal{F}_{00} &= \sum_{n,m}^N \int \prob(\bs{x}) \,\text{d}^N\bs{x}\,
  \qfif{\rho_{\text{s}},j_z^{(n)},j_z^{(m)}}\\
  &= \sum_{n,m}^N \qfif{\rho_{\text{s}},j_z^{(n)},j_z^{(m)}}\\
  &= \qfif{\rho_{\text{s}},J_z,J_z} = \qfif{\rho_{\text{s}},J_z},
\end{split}
\label{eq:app-computing-f00}
\ee
where we simplify the integral using the normalization of the PDF which is equal to $1$.
Note that we obtain the QFI for the homogeneous field as expected.